\documentclass[12pt,letterpaper]{article}%
\usepackage{amsmath}
\usepackage{rotating}
\usepackage{version}
\usepackage{amsfonts}
\usepackage{amssymb}
\usepackage{graphicx}
\usepackage{float}
\usepackage{url}
\usepackage[round,authoryear,compress]{natbib}
\usepackage{color}
\usepackage{xr}
\usepackage[normalem]{ulem}
\usepackage[labelfont={bf}]{caption}
\usepackage{wrapfig}
\usepackage{titletoc}
\usepackage[singlespacing,nodisplayskipstretch]{setspace}
\usepackage[pdftex,letterpaper,portrait,twocolumn=false,nomarginpar]%
{geometry}
\usepackage{tabularx}
\usepackage{fancyhdr}%
\providecommand{\U}[1]{\protect\rule{.1in}{.1in}}

\makeatletter
\def\ps@myheader{
\renewcommand{\@oddhead}{\hfill\thepage} \renewcommand{\@evenhead}{\hfill\thepage} \renewcommand{\@oddfoot}{} \renewcommand{\@evenfoot}{} }
\makeatother
\ifx\pdfoutput\relax\let\pdfoutput=\undefined\fi
\newcount\msipdfoutput
\ifx\pdfoutput\undefined\else
\ifcase\pdfoutput\else
\ifx\paperwidth\undefined\else
\ifdim\paperheight=0pt\relax\else\pdfpageheight\paperheight\fi
\ifdim\paperwidth=0pt\relax\else\pdfpagewidth\paperwidth\fi
\fi\fi\fi
\begin{document}

\title{High-dimensional forecasting with known knowns and known unknowns\thanks{This
paper developed from Pesaran's Deane-Stone Lecture at the National Institute
of Economic and Social Research, London. We greatly benefitted from comments
when earlier versions of this paper were presented in 2023 at NIESR on 21
June, the International Association of Applied Econometrics Annual Conference
in Oslo, June 27-30, 2023, Bayes Business School, City University, 22 November
2023, and at Economics Department-Wide Seminar, Emory University, 16 February,
2024. We are also grateful for comments from Alex Chudik, Anthony Garratt,
George Kapetanios, Essie Maasoumi, Alessio Sancetta, Mahrad Sharifvaghefi,
Allan Timmermann and Stephen Wright. We particularly thank Hayun Song, our
research assistant, for his work in coding, empirical implementation, and the
tabulation of results, along with his invaluable research assistance.}}
\author{M. Hashem Pesaran\\University of Southern California,\ and Trinity College, Cambridge
\and Ron P. Smith\\Birkbeck, University of London}
\maketitle

\begin{abstract}
Forecasts play a central role in decision making under uncertainty. After a
brief review of the general issues, this paper considers ways of using
high-dimensional data in forecasting. We consider selecting variables from a
known active set, known knowns, using Lasso and OCMT, and approximating
unobserved latent factors, known unknowns, by various means. This combines
both sparse and dense approaches to forecasting. We demonstrate the various
issues involved in variable selection in a high-dimensional setting with an
application to forecasting UK inflation at different horizons over the period
2020q1-2023q1. This application shows both the power of parsimonious models
and the importance of allowing for global variables.

\bigskip



\end{abstract}

%

\thispagestyle{empty}%

\pagebreak

"\textbf{All the business of war, indeed all the business of life, is
endeavour to find out what you don't know by what you do; that's what I called
`\textit{guess what was the other side of the hill'} "\ }

\textbf{Duke of Wellington }%

\pagenumbering{arabic}%
\onehalfspacing
\setlength{\abovedisplayskip}{0pt}
\setlength{\belowdisplayskip}{0pt}
\setlength{\abovedisplayshortskip}{0pt}
\setlength{\belowdisplayshortskip}{0pt}%

\section{Introduction\label{Intro}}

Forecasts play a central role in decision making under uncertainty. Good
forecasts are those that lead to good decisions, in the sense that the
expected payoff to the decision maker using the forecast is greater than it
would be otherwise.\footnote{The linkage between forecasting and decision
making is discussed in
\citet{granger2000economic,granger2000decision}%
, who argue in favour of a closer link between the decision and forecast
evaluation problems. \cite{pesaran2002decision} provide a more general survey
of decision theoretic approaches to forecast evaluation.} In the case of
inflation forecasts, which we consider below, the Bank of England makes
forecasts to help it set monetary policy to keep inflation within a target
range.\footnote{The letter of 26 June 2023 by Huw Pill, Bank chief economist
to the chair of the House of Commons Treasury Committee sets out his
assessment of the role played by the forecasts in the policy process.} The
payoff is the variation of inflation around target. However, it is not clear
how one would quantify the contribution of the forecast to the payoff in terms
of a specific central bank loss function.

Since forecasts are designed to inform decisions, they are inherently linked
to policy making. However, there is an issue as to whether one should use the
same model for both forecasting and setting the policy instruments. Different
questions require different types of model to answer them. A policy model
might be quite large, while a forecasting model might be quite small. There is
also an issue of how transparent the model should be. It may be difficult to
interpret why a machine learning statistical model makes the predictions it
does and this can be a major disadvantage when policy requires communication
of a persuasive narrative.

In recent years forecasting has been influenced by the increasing availability
of high-dimensional data, improvements in computational power and advances in
econometrics and machine learning techniques. In some areas, such as
meteorology, this has resulted in improved forecasts, increasing the number of
hours ahead for which accurate predictions can be made. The improved forecasts
lead to better decision making as people change their behaviour in response to
the predictions and the effect of such responses on mortality from heat and
cold is examined in \cite{Shrader2023Fatal}. Despite advances in data,
computation and technique, the improvement in accuracy of weather forecasts
has not been matched by economic forecasts.\textbf{ }This is a cause for
concern, since as emphasised in the classic analysis of
\cite{Whittle1983Prediction}, prediction and control are inherently linked and
decisions over such elements of economic management as monetary policy are
dependent on a view of the future.

Macroeconomic forecasting is challenging because lags in responses to policies
or shocks are long and variable and the economic system is responsive, events
prompting changes in the structure of the economy. Forecasting tends to be
relatively successful during normal times, but in times of crises and change,
in the face of large shocks or structural changes, when accurate predictions
are most needed, forecasters tend to fail. For instance, inflation was 5.4\%
in December 2021. This was the last figure they had when, in February 2022,
the Bank forecast that inflation would peak at 7.1\% in 2022q2, and fall to
5.5\% in 2023q1, and be back within target at 2.6\% in 2024q1. The 2023q1
actual was 10.2\%, almost 5 percentage points higher than forecast. This burst
in inflation was a global phenomenon and other central banks made similar errors.

Economic forecasters may use purely statistical models or more structural
economic models, which include the policy variables and important economic
linkages. We will call these more structural models `policy models', given the
way structural has a number of interpretations. The statistical models will
typically be conditional on information available at the time of the forecast,
which may be inaccurate: knowing where one is at the time of forecast,
nowcasting, is an important element. Policy models may also be conditional on
assumed future values. The Bank of England makes forecasts conditional on
market expectations of future interest rates, assumptions about future energy
prices and government announcements about future fiscal policy as well as
other measures. Wrong assumptions about those future values may cause problems
as it did for the Bank in August 2022, when it anticipated a rise in energy
prices but not the government response to them, so over-estimated inflation.
There is also a policy issue as to whether fiscal and monetary policy should
be determined independently by different institutions or jointly.

If both statistical and policy models are used, there is an issue as to how to
integrate them. Forecast averaging has been widely shown to improve forecast
performance, but forming averages on many variables may lack coherence and
consistency. In September 2018, the Bank of England Independent Evaluation
Office, IEO reported back to Court of the Bank on the implementation and
impact of the 2015 IEO review of forecasting performance. IEO reported that
some 'non-structural' models had been introduced as a source of challenge, and
outputs were routinely shown to the Monetary Policy Committee as a way of
cross checking the main forecast. While, some but not all members found them
helpful, but there was no desire to develop more models of this sort and
internally they had not been integrated into the forecast process as a source
of challenge.

Forecasts by central banks fulfill multiple purposes including as a means of
communication to influence expectations in the wider economy. This also makes
it difficult to choose a loss function to evaluate forecasts. For instance,
Bank of England forecasts for inflation at a 2 year horizon are always close
to the target of 2 per cent. Even were the Bank to think it unlikely that it
could get back to target within 2 years, it might feel that its credibility
might be damaged were it to admit that. An institutional issue is who "owns"
the forecast. The Bank of England forecast is the responsibility of the 9
member Monetary Policy Committee (MPC); other central banks have different
systems. For instance the US Federal Reserve has a staff forecast not
necessarily endorsed by the decision makers.

There is an issue about the optimal amount of information to use: both with
respect to breadth, how many variables, and length, how long a run of data.
With respect to breadth, in principle, one should use information on as many
variables as possible and not just for the country being forecast, since in a
networked world foreign variables contain information. This is the information
that is used in the Global VAR (GVAR) whose use is surveyed in
\cite{chudik2016theory}. While the use of many variables might imply a large
model, in practice quite parsimonious small models tend to be difficult to
beat in forecasting competitions.

With respect to length, in a May 2023 hearing, the Chair of the House of
Commons Treasury Select Committee asked the chief economist of the Bank of
England, Huw Pill: "Are you saying that, despite the Bank of England having
been in existence for over 300 years, you look at only the last 30 years when
you think about what the risks are to inflation?". Pill emphasised the
importance of the policy regime, which had been different in the past 30 years
of inflation targeting than in earlier high inflation periods. The 30 years up
to 2019 had also been different in terms of the absence of large real shocks,
like Covid-19 and the effects of the Russian invasion of Ukraine.

Whether it is statistical or policy, the model will typically be supplemented
by a judgemental input, justified by the argument that the forecaster has a
larger information set than the model. In evidence to the Treasury Select
Committee in September 2023, Sir Jon Cunliffe said: "We start with the model.
All models are caricatures of real life. There is a suite of models; that is
the starting point. But then the MPC itself puts judgments that change the
model, and we have made some quite big judgments in the past about inflation
persistence and the like. Finally, when we have the best collective view of
the committee, which is our judgment on top of the model, the model keeps us
honest. It ensures that there is a general equilibrium and we cannot just move
things around."

In short, macroeconomic forecasting faces important challenges. It depends on
how forecasts are announced and used in the decision making process. To deal
with a constantly changing economic environment, forecasts must continually
adapt to new data sets, statistical techniques and theory-based economic
insights, knowing that there are still key variables that might have been left
out, either due to difficulties in measurement, oversight or ignorance.
Forecasters must answer a range of difficult questions. What sample periods
and potential variables to consider? How to decide which variables to use for
forecasting, and whether to use the same sample periods for variable selection
and for forecasting? Should one use ensemble forecasting from forecasts
obtained either from different models or from the same model estimated over
different sample sizes or with different degrees of down-weighting? One must
only be humbled by the sheer extent of the uncertainty that these choices
entail. It is within this wider context that this paper tries to formalize
some elements of the problem of forecasting with high-dimensional data and
illustrates the various issues involved with an application to forecasting UK inflation.

The rest of this paper is organized as follows. Section \ref{FRAME} sets out
the high-dimensional forecasting framework we will be considering. Section
\ref{Known} considers "known knowns", selecting relevant variables from a
known active set. Section \ref{Uknowns} considers "known unknowns" where there
are known to be unobserved latent variables. Section \ref{Inflation} presents
the empirical application on forecasting UK\ inflation. Section \ref{Conc}
contains some concluding comments.

\section{The high-dimensional forecasting problem\label{FRAME}}

Suppose the aim is to forecast a scalar target variable, denoted by $y_{T+h},$
at time $T$, for the future dates, $T+h$, $h=1,2,...,H$. Given the historical
observations the optimal forecast of the target variable, $y_{T+h},$ depends
on how the forecasts are used, namely the underlying decision problem. In
practice, specifying loss functions associated with decision problems is hard,
hence the tendency to fall back on mean squared error loss. Under this loss
function the optimal forecasts are given by conditional expectations,
$E\left(  y_{T+h}\left\vert \mathcal{I}_{T}\right.  \right)  $, where
$\mathcal{I}_{T}$ is the set of available information, and expectations are
formed with respect to the joint probability distribution of the target
variable and the set of potential predictors under consideration. But when the
number of potential predictors, say $K$, is large even this result is too
general to be of much use in practice.

The high-dimensional nature of the forecasting problem also presents a
challenge of its own when we come to multi-step ahead forecasting when
forecasts of the target variable are required for different horizons,
$h=1,2,...,H$. Many decision problems require having forecasts many periods
ahead, months, years and even decades ahead. Monetary policy is often
conducted over the business cycle, at least 2-3 years ahead of the policy
formulation. Climate change policy requires forecasts over many decades ahead.
In interpreting Pharaoh's dreams, Joseph considered a two-period decision
problem whereby seven years of plenty are predicted to be followed by seven
years of drought. Multi-horizon forecasting is relatively straightforward when
the number of potential predictors is small and a complete system of
equations, such as a vector autoregression (VAR), can be used to generate
forecasts for different horizons from the same forecasting model in an
iterative manner. Such an \textbf{iterated} approach is not feasible, and
might not even be desirable, when the number of potential predictors is too
large, since future forecasts of predictors are also needed to generate
forecasts of $y_{T+h}$ for $h\geq2$. This is why in high-dimensional set ups
multi-period ahead forecasts are typically formed using different models for
different horizons. This is known as the \textbf{direct} approach and avoids
the need for forward iteration by directly regressing the target variable
$y_{t+h}$ on the predictors at time $t$, thus possibly ending up with
different models and/or estimates for each $h$%
.\footnote{\cite{marcellino2006comparison} discuss the pros and cons of
iterated and direct approaches to forecasting when $K$ is small, and the
target variable and the predictors can be jointly modelled as low-dimensional
VARs or VARMAs. It is shown that if the underlying VAR model is correctly
specified then iterated forecasts, being coherent, are preferred to direct
forecasts. However, under misspecification direct forecasts could perform
better. \cite{pesaran2011variable} reconsider the comparison of iterated and
direct forecasts to factor augmented VARs.}

To be more specific, ignoring intercepts and factors which we introduce below,
suppose $y_{t}$ is the first element of the high-dimensional vector
$\boldsymbol{w}_{t}$, assumed to follow the first order VAR model,
\begin{equation}
\boldsymbol{w}_{t}=\boldsymbol{\Phi w}_{t-1}+\mathbf{u}_{t}. \label{vlvar}%
\end{equation}
Higher order VARs can be written as first order VARs using the companion form.
The error vector, $\mathbf{u}_{t}$, satisfies the orthogonality condition
$E\left(  \mathbf{u}_{t}\left\vert \mathcal{I}_{t-1}\right.  \right)
=\mathbf{0}$, where $\mathcal{I}_{t-1}=(\boldsymbol{w}_{t-1},\boldsymbol{w}%
_{t-2},...)$. Then%

\begin{equation}
\boldsymbol{w}_{T+h}=\boldsymbol{\Phi}^{h}\boldsymbol{w}_{T}+\mathbf{u}%
_{h,T+h}, \label{itvar}%
\end{equation}
where, except for $h=1,$ the overlapping observations cause the error in
(\ref{itvar}) to have the moving average structure of order $h-1$:
\[
\mathbf{u}_{h,t+h}=\mathbf{u}_{t+h}+\boldsymbol{\Phi}\mathbf{u}_{t+h-1}%
+\boldsymbol{\Phi}^{2}\mathbf{u}_{t+h-2}+...+\boldsymbol{\Phi}^{h-1}%
\mathbf{u}_{t+1}.
\]
Under the VAR specification $E\left(  \mathbf{u}_{h,T+h}\left\vert
\mathcal{I}_{T}\right.  \right)  =0$, for $h=1,2,...$ and the optimal (in the
mean squared error sense) $h$-step ahead forecast of $\boldsymbol{w}_{T+h}$ is
$E\left(  \boldsymbol{w}_{T+h}\left\vert \mathcal{I}_{T}\right.  \right)
=\boldsymbol{\Phi}^{h}\boldsymbol{w}_{T}$. But given that in most forecasting
applications the dimension of $\boldsymbol{w}_{t}$ is large, it is not
feasible to estimate $\boldsymbol{\Phi}$ directly without imposing strong
sparsity restrictions. Instead we take the target variable, $y_{T+h}$, to be
the first element of $\boldsymbol{w}_{T+h}$ and consider the direct regression%
\[
y_{t+h}=\boldsymbol{\phi}_{h}^{\prime}\boldsymbol{w}_{t}+u_{h,t+h\text{ }},
\]
where $\boldsymbol{\phi}_{h}^{\prime}$ is the first row of $\boldsymbol{\Phi
}^{h},$ and $u_{h,t+h}$ is the first element of $\mathbf{u}_{h,t+h}$. We still
face a high-dimensional problem since there are a large number of potential
covariates in $\boldsymbol{w}_{t}$. We consider the implementation of the
direct approach under two scenarios concerning the potential predictors.
First, when it is known that the target variable $y_{t+h}$ is a sparse linear
function of a large set of observed variables $\boldsymbol{x}_{t}$ (a subset
of $\boldsymbol{w}_{t}$) known as the `active set'. The model is sparse in the
sense that $y_{t+h}$ depends on a small number of covariates, that are
\textit{known} to be a subset of the much larger active set. The machine
learning literature focuses on this case, which we refer to as the case of
"known knowns". Second, when $y_{t+h}$ could also depend on a few latent
(unobserved) factors, $\mathbf{f}_{t}$, not directly included in the active
set, which we call the case of "known unknowns".

Specifically, we suppose that for each $h$, $y_{t+h}$ can be approximated by
the following linear model, where the predictors are also elements of
$\boldsymbol{w}_{t}$ in the high-dimensional VAR (\ref{vlvar}),
\begin{equation}
y_{t+h}=c_{h}+\mathbf{a}_{h}^{\prime}\mathbf{z}_{t}+\sum_{j=1}^{K}\beta
_{jh}I(j\in DGP)x_{jt}+\mathbf{\psi}_{h}^{\prime}\mathbf{f}_{t}+u_{h,t+h\text{
}}, \label{general}%
\end{equation}
for $t=1,2,...,T-h$, where $c_{h}$ is the intercept, $\mathbf{z}_{t}$ is a
vector of small number, $p$, of pre-selected covariates included across all
horizons $h$. Obvious examples, include lagged values of the target variable
$(y_{t},y_{t-1},...)$. Other variables can also be included in $\mathbf{z}%
_{t}$ on the basis of \textit{a priori }theory or strong beliefs.\ The third
component of $y_{t+h}$ specifies the subset of variables in the active set
$\boldsymbol{x}_{Kt}=\left(  x_{1t},x_{2t},....,x_{Kt}\right)  ^{\prime}$.
$I(j\in DGP)$ is an indicator variable which takes the value of unity if
$x_{jt}$ is included in the data generating process (DGP) for $y_{t+h}$ and
zero otherwise. It is only if $I(j\in DGP)=1$ that $\beta_{jh}$ will be
identified. We discuss ways to determine the selection indicator $I(j\in DGP)$
below. The number of variables included in the DGP is given by $k=\sum
_{j=1}^{K}I(j\in DGP)$, which is supposed to be small and fixed as $T$ (and
possibly $K$) become large. This assumption imposes sparsity on the
relationship between the target and the variables in the active set. In
addition, we allow for a small number of latent factors, $\mathbf{f}_{t},$
that represent other variables influencing $y_{t+h}$ that are not observed
directly, but known to be present - the known unknowns.

\cite{giannone2021economic} contrast \textbf{sparse methods}, that select a
few variables from the active set as predictors, such as Lasso and OCMT
discussed below, and \textbf{dense methods}, that select all the variables in
the active set but attach small weights to many of them, such as principal
components (PC), ridge regression and other shrinkage techniques. Rather than
having to choose between sparse and dense predictors, we consider approaches
that combine the two. We apply sparse selection methods to the variables in
the active set, and use dense shrinkage methods to approximate $\mathbf{f}%
_{t}$ from a wider set of variables with $\boldsymbol{x}_{t}$ included as a
subset. We first consider the selection problem, known knowns, where we know
the active set of potential covariates, and we then consider known unknowns
where there are unobserved factors. The elastic net regression of
\cite{ZouHastie} discussed below also combines sparse and dense techniques.

Throughout, we shall assume that the errors, $u_{h,t+h\text{ }}$, in
(\ref{general}) satisfy the orthogonality condition $E\left(  u_{h,t+h\text{
}}\left\vert \mathbf{z}_{t},\boldsymbol{x}_{t},\mathbf{f}_{t}\right.  \right)
=0,$ for $h\geq1$. In the context of the high-dimensional VAR model discussed
above, this orthogonality condition holds so long as the underlying errors,
$\mathbf{u}_{t}$, are serially uncorrelated. This is so despite the fact that
due to the use of overlapping observations $u_{h,t+h\text{ }}$ will be
serially correlated when $h>1$. This is an important consideration when
high-dimensional techniques are applied to select predictors for multi-step
ahead forecasting; an issue to which we will return.

\section{Known knowns\label{Known}}

In the case of known knowns, forecasts are obtained assuming that $y_{t+h}$ is
a linear function of $\boldsymbol{x}_{t}$%
\begin{equation}
y_{t+h}=c_{h}+\mathbf{a}_{h}^{\prime}\boldsymbol{z}_{t}+%
{\displaystyle\sum\limits_{j=1}^{K}}
\beta_{hj}x_{jt}+u_{h,t+h}\text{, for }t=1,2,...,T, \label{BM}%
\end{equation}
subject to some penalty condition on $\{\beta_{hj}\}$. Some of the covariates,
$x_{jt}$, could be transformations of other covariates, such as
interaction\ terms. It is assumed that the model is correctly specified, in
the sense that, apart from $\boldsymbol{z}_{t}$, the variables that drive
$y_{t+h}$ are all included in the active set, $\mathbf{x}_{Kt}$.

Penalized regressions estimate $\boldsymbol{\beta}$ by solving the following
optimization problem:\textbf{\ }%
\[
\min_{\mathbf{a}_{h},\mathbf{\beta}_{h}}\left\{
{\displaystyle\sum\limits_{t=1}^{T}}
\left(  y_{t+h}-\mathbf{a}_{h}^{\prime}\boldsymbol{z}_{t}-\boldsymbol{\beta
}_{h}^{\prime}\mathbf{x}_{Kt}\right)  ^{2}+\lambda_{hT}%
{\displaystyle\sum\limits_{i=1}^{K}}
\left[  (1-\alpha){\left\vert \beta_{hi}\right\vert }+\alpha\beta_{hi}%
^{2}\right]  \right\}  ,
\]
where $\boldsymbol{\beta}_{h}=(\beta_{h1},\beta_{h2},...,\beta_{hK})^{\prime}%
$, for given values of the "tuning" parameters $\lambda$ and $\alpha$. When
$\alpha=1,$ we have ridge regression. When $\alpha=0$, $\lambda_{hT}\neq0,$ we
have the \textbf{Lasso} regression, which is better suited for variable
selection. When $\lambda_{hT}\neq0$ and $\alpha\neq0$ we have the
\cite{ZouHastie} \textbf{elastic net} regression, which also mixes sparse and
dense approaches.

\begin{center}%
\begin{figure}[h]%
\centering

\includegraphics[height=2.5486in,width=4.7383in]{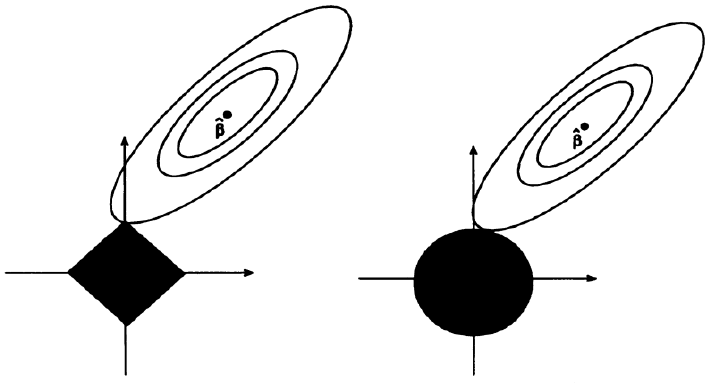}

\caption{\textbf{Estimation for the Lasso (left) and ridge (right)
regression}}%
\label{fig:lasso_ridge}%
\end{figure}

\end{center}

Many standard forecasting techniques result from the particular choice of the
penalty function. Shrinkage estimators such as ridge or some Bayesian
forecasts can be derived using the $\ell_{2}$ norm $\sum_{i=1}^{K}$
$\beta_{hi}^{2}<C_{h}<\infty.$ Lasso (least absolute shrinkage and selection
operator) follows when the $\ell_{1}$ norm is used $\sum_{i=1}^{K}\left\vert
\beta_{hi}\right\vert <C_{h}<\infty.$ The difference is shown in Figure
\ref{fig:lasso_ridge}\ below in \cite{tibshirani1996regression} where the
$\ell_{1}$ norm yields corner solutions with many of the coefficients,
$\beta_{hj}$, estimated to be zero. In contrast, the use of $\ell_{2}$ norm
yields non-zero estimates for all the coefficients with many very close to zero.

There are also a large number of variants of Lasso, including adaptive Lasso,
group Lasso, double Lasso, fused Lasso and prior Lasso. We will focus on Lasso
itself, which we use in our empirical application and which we will compare to
OCMT as an alternative procedure which is based on inferential rather than
penalized procedures.

\subsection{Lasso}

In this paper we focus on Lasso, but acknowledge that are many variations on
Lasso such as: adaptive Lasso, group Lasso, fused Lasso, and prior Lasso.
Lasso estimates $\boldsymbol{\beta}_{h}$ by solving the following optimization
problem:\textbf{\ }%
\begin{equation}
\min_{\boldsymbol{\beta}_{h}}\left\{
{\displaystyle\sum\limits_{t=1}^{T}}
\left(  y_{t+h}-c_{h}-\boldsymbol{\beta}_{h}^{\prime}\boldsymbol{x}%
_{t}\right)  ^{2}+\lambda_{hT}%
{\displaystyle\sum\limits_{i=1}^{K}}
{\left\vert \beta_{hi}\right\vert }\right\}  \text{ }, \label{BP}%
\end{equation}
where $\boldsymbol{\beta}_{h}=(\beta_{h1},\beta_{h2},...,\beta_{hK})^{\prime}%
$, $\boldsymbol{x}_{Kt}=\left(  x_{1t},x_{2t},....,x_{Kt}\right)  ^{\prime}$
for a given choice of the "tuning" parameter, $\lambda_{hT}$. The variable
selection consistency of Lasso has been investigated by \cite{zhao2006model},
\cite{meinshausen2006variable} and more recently by \cite{lahiri2021necessary}%
. The key condition is the so-called "Irrepresentable Condition, IRC" that
places restrictions on the magnitudes of the correlations between the signals
($\mathbf{X}_{1h},$ standardized) and the rest of the covariates
($\mathbf{X}_{2h}$, standardized), taken as given (deterministic). The IRC
is:
\begin{equation}
\text{IRC: }\left\Vert (T^{-1}\mathbf{X}_{2h}^{\prime}\mathbf{X}_{1h})\left(
T^{-1}\mathbf{X}_{1h}^{\prime}\mathbf{X}_{1h}\right)  ^{-1}%
sign(\boldsymbol{\beta}_{h}^{0})\right\Vert _{\infty}<1, \label{IRC}%
\end{equation}
where $\boldsymbol{\beta}_{h}^{0}=(\beta_{1h}^{0},\beta_{2h}^{0}%
,...,\beta_{k_{h}h}^{0})^{\prime}$ denotes the vector of true signal
coefficients.\footnote{The number of signals, $k_{h}$, could vary with $h$.}
The IRC condition is met for pure noise variables, but need not hold for proxy
variables, noise variables that are correlated with the true signals.

To appreciate the significance of the IRC, suppose the DGP contains $x_{1t}$
and $x_{2t}$ and the rest of the covariates in the active set are
$x_{3t},x_{4t},...,x_{Kt}.$ Denote the sample correlation coefficient between
$x_{1t}$ and $x_{2t}$ by $\hat{\rho}$ ($\hat{\rho}^{2}<1$) and the sample
correlation coefficient of $x_{1t}$ and $x_{2t}$ with the rest of the
covariates in the active set by $\hat{\rho}_{1s},\hat{\rho}_{2s},$ for
$s=3,4,...,K.$ Then, dropping the subscript $h,$ the IRC for the $s^{th}$
covariate is given by
\[
\left\vert \left[  sign\left(  \beta_{01}\right)  ,sign\left(  \beta
_{02}\right)  \right]  ^{\prime}\left(
\begin{array}
[c]{cc}%
1 & \hat{\rho}\\
\hat{\rho} & 1
\end{array}
\right)  ^{-1}\left(
\begin{array}
[c]{c}%
\hat{\rho}_{1s}\\
\hat{\rho}_{2s}%
\end{array}
\right)  \right\vert <1
\]
which yields
\[
\left\vert sign\left(  \beta_{01}\right)  \left(  \hat{\rho}_{1s}-\hat{\rho
}\hat{\rho}_{2s}\right)  +sign\left(  \beta_{02}\right)  \left(  \hat{\rho
}_{2s}-\hat{\rho}\hat{\rho}_{1s}\right)  \right\vert <1-\hat{\rho}^{2}%
\]
for $s=3,4,...,K.$ In this example there are two cases to consider: A:
$sign\left(  \beta_{02}\right)  =sign\left(  \beta_{01}\right)  ;$ and B:
$sign\left(  \beta_{02}\right)  =-sign\left(  \beta_{01}\right)  .$ For case A
$\sup_{s}\left\vert \hat{\rho}_{1s}+\hat{\rho}_{2s}\right\vert <1+\hat{\rho}$,
and for case B $\sup_{s}\left\vert \hat{\rho}_{1s}-\hat{\rho}_{2s}\right\vert
<1-\hat{\rho}$. Since the signs of the coefficients are unknown, for all
possible values of $\hat{\rho}$, $\hat{\rho}_{1s}$ and $\hat{\rho}_{2s}$, we
can ensure the IRC condition is met if $\left\vert \hat{\rho}\right\vert
+\sup_{s}\left\vert \hat{\rho}_{1s}\right\vert +\sup_{s}\left\vert \hat{\rho
}_{2s}\right\vert <1.$ This example shows the importance of the correlations
between the true covariates in the DGP as well as between the true covariates
and the other members of the active set that do not belong to the DGP. The IRC
is quite a stringent condition and it is not just when one has proxies in the
active set that are highly correlated with the true covariates that Lasso will
tend to choose too many variables. In practice one cannot check the IRC
condition since one does not know which variables are the true signals.

In addition to the IRC it is also required that
\[
\text{MinC: min}_{j=1,2,...,k}\left\vert \beta_{jh}^{0}\right\vert
>(2T)^{-1}\lambda_{hT}\left\vert \left(  T^{-1}\mathbf{X}_{1h}^{\prime
}\mathbf{X}_{1h}\right)  ^{-1}sign(\mathbf{\beta}_{h}^{0})\right\vert
_{j}\text{, }%
\]%
\[
\text{Penalty Condition : }T^{-1}\lambda_{hT}=o(1).
\]
The penalty condition, which follows from MinC, says that the penalty has to
rise with $T$, but not too fast and not too slowly. The expansion rate of
$\lambda_{hT}$ depends on the magnitude and the sign of $\beta_{jh}^{0}$, and
the correlations of signals with the proxy variables.
\cite{lahiri2021necessary} shows that the penalty condition can be relaxed to
$\lim_{T\rightarrow\infty}T^{-1}\lambda_{hT}<\lim\inf_{T\rightarrow\infty
}d_{hT},$ where%
\[
d_{hT}=2\min_{j}\left\vert \beta_{jh}^{0}\right\vert /\left\vert \left(
T^{-1}\mathbf{X}_{1h}^{\prime}\mathbf{X}_{1h}\right)  ^{-1}sign(\mathbf{\beta
}_{h}^{0})\right\vert _{j}.
\]

The above conditions do not restrict the choice of $\lambda_{hT}$ very much,
hence the recourse to cross-validation to determine it. In practice,
$\lambda_{hT}$ is calibrated using $M$-fold cross-validation techniques. The
observations, $t=1,2,...,T$, are partitioned into $M$ disjoint subsets
(folds), of size approximately $m=T/M.$ Then $M-1$ subsets are used for
training and one for evaluation. This is repeated with each fold being used in
turn for evaluation. $M$ is typically set to $5$ or $10$. Cross-validation
methods are often justified in machine learning literature under strong
assumptions, such as independence and parameter stability across the
sub-samples used in cross-validation. These assumptions are rarely met in the
case of economic time series data, an issue that is discussed further in the
context of the empirical example in Section \ref{Inflation}.

\subsection{OCMT}

The need for cross-validation is avoided in the procedure proposed by
\cite{chudik2018one}, (CKP). This is the "\textbf{O}ne \textbf{C}ovariate at a
time\textbf{\ M}ultiple \textbf{T}esting" (\textbf{OCMT}) procedure, where
covariates are selected \textbf{one at a time}, using $t-$statistic for
testing the significance of the variables in the active set,
\textbf{individually.}\footnote{Since $t$\textbf{-}ratios are invariant to
scale, no pre-standardization of the covariates in the active set is required.
This is in contrast to Lasso which is typically implemented after in-sample
standardization of the covariates.} Ideas from the multiple testing literature
are used to control the false discovery rate, and ensure the selected
covariates encompass the true covariates (signals) with probability tending to
unity, under certain regularity conditions. Like Lasso, OCMT has no difficulty
in dealing with (pure) noise variables, and is very effective at eliminating
them. Also, like Lasso, it requires some $min$ condition such as $\left\vert
\beta_{jh}^{0}\right\vert >>\sqrt{\frac{k\log(K)}{T}}$, for $j=1,2,...,k$%
.\footnote{To simplify the exposition we have dropped explicit reference to
the forecast horizon, $h$. But in practice, and as we shall see from the
empirical applications below, the number and identity of selected signals
could differ with $h$.} But because it considers a single variable at a time,
OCMT\ does not require the IRC condition to hold and is not affected by the
correlation between the members of the DGP as Lasso is. Instead it requires
the number of proxies signals, say $k_{T}^{\ast}$, to rise no faster than
$\sqrt{T}$. \ \cite{chudik2023variable}, discussed below, is primarily
concerned with parameter instability, but Section 4 of that paper has a
detailed comparison of the assumptions required for Lasso and OCMT under
parameter stability.

OCMT's condition on $k_{T}^{\ast}$ has been recently relaxed by
\cite{sharifvaghefi2022variable} who allows $k_{T}^{\ast}\rightarrow\infty$,
with $T$. He considers the following DGP
\begin{equation}
y_{t+h}=c_{h}+\mathbf{a}_{h}^{\prime}\boldsymbol{z}_{t}+\sum_{j=1}^{K}%
\beta_{jh}I(j\in DGP)x_{jt}+u_{h,t+h}\text{ ,} \label{DGP}%
\end{equation}
where as before $\boldsymbol{z}_{t}$ is a known vector of pre-selected
variables, and it is assumed that the $k$ signals are contained in the known
active set $\mathcal{S}_{K,t}=\left\{  x_{jt},j=1,2,...,K\right\}  $. Note
that for now the DGP\ in (\ref{DGP}) does not include the additional latent
factors, $\mathbf{f}_{t}$, introduced in (\ref{general}). Without loss of
generality, consider the extreme case where there are no noise variables and
\textbf{all} proxy or pseudo signal variables ($x_{jt}$, for
$j=k+1,k+2,...,K)$ are correlated with the signals, $\mathbf{x}_{1t}%
=(x_{1t,}x_{2t},...,x_{kt})^{\prime}$. In this case $k_{T}^{\ast}$ rises with
$K$ and OCMT is no longer applicable. However, in this case because of the
correlation with the proxies,\ the signals, $\mathbf{x}_{1t}$, become latent
factors for the proxy variables and we have \
\[
x_{jt}=\phi_{j0}+\sum_{i=1}^{k}\phi_{ji}x_{it}+\varepsilon_{jt}=\phi
_{j0}+\mathbf{\phi}_{j}^{\prime}\mathbf{x}_{1t}+\varepsilon_{jt}\text{,}%
\]
for $j=k+1,k+2,....,K$. Although the identity of these common factors are
unknown, because we do not know the true signals, they can be approximated by
the principal components of the variables in the active set.

Specifically, following \cite{sharifvaghefi2022variable}, denote the latent
factors that result in non-zero correlations between the noise variables in
the active set and the signals by $\boldsymbol{\varkappa}_{t}$ and consider
the factor model
\begin{equation}
x_{jt}=\boldsymbol{\kappa}_{j}^{\prime}\boldsymbol{\varkappa}_{t}%
\mathbf{+}v_{jt}\text{, for }j=1,2,...,K\text{ ,} \label{FM}%
\end{equation}
where $\boldsymbol{\kappa}_{j}$, for $j=1,2,...,K$ are the factor loadings and
the errors, $v_{jt}$, are weakly cross-correlated and distributed
independently of the factors and their loadings. Under (\ref{FM}), the DGP,
(\ref{DGP}), can be written equivalently as
\begin{equation}
y_{t+h}=c_{h}+\mathbf{a}_{h}^{\prime}\boldsymbol{z}_{t}+\mathbf{b}_{h}%
^{\prime}\boldsymbol{\varkappa}_{t}+\sum_{j=1}^{K}\beta_{jh}I(j\in
DGP)v_{jt}+u_{h,t+h}\text{ }, \label{GOCMT}%
\end{equation}
where $\mathbf{b}_{h}=\sum_{j=1}^{K}I(j\in DGP)\beta_{jh}\boldsymbol{\kappa
}_{j}$. When $\boldsymbol{\varkappa}_{t}$ and $v_{jt}$ are known, the problem
reduces to selecting $v_{jt}$ from $\mathcal{S}_{K,t}^{v}=\left\{
v_{jt},j=1,2,...,K\right\}  ,$ conditional on $\mathbf{z}_{t}$ and
$\boldsymbol{\varkappa}_{t}$. Sharifvaghefi shows that the OCMT selection can
be carried out using the principal component estimators of
$\boldsymbol{\varkappa}_{t}$ and $\mathbf{v}_{jt}$ - denoted by
$\boldsymbol{\hat{\varkappa}}_{t}$ and $\mathbf{\hat{v}}_{jt}$, if both $K$
and $T$ are large. He labels this procedure as generalized OCMT (GOCMT). Note
that at the moment it is assumed that $\boldsymbol{\varkappa}_{t}$ does not
\textbf{directly} affect $y_{t+h},$ it only enters through the $x_{jt}.$ It
represents the signals, the common factors correlated with the proxies, and
provides a way of filtering the correlations in the first step.

\subsection{GOCMT}

The GOCMT procedure simply augments the OCMT\ regressions with the PCs,
$\boldsymbol{\hat{\varkappa}}_{t}$, and considers the statistical significance
of $\mathbf{\hat{v}}_{jt}$ for each $j$, \emph{one at the time}. Lasso-factor
models have also been considered by \cite{fan2020factor} and
\cite{hansen2019factor}. In practice, since $\mathbf{x}_{j}=\mathbf{\hat{\Xi
}\hat{\psi}}_{j}+\mathbf{\hat{v}}_{j}$, where $\mathbf{\hat{\Xi}=}\left(
\boldsymbol{\hat{\varkappa}}_{1},\boldsymbol{\hat{\varkappa}}_{2}%
,...,\boldsymbol{\hat{\varkappa}}_{T}\right)  ^{\prime}$, then $\mathbf{M}%
_{\mathbf{\hat{\Xi}}}\mathbf{x}_{j}=\mathbf{M}_{\mathbf{\hat{\Xi}}%
}\mathbf{\hat{v}}_{j}$, where $\mathbf{M}_{\mathbf{\hat{\Xi}}}=\mathbf{I}%
_{T}-\mathbf{\hat{\Xi}}\left(  \mathbf{\hat{\Xi}}^{\prime}\mathbf{\hat{\Xi}%
}\right)  ^{-1}\mathbf{\hat{\Xi}}^{\prime}$, and GOCMT\ reduces to OCMT\ when
$\mathbf{z}_{t}$ is augmented with $\boldsymbol{\hat{\varkappa}}_{t}$, where
the statistical significance of $x_{jt}$ as a predictor of $y_{t+h}$ is
evaluated for each $j$, one at a time. Like OCMT, GOCMT allows for the
multiple testing nature of the procedure ($K$ separate tests - with $K$ large)
by increasing the level of significance with $K$. The number of PCs,
$dim(\boldsymbol{\hat{\varkappa}}_{t})$, can be determined using one of the
criteria suggested in the factor literature.

In the first stage, $K$ \textbf{separate} OLS regressions are computed, where
the variables in the active set are entered one at a time:
\begin{equation}
y_{t+h}=c_{h}+\mathbf{a}_{h}^{\prime}\boldsymbol{z}_{t}+\mathbf{b}_{h}%
^{\prime}\boldsymbol{\hat{\varkappa}}_{t}+\phi_{jh}x_{jt}+e_{j,h,t+h}\text{,
}t=1,2,...,T\text{, for }j=1,2,...,K, \label{br}%
\end{equation}
Denote the $t$-ratio of $\phi_{jh}$ by $t_{\hat{\phi}_{j,\left(  1\right)  }}%
$. Then variable $j$ is selected if%
\begin{equation}
\widehat{\mathcal{J}}_{j,(1)}=I\left[  \left\vert t_{\hat{\phi}_{j,\left(
1\right)  }}\right\vert >c_{p}(K,\delta)\right]  ,\text{ for }%
j=1,2,...,K\text{ }, \label{btild1}%
\end{equation}
where $c_{p}(K,\delta)$ is a critical value function given by
\begin{equation}
c_{p}(K,\delta)=\Phi^{-1}\left(  1-\frac{p}{2K^{\delta}}\right)  ,
\label{cp(n)}%
\end{equation}
$p$ is the nominal size (usually set to $5\%$), $\Phi^{-1}(\cdot)$ is the
inverse of a standard normal distribution function and $\delta$ is a fixed
constant set in the interval $[1,1.5]$. In the second step a multivariate
regression of $y_{t+h}$ on $\boldsymbol{z}_{t}$ and all the selected
regressors is considered for inference and forecasting. Serial correlation
will arise with OCMT when selection is based on one variable at the time, and
the omitted variables are mixing (serially correlated). CKP discuss this in
section C of the online theory supplement to their paper and suggest using a
more conservative (higher) critical value -- namely using $\delta=1.5$ rather
than $\delta=1.5$.

When the covariates are \textbf{not }highly correlated, OCMT applies
irrespective of whether $K$ is small or large relative to $T$, so long as
$T=\ominus\left(  K^{c}\right)  $, for some finite $c>0$.\ \ But to allow for
highly correlated covariates, GOCMT requires $K$ to be sufficiently large to
enable the identification of the latent factor, $\boldsymbol{\varkappa}_{t}$.
In cases where $K$ is not that large, it might be a good idea to augment the
active set for the target variable, $y_{t+h}$, $\mathcal{S}_{K,t}=\left\{
x_{jt},\text{ }j=1,2,...,K\right\}  $, with covariates for other variables
determined simultaneously with $y_{t+h}$, ending up with $\overline
{\mathcal{K}}>K$ covariates for identification of $\boldsymbol{\varkappa}_{t}%
$. GOCMT does not impose any restriction on the correlations between the
variables other than that they cannot be perfectly collinear.

\subsection{High-dimensional variable selection in presence of parameter
instability}

OCMT has also been recently generalized by \cite{chudik2023variable} to deal
with parameter instability. Under parameter instability OCMT correctly selects
the covariates with non-zero average (over time) effects, using the full
sample. However the adverse effects of changing parameters on the forecast may
mean that while the full sample is the best to use for selection, it need not
be the best to use for estimating the forecasting model. Instead, it may be
better to use shorter windows or weight the observations in the light of the
evidence on break points and break sizes.

Determining the appropriate window or weighting for the observations before
estimation is a difficult problem and no fully satisfactory procedure seems to
be available. It is common in finance to use rolling windows of 60 or 120
months, but one problem with shorter windows is that if you have periods of
instability interspersed with periods of stability, like the Great Moderation,
estimates using a short window from the stable period may understate the
degree of uncertainty. This happened during the financial crisis when the
short windows used for estimation did not reflect past turbulence. Similarly,
the Bank of England estimating their models using the low inflation regime of
the past 30 years discounted the evidence from the high inflation regime of
the 1970s and 1980s.

While identifying the date of a break might not be difficult, identifying the
size of the break may be problematic if the break point is quite recent. If
there is a short time since the break, there is little data on which to
estimate the post-break coefficient with any degree of precision. If there is
a long time since the break, then using post break data is sensible.
\cite{pesaran2013optimal} examine optimal forecasts in the presence of
continuous and discrete structural breaks. These present quite different sorts
of challenges. With continuous breaks the parameters change often by small
amounts. With discrete breaks the parameters change rarely but by large
amounts. They propose weighting observations to obtain optimal forecasts in
the MSFE sense and derive optimal weights for one step ahead forecasts for the
two types of break. Under continuous breaks, their approach largely recovers
exponential smoothing weights. Under discrete breaks between two regimes the
optimal weights follow a step function that allocates constant weights within
regimes but different weights in different regimes. In practice, the time and
size of the break is uncertain and they investigate robust optimal weights.
Averaging forecasts with different weighting schemes, for instance with
exponential smoothing parameters between 0.96 and 0.99, may also be a way to
produce more robust forecasts.

\section{Known unknowns\label{Uknowns}}

So far we have considered techniques (penalized regressions and OCMT) that
assume $y_{t+h}$ depends on $\mathbf{z}_{t}$ and a subset of a\textbf{ }set of
covariates - the active set - which is assumed \textbf{known}. In contrast,
shrinkage type techniques such as PCs, (implicitly) assume that $y_{t+h}$
depends on $\mathbf{z}_{t}$ and the $m\times1$ vector of \textbf{unknown}
factors $\mathbf{f}_{t}$
\[
y_{t+h}=c_{h}+\mathbf{a}_{h}^{\prime}\boldsymbol{z}_{t}+\mathbf{\theta}%
_{h}^{\prime}\mathbf{f}_{t}+u_{h,t+h}.
\]
This is a simple example of techniques that in our terminology can be viewed
as belonging to a class of forecasting models based on \textbf{known
unknowns}. The uncertainty about $\mathbf{f}_{t}$ is resolved assuming it can
be identified from a \textbf{known} active set, such as $\mathcal{S}%
_{K,t}=\left\{  x_{jt},j=1,2,...,K\right\}  $. Individual covariates in
$\mathcal{S}_{K,t}$ are not considered for selection (although a few could be
pre-selected and included in $\boldsymbol{z}_{t}$). To forecast $y_{t+h}$ one
still requires to forecast the PCs and to allow for the uncertainty regarding
$m=\dim(\mathbf{f}_{t})$.

Factor augmented VARs (FAVAR), initially proposed by \cite[BBE]%
{bernanke2005measuring}, augment the standard VAR models with a set of
unobserved common factors. In the context of our set up, FAVAR can be viewed
as a generalized version of (\ref{general}), where $\boldsymbol{y}_{t+h}$ is a
vector and $\boldsymbol{z}_{t}=\left\{  \boldsymbol{y}_{t},\boldsymbol{y}%
_{t-1},...,\boldsymbol{y}_{t-p}\right\}  $. BBE argue that small VARs gave
implausible impulse response functions, such as the "price puzzle", which were
interpreted as reflecting omitted variables. One response was to add variables
and use larger VARs, but this route rapidly runs out of degrees of freedom,
since Central Bankers monitor hundreds of variables. The FAVAR was presented
as a solution to this problem. Big Bayesian VARs are an alternative solution.

The assumptions that underlie both penalized regression and PC shrinkage are
rather strong. The former assumes that $\mathbf{f}_{t}$ can affect $y_{t+h}$
only indirectly through $x_{jt}$, $j=1,2,...,K$, and the latter does not allow
for individual variable selection. Suppose that $\mathbf{f}_{t}$ also enters
(\ref{DGP}) then the model can be written as (\ref{general}) above, repeated
here for convenience:
\[
y_{t+h}=c_{h}+\mathbf{a}_{h}^{\prime}\boldsymbol{z}_{t}+\sum_{j=1}^{K}%
\beta_{jh}I(j\in DGP)x_{jt}+\mathbf{\psi}_{h}^{\prime}\mathbf{f}_{t}%
+u_{h,t+h}.
\]
The forecasting problem now involves both selection and shrinkage. The
$\mathbf{f}_{t}$ can be identified by, for instance, the PCs of the augmented
active set $x_{jt},$ $j=1,2...,K,K+1,...,\overline{\mathcal{K}}$ which can be
wider than the active set of covariates used to predict $y_{t}$. There are
various other ways that the unobserved $\mathbf{f}_{t}$ could be estimated,
but we use PCs as an example, since they are widely used.

The latent factors are unlikely to be only specific to the target variable
under consideration. Observed global factors, such as oil and raw material
prices or inflation and output growth of major countries such as US can be
included in the active set. The main issue is how to deal with global factors,
such as technology, political change and so on that are unobserved and tend to
affect many countries in the world economy. Call this vector of global factors
$\mathbf{g}_{t}$. A natural extension is to introduce forecast equations for
other countries (entities) who have close trading relationships with UK and
use penalized panel regressions, where the panel dimension allows
identification of the known unknowns.

More specifically, suppose there are $N$ other units (countries) that are
affected by observed country specific covariates, $\mathbf{z}_{it},$
$i=1,2,...,N,$ and $x_{i,jt}$ for $j=1,2,...,K_{i},$ plus domestic latent
factors $\mathbf{f}_{it}$, and global latent factors, $\mathbf{g}_{t}$. The
forecasting equations are now generalized as%
\begin{align*}
y_{i,t+h}  &  =c_{h}+\mathbf{a}_{ih}^{\prime}\boldsymbol{z}_{it}+\sum
_{j=1}^{K_{i}}\beta_{ijh}I(j\in DGP_{i})x_{i,jt}+\\
&  \mathbf{\theta}_{ih}^{\prime}\mathbf{f}_{it}+\mathbf{\psi}_{ih}^{\prime
}\mathbf{g}_{t}+u_{i,h,t+h}\text{,}%
\end{align*}
where $k_{i}=\sum_{j=1}^{K_{i}}I(j\in DGP_{i})$ is finite as $K_{i}%
\rightarrow\infty$, for $i=0,1,2,...,N$. For the country-specific covariates
we postulate that there is an augmented active set%
\[
x_{i,jt}=\mathbf{\gamma}_{ij}^{\prime}\mathbf{f}_{it}+v_{i,jt},\text{
}j=1,2,...,K_{i}\text{, }K_{i+1},....,\overline{\mathcal{K}}_{i},
\]
where $\mathbf{f}_{it}$ are latent factors. The global factors are then
identified as the common components of the country-specific factors, namely%
\[
\mathbf{f}_{it}=\boldsymbol{\Psi}_{i}\mathbf{g}_{t}+\boldsymbol{\xi}%
_{it}\text{ },
\]
for $i=1,2,..,N$, with $N$ large.

Variable selection for the target variable (say UK inflation) can now proceed
by applying GOCMT, with the UK model augmented with UK-specific PCs,
$\mathbf{\hat{f}}_{it}$ as well as the PC estimator of the global factor,
$\mathbf{g}_{t},$ that drives the country specific factors. This can be
extracted from $\mathbf{\hat{f}}_{it}$ as PCs of the country-specific PCs. In
addition to common factor dependence, countries are also linked through trade
and other more local features (culture, language). Such "network" effects can
be captured by using "starred" variables, to use the GVAR terminology. A
simple example would be$\ ($for $i=0,1,....,N)$%
\begin{align}
y_{i,t+h}  &  =c_{ih}+\delta_{ih}y_{it}^{\ast}+\mathbf{a}_{ih}^{\prime
}\boldsymbol{z}_{it}+\sum_{j=1}^{K_{i}}\beta_{ijh}I(j\in DGP_{i}%
)x_{i,jt}+\label{main}\\
&  \mathbf{\theta}_{ih}^{\prime}\mathbf{f}_{it}+\mathbf{\psi}_{ih}^{\prime
}\mathbf{g}_{t}+u_{i,h,t+h}\text{ ,}\nonumber
\end{align}
where $i=0$ represents UK, and $y_{it}^{\ast}=\sum_{j=1}^{N}w_{ij}y_{jt},$
$w_{ij}$ (trade weights) measures the relative importance of country $j$ in
determination of country $i^{th}$ target variable. Similarly, $\boldsymbol{z}%
_{it}^{\ast}=\sum_{j=1}^{N}w_{ij}^{\ast}\boldsymbol{z}_{jt}$ can also be added
to the model if deemed necessary.

The network effects can be included either as an element of $\boldsymbol{z}%
_{it}$ or could be made subject to variable selection. The problem becomes
much more complicated if we try to relate $y_{i,t+h}$ simultaneously to
$y_{i,t+h}^{\ast}$. Further, for forecasting, following \cite{chudik2016multi}%
, one might also need to augment the UK regressions with time series,
forecasting models for the common factors.\textbf{\ }

Equation (\ref{main}) allows for a number of different approaches to dimension
reduction. \textbf{ }As has been pointed out by \cite{wainwright2019high}:
\textquotedblleft Much of high-dimensional statistics involves constructing
models of high-dimensional phenomena that involve some implicit form of
low-dimensional structure, and then studying the statistical and computational
gains afforded by exploiting this structure\textquotedblright. Shrinkage
methods, like PCs, assume a low dimensional factor structure. The two
selection procedures that we have considered, Lasso and OCMT, exploit
different aspects of the low-dimensional sparsity structure assumed for the
underlying data generating process. Lasso restricts the magnitude of the
correlations within and between the signals and the noise variables. OCMT
limits the rate at which the number of proxy variables rises with the sample
size. GOCMT relaxes this restriction by filtering out the effects of latent
factors that bind the proxies to the true signals before implementing the OCMT procedure.

\section{Forecasting UK inflation\label{Inflation}}

\subsection{Introduction}

We apply the procedures proposed above to the problem of forecasting quarterly
UK inflation at horizons $h=1,2$ and $4$. The target variable is the headline
rate, average annual UK inflation, which is also forecast by the Bank of
England. It is labelled DPUK4, defined as $\pi_{t+h}=100\times\log
(p_{t+h}/p_{t+h-4}),$ where $p_{t}$ is the UK consumer price index taken from
the IMF International Financial Statistics. Forecasting annual rates of
inflation at quarterly frequencies is subject to the overlapping observations
problem when $h>1$, and it is important that the pre-selected variables in
$\boldsymbol{z}_{t}$, or the variables include in the active set
$\mathcal{S}_{K,t}=\left\{  x_{jt},j=1,2,...,K\right\}  $ are all
pre-determined (known) at time $t$.\ Furthermore, as discussed earlier, the
variables selected for forecasting inflation at different horizons need not be
the same, and the selected variables are also likely to change over time.

Since we have emphasised the importance of international network effects, we
need to use a quarterly dataset that includes a large number of countries to
estimate global factors and to allow the construction of the $y_{it}^{\ast}$
variables that appear in (\ref{main}). The Global VAR (GVAR) data set provides
such a source. The publicly available data set compiled by
\cite{mohaddes2024compilation} covers 1979q1-2023q3. We are very grateful to
them for extending the data. While the latest GVAR dataset goes up to 2023q3
we only had access to data till 2023q1 when we started the forecasting
exercise, the results of which are reported in this paper, but the data that
we used matches that GVAR 2023 vintage which was released in January
2024.\footnote{GVAR Data 1979q1-2023q3 (2023 Vintage) is available at
https://www.mohaddes.org/gvar, further material on the GVAR is provided at
https://sites.google.com/site/gvarmodelling/gvar-toolbox}

The data base includes quarterly macroeconomic data for 6 variables (log real
GDP, y, the rate of inflation, dp, short-term interest rate, r, long-term
interest rate, lr, the log deflated exchange rate, ep, and log real equity
prices, eq), for 33 economies as well as data on commodity prices (oil prices,
poil, agricultural raw material, pmat, and metals prices, pmetal). These 33
countries cover more than 90\% of world GDP. The GVAR data was supplemented
with other specific UK data on money, wages, employment and vacancies, in the
construction of the active set discussed below.

In the light of the argument in \cite{chudik2023variable}, we use the full
sample beginning in\textbf{ }1979q1 for variable selection. There are
arguments for down-weighting earlier data for estimation when there have been
structural changes, as discussed by \cite{pesaran2013optimal}. However, the
full sample was used both for variable selection and estimation of the
forecasting model in order to allow evidence from the earlier higher inflation
regime to inform both aspects.

Two sets of variables are considered for inclusion in $\boldsymbol{z}_{t}$.
The first set, which we label AR2, includes lags of the target variable
$\pi_{t},\pi_{t-1}$ (or equivalently $\pi_{t}$ and $\Delta\pi_{t}).$ Given the
importance we attach to global variables and network effects, the second set,
which we label ARX2, also includes $\pi_{t}^{\ast}$, $\pi_{t-1}^{\ast}$ (or
equivalently $\pi_{t}^{\ast}$ and $\Delta\pi_{t}^{\ast})$ where $\pi_{t}%
^{\ast}$ is a measure of UK\ specific foreign inflation constructed using UK
trade weights with the other countries.\footnote{Specifically, $\pi_{t}^{\ast
}=\sum_{j}w_{j}\pi_{jt},$ where $\pi_{jt}$ is the inflation rate in country
$j$ and $w_{j}$ is the trade weight of country $j$ with UK.}

\begin{center}%
\begin{figure}[h]%
\centering

\includegraphics[height=3.5129in,width=5.8219in]{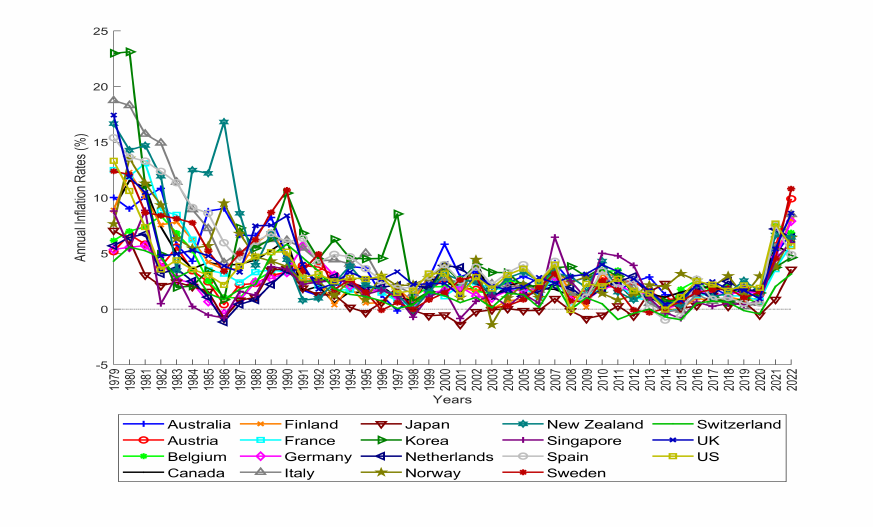}

\caption{\textbf{Inflation Across Advanced Economies}}%
\label{fig:InfAdv}%
\end{figure}

\end{center}

If there is a global factor in inflation, the inflation rates of different
countries will be highly correlated and tend to move together. Figure
\ref{fig:InfAdv} demonstrates that this is in fact the case. It plots the
inflation rates for 19 countries over the period 1979-2022. It is clear that
they do move together, reflecting a strong common factor. The dispersion is
somewhat greater in the high inflation 1980s. At times individual countries
break away from the herd with idiosyncratic bursts of inflation, like New
Zealand in the mid 1980s. But it is striking that inflation in every country
increases from 2020.

\begin{figure}[ptb]%
\centering

\includegraphics[height=3.5129in,width=5.8219in]{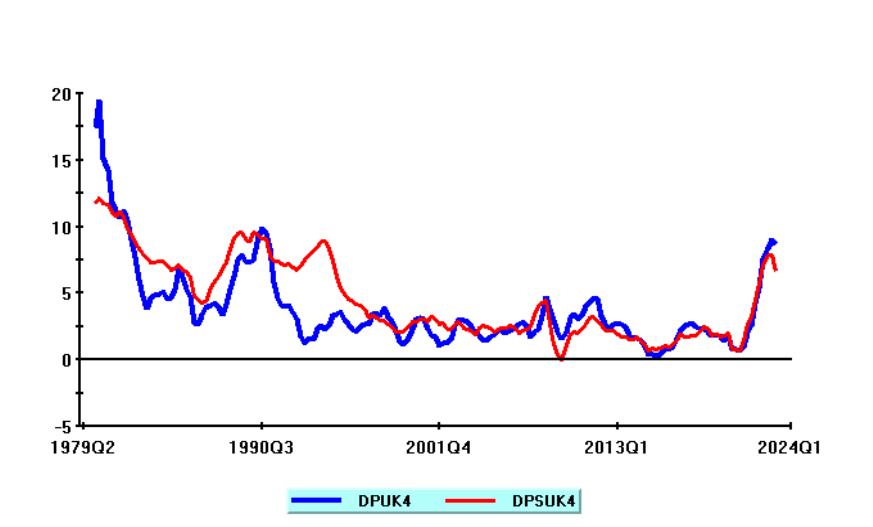}

\caption{\textbf{UK and UK-specific global inflation}}%
\label{fig:ukglobinf}%
\end{figure}

To demonstrate the importance of the global factor for the UK, Figure
\ref{fig:ukglobinf} plots $\pi_{t}$, and $\pi_{t}^{\ast},$ UK inflation and UK
specific foreign inflation. The two series move together, and from the mid
1990s they are very close. This indicates that not only is one unlikely to be
able to explain UK inflation just by UK\ variables but that there are good
reasons to include this UK specific measure of foreign inflation in one of our
specifications for $\boldsymbol{z}_{t}.$ The GVAR estimates also indicate the
higher sensitivity of the UK to foreign variables than the US or euro area.
This is not surprising, they are larger, less open, economies.%

\subsection{Active set}

We now turn to the choice of the members of the active set, $\mathcal{S}%
_{K,t}=\left\{  x_{jt},j=1,2,...,K\right\}  ,$ some of which may be included
in $\boldsymbol{z}_{t}.$ While our focus is on forecasting not on building a
coherent economic model, our choice of the covariates in the active set is
motivated by the large Phillips curve literature, which suggests important
roles for demand, supply and expectations variables. The demand and supply
variables in the active set are both domestic and foreign from both product
and labour markets. Expectations are captured by financial variables.
Interaction terms were not included, but the non-linearities that have been
investigated in the literature may be picked up by the latent foreign
variables. These are represented by UK specific measures of foreign inflation
and output that could be viewed as estimates of $\mathbf{g}_{t}$ that are
tailored to the UK in relation to her trading partners.

Accordingly, we consider 26 covariates ($x_{jt}$) listed in Table
\ref{tab:Active_Set_Variables}, and their changes $\Delta x_{jt}%
=x_{jt}-x_{j,t-1}$, giving an active set with $K=52$ variables to select from.
Whereas in a regression including current and lagged values of a regressor
(say $x_{t}$ and $x_{t-1})$ is equivalent to including current and change
(namely $x_{t}$ and $\Delta x_{t}$), in selection the two specifications can
result in different outcomes. Including $\Delta x_{t}$ is better since, as
compared to $x_{t-1}$, it is less correlated with the level of the other
variables in the active set. The 26 included covariates are measured as
four-quarter rates of change, changes or averages to match the definition of
the target variable. The rates of change are per cent per annum.

\textbf{UK goods market demand indicators}: rate of change of output, two
measures of the output gap: log output minus either a $P=8$ or $P=12$ quarter
moving average of log output, $Gap(y_{t},P)=y_{t}-P^{-1}\sum_{p=1}^{P}y_{t-p}$;

\textbf{UK labour market demand indicators}: rate of change of UK employment,
vacancies, and average weekly earnings and the change in unemployment;

\textbf{UK financial indicators}: annual averages of UK short and long
interest rates, the rate of change of money, UK M4, and of UK real equity prices;

\textbf{Global cost pressures on the UK}: rate of change of the price of oil,
metals, materials, UK import prices and deflated dollar exchange rate;

\textbf{Foreign demand and supply variables:} UK specific global measures,
foreign inflation, rate of change of foreign output and two measures of the
foreign output gap: log foreign output minus either an 8 or 12 quarter moving
average of log foreign output. In addition, large country variables were
added: annual average of US short and long interest rates, rates of change of
US output and prices, and of Chinese output.%

\begin{table}[H]%
\caption{List of covariates for UK inflation forecasting}%

\begin{center}%
\begin{tabular}
[c]{ccl}\hline\hline
{\small Variable} &  & {\small Description}\\\hline
\multicolumn{1}{l}{{\small DPUK4}} &  & {\small 4} {\small quarter UK rate of
inflation, }\\
\multicolumn{1}{l}{{\small DYUK4}} &  & {\small 4} {\small quarter rate of
change of} {\small UK real GDP}\\
\multicolumn{1}{l}{{\small GAPUK8}} &  & {\small UK log real GDP relative to
its 8 quarter moving average}\\
\multicolumn{1}{l}{{\small GAPUK12}} &  & {\small UK log real GDP relative to
its 12 quarter moving average}\\
\multicolumn{1}{l}{{\small DEMUK4}} &  & {\small 4 quarter rate of change of
UK employment }\\
\multicolumn{1}{l}{{\small DVUK4}} &  & {\small 4 quarter rate of change of UK
vacancies}\\
\multicolumn{1}{l}{{\small DUUK}} &  & {\small 4 quarter change in UK
unemployment}\\
\multicolumn{1}{l}{{\small DWUK4}} &  & {\small 4 quarter rate of change of
average weekly earnings,}\\
\multicolumn{1}{l}{{\small RUK4}} &  & {\small 4 quarter average UK short
interest rate }\\
\multicolumn{1}{l}{{\small LRUK4}} &  & {\small 4 quarter average UK long
interest rate }\\
\multicolumn{1}{l}{{\small DMUK4}} &  & {\small 4 quarter rate of change of UK
M4 money}\\
\multicolumn{1}{l}{{\small DEQUK4}} &  & {\small 4 quarter rate of change of
UK real equity prices }\\
\multicolumn{1}{l}{{\small DPOIL4}} &  & {\small 4 quarter rate of change of
oil prices}\\
\multicolumn{1}{l}{{\small DPMAT4}} &  & {\small 4 quarter rate of change of
material prices}\\
\multicolumn{1}{l}{{\small DPMETAL4}} &  & {\small 4 quarter rate of change of
metal prices}\\
\multicolumn{1}{l}{{\small DPMUK4}} &  & {\small 4 quarter rate of change of
UK import prices}\\
\multicolumn{1}{l}{{\small DEPUK4}} &  & {\small 4 quarter rate of change of
UK deflated dollar exchange rate}\\
\multicolumn{1}{l}{{\small DPSUK4}} &  & {\small UK specific measure of 4
quarter foreign inflation}\\
\multicolumn{1}{l}{{\small DYSUK4}} &  & {\small 4 quarter rate of change of
UK specific foreign real GDP\ }\\
\multicolumn{1}{l}{{\small RUS4}} &  & {\small 4 quarter average US short
interest rate}\\
\multicolumn{1}{l}{{\small LRUS4}} &  & {\small 4 quarter average US long
interest rate}\\
\multicolumn{1}{l}{{\small DYCHINA4}} &  & {\small 4 quarter rate of change of
Chinese real GDP}\\
\multicolumn{1}{l}{{\small GAPSUK8}} &  & {\small UK specific log foreign real
GDP relative to 8 quarter moving average}\\
\multicolumn{1}{l}{{\small GAPSUK12}} &  & {\small UK specific log foreign
real GDP relative to 12 quarter moving average}\\
\multicolumn{1}{l}{{\small DYUS4}} &  & {\small 4 quarter rate of change of US
real GDP}\\
\multicolumn{1}{l}{{\small DPUS4}} &  & {\small 4 quarter US rate of
inflation}\\\hline\hline
\end{tabular}
\bigskip
\end{center}

\label{tab:Active_Set_Variables}%

\end{table}%

\subsection{Variable selection}

\subsubsection{Variable selection procedures}

We consider Lasso, Lasso conditional on $\boldsymbol{z}_{t}$, and GOCMT
conditional on $\boldsymbol{z}_{t}$. With Lasso, the variables are
standardized in-sample before implementing variable selection. The Lasso
penalty parameter, $\lambda_{T},$ is estimated using 10 fold cross-validation
(CV), across subsets of the observations. As noted above, the assumptions
needed for standard CV procedures, for instance those used in the program
cv.glmnet, are not appropriate for time series. Time series show features such
as persistence and changing variance that are incompatible with those
assumptions. In the standard procedure the CV subsets (folds) are typically
chosen randomly. This is appropriate if the observations are independent draws
from a common distribution, but this is not the case with time series. Since
order matters in time series, we retain the time order of the data within each
subset. See \cite{bergmeir2018note} who provide Monte Carlo evidence on
various procedures suggested for the case of serially correlated data. We use
all the data, and do not leave gaps between subsets. In addition, the standard
procedure chooses the $\hat{\lambda}_{hT}$ that minimises the pooled MSE over
the ten subsets. But when variances differ substantially over subsets pooling
is not appropriate, instead we follow Chudik, Kapetanios, and Pesaran (2018,
CKP), and use the average of the $\hat{\lambda}_{hT}$ chosen in each subset.
Full details are provided in the online simulation appendix to CKP (2018).

As well as standard Lasso, for consistency with OCMT, we also generated Lasso
forecasts conditional on $\boldsymbol{z}_{t}$ by including a pre-selected set
of variables $\boldsymbol{z}_{t}$ in the optimization problem (\ref{BP}). This
generalized Lasso procedure solves the following optimization problem:
\begin{equation}
\min_{\mathbf{a}_{h},\mathbf{\beta}_{h}}\left\{
{\displaystyle\sum\limits_{t=1}^{T}}
\left(  y_{t+h}-c_{h}-\mathbf{a}_{h}^{\prime}\boldsymbol{z}_{t}%
-\boldsymbol{\beta}_{h}^{\prime}\boldsymbol{x}_{t}\right)  ^{2}+\lambda_{T}%
{\displaystyle\sum\limits_{i=1}^{K}}
{\left\vert \beta_{hi}\right\vert }\right\}  \text{ ,} \label{lasso_z}%
\end{equation}
where the penalty is applied only to the variables in the active set,
$\boldsymbol{x}_{t},$ and not to the pre-selected variables, $\boldsymbol{z}%
_{t}$. The above optimization problem can be solved in two stages. In the
first stage the common effects of $\boldsymbol{z}_{t}$ are filtered out by
regressing $y_{t+h}$ and $\boldsymbol{x}_{t}$ on the pre-selected variables
$\boldsymbol{z}_{t}$ and saving the residuals $e_{y.z}$ and $e_{xj.z}$,
$j=1,2,..,K$. In the second stage Lasso is applied to these residuals. A proof
that this two-step procedure solves the constrained minimization problem in
(\ref{lasso_z}) is provided by Sharifvaghefi and reproduced in the Appendix.

In the OCMT critical value function, $c_{p}(K,\delta)=\Phi^{-1}\left(
1-\frac{p}{2K^{\delta}}\right)  ,$ we set $p=0.05$ and $\delta=1$. With $K=52$
this means that we only retain variables with t-ratios (in absolute value)
exceeding $c_{0.05}(52,1)=3.3$. To allow for possible serieal correlation, we
also experimented with setting $\delta=1.5,$ which yields, $c_{0.05}%
(52,1.5)=3.82$. The results were reasonably robust and we focus on the
baseline choice of $\delta=1,$ also recommended by CKP.

We implement Lasso and OCMT conditional on two pre-selected sets of variables,
either an AR2 written as level and change $\mathbf{z}_{t}=$ ($\pi_{t}$,
$\Delta\pi_{t})^{\prime}$ or given the role of foreign inflation, shown above,
the AR2 augmented by the level and change of the UK specific measure of
foreign inflation, denoted ARX, $\mathbf{z}_{t}=$ ($\pi_{t}$, $\Delta\pi_{t}%
$,$\pi_{t}^{\ast}$,$\Delta\pi_{t}^{\ast})^{\prime}$. As noted above, for
selection including current and change is better than including current and
lag. For comparative purposes we also generated forecasts with the
pre-selected variables only, namely the AR2 forecasts generated from the
regressions%
\[
AR2:\text{ \ \ \ \ }\pi_{t+h}=c_{h}+a_{1}\pi_{t}+a_{2h}\Delta\pi_{t}%
+u_{h,t+h}\text{,}%
\]
and the ARX forecasts generated from
\[
ARX:\text{ \ \ \ \ \ \ \ \ }\pi_{t+h}=c_{h}+a_{1h}\pi_{t}+a_{2h}\Delta\pi
_{t}+a_{3h}\pi_{t}^{\ast}+a_{4h}\Delta\pi_{t}^{\ast}+u_{h,t+h}.
\]

Variable selection is carried out recursively, for each forecast horizon $h$
separately, using an expanding windows approach. All data samples start in
$1979q2$ and end in the quarter that forecasts are made. To forecast the
average inflation over the four quarters to $2020q1$ using a forecast horizon
of $h=4$, the sample used for selection and estimation ends in $2019q1$. The
end of the sample is then moved to $2019q2$ to forecast the average inflation
over the four quarters to $2020q2,$ and so on. Similarly, to forecast the
average inflation over the four quarters to $2020q1$ using $h=2$, the sample
ends in $2019q3$, and using $h=1$ the sample ends in $2019q4.$ These sequences
continue one quarter at a time until the models are selected and estimated to
forecast inflation over the four quarters to $2023q1$. Thus for $h=4,$ there
are $17$ samples used for variable selection, while for $h=2$ and $h=1$ there
are $15$ and $14$ such variable selection samples. This process of recursive
model selection and estimation means that the variables selected can change
from quarter to quarter, and for each forecast horizon, $h$.

Section \ref{Select} of the online supplement list the variables selected by
each of the procedures, for each quarter and each forecast horizon. The main
features are summarised here.

\subsubsection{Number of variables selected}

Table \ref{tab:Variables selected} gives the minimum, maximum, and average
number of variables selected for the 3 forecast horizons and 5 variable
selection procedures. Except for AR2-OCMT, at $h=4,$ OCMT chooses fewer
variables than Lasso. Lasso conditional on the pre-selected variables selects
a larger number of variables in total than the standard Lasso without
conditioning. Conditioning on pre-selected variables is much more important
for OCMT as compared to Lasso. This finding is in line with the theoretical
results obtained by \cite{sharifvaghefi2022variable} who establishes the
importance of conditioning on the latent factors when applied to an active set
with highly correlated covariates. The number of variables Lasso selects falls
with the forecast horizon,\footnote{This is possibly because, as $h$ increases
the $\mathbf{\beta}_{h}^{0}$ get smaller in (\ref{IRC}), the IRC is more
likely to be satisfied and Lasso is less likely to falsely select additional
variables.} whilst the number of variables selected by OCMT rises with the
horizon. These results show that Lasso and OCMT could select very different
models for forecasting.%

\begin{table}[H]%
\caption{Number of variables selected by Lasso and OCMT including preselected}%

\begin{center}%
\begin{tabular}
[c]{lllllllllllll}\hline\hline
& \multicolumn{12}{l}{Forecast horizon, $h$, in quarters}\\\hline
& \multicolumn{12}{l}{Total number of pre-selected and selected variables}%
\\\hline
&  & \multicolumn{3}{l}{\ \ $\ h=1$} &  & \multicolumn{3}{l}{\ \ \ $\ h=2$} &
& \multicolumn{3}{l}{\ \ \ \ $h=4$}\\
&  & Min & Max & Mean & .. & Min & Max & Mean & .. & Min & Max & Mean\\\hline
Lasso &  & 7 & 12 & 8.1 &  & 5 & 9 & 6.1 &  & 3 & 6 & 5.2\\
AR2-Lasso &  & 5 & 11 & 8.2 &  & 9 & 16 & 13.5 &  & 8 & 11 & 9.5\\
AR2-OCMT &  & 2 & 3 & 2.2 &  & 4 & 5 & 4.5 &  & 5 & 14 & 6.2\\
ARX-Lasso &  & 8 & 16 & 12.4 &  & 12 & 19 & 16.3 &  & 2 & 15 & 9.9\\
ARX-OCMT &  & 4 & 4 & 4 &  & 5 & 5 & 5 &  & 5 & 8 & 5.8\\\hline\hline
\end{tabular}

\end{center}

{\footnotesize \emph{{}Note}: The reported results are based on 14, 15 and 17
variable selection samples for 1, 2 and 4 quarter ahead models, respectively.
The AR2 and ARX components include 2 and 4 pre-selected variables,
respectively.}

\label{tab:Variables selected}%

\end{table}%

As expected, the number of variables selected by Lasso correlates with the
estimates of the penalty parameter $\hat{\lambda}_{hT}$, computed by
cross-validation. These values are summarized in Table \ref{tab:penalty}. For
all three Lasso applications the mean of the estimated penalty parameter
increases with the forecast horizon, though more slowly for the specifications
that include pre-selected variables. For Lasso (without pre-selection) the
number of variables selected falls with the forecast horizon because of the
increasing penalty parameter. This is not as clear cut for the specifications
including pre-selected variables.%

\begin{table}[H]%
\caption
{Estimates of the Lasso penalty parameter computed by 10-fold cross-validation procedure. }%

\begin{center}%
\begin{tabular}
[c]{llllllllllll}\hline\hline
&  &  & \multicolumn{7}{l}{Forecast horizon, $h$, in quarters} &  & \\\hline
& \multicolumn{3}{l}{$\ \ \ h=1$} &  & \multicolumn{3}{l}{$\ \ \ \ h=2$} &  &
\multicolumn{3}{l}{$\ \ \ \ h=4$}\\
& Min & Max & Mean & .. & Min & Max & Mean & .. & Min & Max & Mean\\\hline
Lasso & 0.07 & 0.12 & 0.11 &  & 0.22 & 0.31 & 0.26 &  & 0.33 & 0.47 & 0.41\\
AR2-Lasso & 0.08 & 0.10 & 0.09 &  & 0.09 & 0.14 & 0.11 &  & 0.18 & 0.27 &
0.22\\
ARX-Lasso & 0.04 & 0.08 & 0.06 &  & 0.07 & 0.12 & 0.09 &  & 0.15 & 0.33 &
0.22\\\hline\hline
\end{tabular}

\end{center}

{\footnotesize \emph{{}Note}: The reported estimates are based on \ Lasso
penalty estimates (obtained from 10-fold cross-validation) for 14, 15 and 17
variable selection samples for 1, 2, and 4 quarter ahead models, respectively.
}

\label{tab:penalty}%

\end{table}%

\subsubsection{OCMT: selected variables by horizon}

OCMT selects only a few variables in addition to the pre-selected UK and
UK-specific foreign inflation ($\pi_{t},\Delta\pi_{t}$,$\pi_{t}^{\ast}$, and
$\Delta\pi_{t}^{\ast}$).\footnote{The OCMT results reported here are based on
the critical value function given by (\ref{cp(n)}) with $\delta=1$. Using the
larger value of $\delta=1.5$ reduces the number of selected variables for a
few of sample periods, but the outcomes are generally robust to the choice on
$\delta$ on the interval $[1,1.5]$. The selection results for OCMT with
$\delta=1.5$ are reported in the online supplement.} For $h=1,$ the variables
selected are given in sub-section \ref{SelOCMTh1} of the online supplement. In
addition to the $2$ pre-selected variables ($\pi_{t},\Delta\pi_{t}$), AR2-OCMT
selects the rate of change of wages (DWUK4) for samples ending in $2021q2$,
$2021q4$ and $2022q1$, and no other variables. ARX-OCMT does not select any
additional variables for any of the $14$ variable selection samples!

For $h=2,$ the variables selected are given in subsection \ref{SelOCMTh2} of
the online supplement. \ AR2-OCMT selects the rates of change of money (DMUK4)
and exchange rate (DEPUK4) from samples ending in $2019q3$ to $2020q3$, then
the rate of change of wages is added till the sample ending in $2022q2$, from
then the rates of change of money (DMUK4) and wages (DWUK4) are selected.
ARX-OCMT selects just the rate of change of money (DMUK4) as an additional
variable in every sample for $h=2$.

For $h=4$, the variables selected are given in subsection \ref{SelOCMTh4} of
the online supplement. AR2-OCMT chooses the same $3$ extra variables - the
rate of change of money (DMUK4) and exchange rate (DEPUK4) as well as the
UK-specific measure of foreign inflation (DPSUK4, $\pi_{t}^{\ast}$) -\textbf{
} for every sample ending from $2019q1$ to $2021q1$. Then, in the sample
ending in $2021q2,$ AR2-OCMT chooses $12$ extra variables. The number of
variables selected then falls to $7$ in $2021q3$, $6$ in $2021q4$, $5$ in
$2022q1$ and $4$ in $2022q2-2022q4.$ These $4$ are the rate of change of money
(DMUK4), of material prices (DPMAT4), of wages (DWUK4), and $\pi_{t}^{\ast}$
(DPSUK4). The number of variables selected falls to $3$ in the sample ending
in $2023q1$ when the foreign inflation measure is no longer selected. ARX-OCMT
chooses the rate of change of employment (DEMUK4) and of money for samples
ending in $2019q1-2021q4$, then adds material prices in $2022q1$, and selects
just the rate of change of money for the last four samples.

\subsubsection{Lasso: selected variables by horizon}

Lasso selections for each sample and horizon are given in the online
supplement, subsection \ref{SelectedLasso}. Lasso tends to select more
variables than OCMT so we give less detail. Table \ref{tab:LassoVarSel} lists
the variables chosen by standard Lasso at each horizon and the number of times
they were chosen out of the maximum number of possible samples: $14,$ for
$h=1,$ $15$ for $h=2,$ and $17$ for $h=4$. UK inflation, $\pi_{t}$ (DPUK4) is
always chosen in every sample at every horizon as is the UK\ measure of
foreign inflation, $\pi_{t}^{\ast}$ (DPSUK4). The change in UK inflation,
$\Delta\pi_{t},$ (DDPUK4) is chosen in every sample in the case of models for
$h=1,$ and $h=2,$ but never for $h=4.$ The change in foreign inflation
$\Delta\pi_{t}^{\ast}$ (DDPSUK4) is chosen in every sample at $h=1,$ in $3$
samples at $h=2$ but never at $h=4$. Thus Lasso provides considerable support
for the choice of pre-selected variables in $\boldsymbol{z}_{t}$ that include
foreign inflation as well as the two lagged inflation variables.

Apart from these variables, the rate of change of wages and of money figure
strongly when using Lasso. The rate of change of wages (DWUK4) is chosen in
all the samples for $h=1$ and $h=2$ and $14$ of the $17$ samples for $h=4.$
The rate of change of money (DMUK) is chosen in $10$ of the $14$ samples for
$h=1,$ and in every sample for $h=2$ and $h=4.$ Money and wages are also
chosen by OCMT but the rate of change of the exchange rate selected by OCMT is
never chosen by Lasso.

When $h=1$ Lasso also always selects two other variables, namely \ the change
in long interest rates (DLRUK4), and import price inflation (DPMUK).%

\begin{table}[H]%
\caption
{The number of times covariates from the active set are selected by Lasso at different forecast horizons}%

\begin{flushleft}
{\footnotesize
\begin{tabular}
[c]{lllll}\hline\hline
& \multicolumn{3}{l}{{\small Horizon}} & {\small Selected covariates}\\\hline
& ${\small h=1}$ & ${\small h=2}$ & ${\small h=4}$ & \\\hline
{\small DPUK4} & {\small 14} & {\small 15} & {\small 17} & {\small 4 quarter
UK rate of inflation}\\
{\small DPSUK4} & {\small 14} & {\small 15} & {\small 17} & {\small UK
specific measure of 4 quarter foreign inflation}\\
{\small DWUK4} & {\small 14} & {\small 15} & {\small 14} & {\small 4 quarter
rate of change of average weekly earnings}\\
{\small DDPUK4} & {\small 14} & {\small 15} & {\small 0} & {\small Change in 4
quarter UK rate of inflation}\\
{\small DPMUK} & {\small 14} & {\small 8} & {\small 0} & {\small 4 quarter
rate of change of UK import prices}\\
{\small DLRUK4} & {\small 14} & {\small 4} & {\small 0} & {\small Change in 4
quarter average UK long interest rate}\\
{\small DDPSUK4} & {\small 14} & {\small 3} & {\small 0} & {\small Change in
UK specific measure of 4 quarter foreign inflation}\\
{\small DMUK} & {\small 10} & {\small 15} & {\small 17} & {\small 4 quarter
rate of change of UK M4 money}\\
{\small DRUK4} & {\small 2} & {\small 0} & {\small 0} & {\small Change in 4
quarter average UK short interest rate}\\
{\small DEMUK4} & {\small 1} & {\small 0} & {\small 6} & {\small 4 quarter
rate of change of UK employment}\\
{\small DDPOIL4} & {\small 1} & {\small 0} & {\small 0} & {\small Change in 4
quarter rate of change of the oil price}\\
{\small DDEQUK4} & {\small 1} & {\small 0} & {\small 0} & {\small Change in 4
quarter rate of change of UK equity prices}\\
{\small DDYSUK4} & {\small 1} & {\small 0} & {\small 0} & {\small Change in 4
quarter rate of change of UK foreign real GDP}\\
{\small DGAPSUK12} & {\small 1} & {\small 0} & {\small 0} & {\small UK log
foreign real GDP relative to 8 quarter moving average}\\
{\small DPUS4} & {\small 0} & {\small 2} & {\small 2} & {\small 4 quarter US
rate of inflation}\\\hline\hline
\end{tabular}
}
\end{flushleft}

{\footnotesize \emph{{}Note}: The number of variable selection samples are 14,
15 and 17 for }$h=1${\footnotesize , }$2$, {\footnotesize and }$4$
{\footnotesize quarter ahead models, respectively.}\bigskip

\label{tab:LassoVarSel}%

\end{table}%

\subsection{Forecasts}

The point forecasts of inflation for $h=1,2$ and $h=4$ for the various
selection procedures are summarized in Section \ref{Forecasts} of the online
supplement. For each forecast horizon we have $13$ forecasts and their
realizations for the quarters $2020q1$ to $2023q1$ inclusive. These are
summarised in Table \ref{tab:rmsfeq1} below. We use root mean square forecast
error (RMSFE) as our forecast evaluation criterion. Since $13$ forecast errors
represent a very short evaluation sample with considerable serial correlation,
testing for the significance of the loss differences using the
\cite{diebold1995paring} test would not be reliable, and is not pursued here.

\subsubsection{One quarter ahead forecasts}

\begin{figure}[h]%
\centering

\includegraphics[height=3.7732in,width=6.0961in]{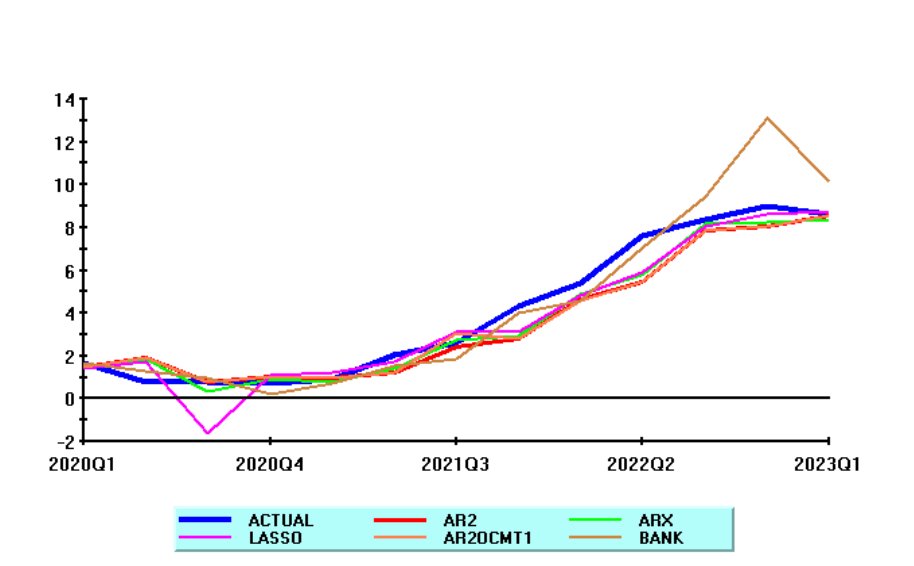}

\caption{\textbf{One quarter ahead forecasts.}}%
\label{fig:1q_ahead}%
\end{figure}

Figure \ref{fig:1q_ahead} gives plots of actual inflation and forecasts one
quarter ahead. Section \ref{Forecasth=1} of the online supplement gives the
point forecasts.\textbf{ }For $h=1$, ARX has the lowest RMSFE, the $\pi
_{t}^{\ast}$ and $\Delta\pi_{t}^{\ast}$ improve forecast performance relative
to the AR2. AR2-OCMT adds wage growth in three periods. Lasso suffers from
choosing too many variables relative to OCMT. The forecasts are very similar,
except Lasso predicted a large drop in $2020q3$ with a subsequent rebound.
This results from selecting an output gap measure, when UK output dropped
sharply in $2020q2$. This sharp drop and rebound was also a feature of Lasso
forecasts at other horizons. The Bank of England over-estimated inflation in
$2022q4$, correctly anticipating higher energy prices but not anticipating the
government energy price guarantees.%

\subsubsection{Two quarter ahead forecasts}

\begin{figure}[h]%
\centering

\includegraphics[height=3.7732in,width=6.0926in]{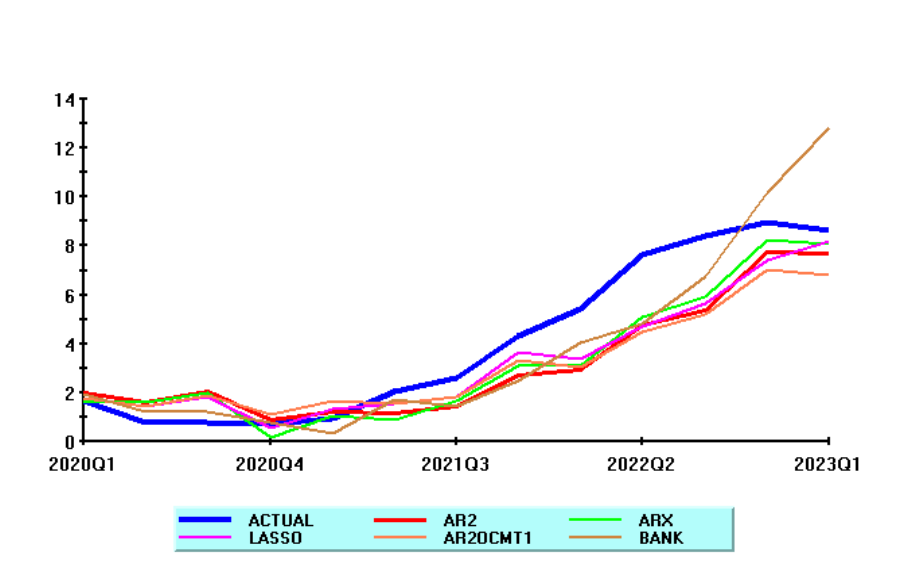}

\caption{\textbf{Plot of forecasts two quarters ahead}}%
\label{fig:2q_ahead}%
\end{figure}

Figure \ref{fig:2q_ahead} gives plots of actual inflation and forecasts two
quarters ahead. Section \ref{Forecasth=2} of the online supplement gives the
values. For $h=2$, ARX again has the lowest RMSFE. ARX-OCMT selects money
growth in every period. Lasso selects between 5 and 9 variables.%

\subsubsection{Four quarter ahead forecasts}

\begin{figure}[h]%
\centering

\includegraphics[height=3.7732in,width=6.0926in]{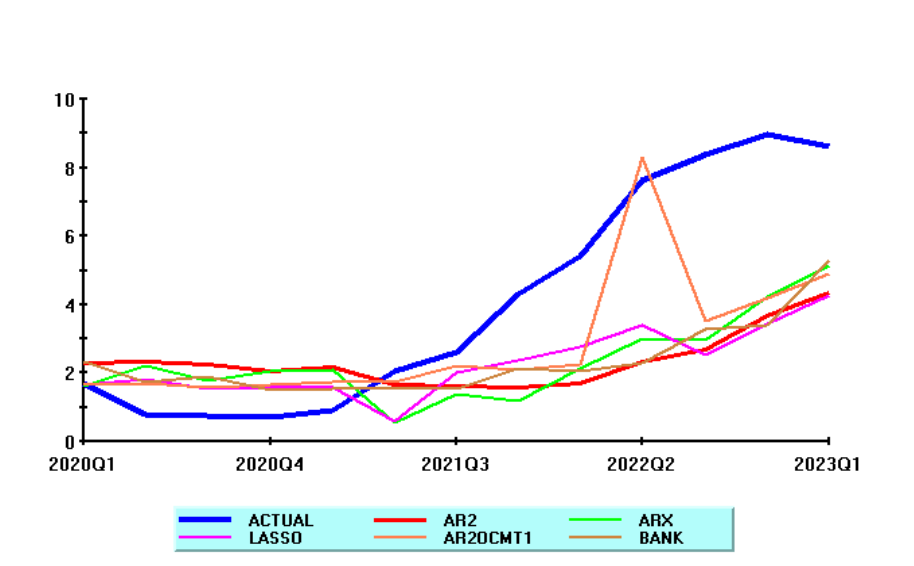}

\caption{\textbf{Plot of forecasts four quarters ahead}}%
\label{fig:4q_ahead}%
\end{figure}

Figure \ref{fig:4q_ahead} gives plots of actual inflation and forecasts four
quarters ahead. Section \ref{Forecasth=4} of the online supplement gives the
values. The case of $h=4$ is the only one where the ARX does not have the
lowest RMSFE. The lowest RMSFE is obtained by AR2-OCMT. It does well by having
a very high inflation forecast in $2022Q2$. This corresponds to the selection
of 12 extra variables in the sample ending in $2021q2$. It then rejoins the
pack in $2022q3$.%

\subsubsection{Summary}

Table \ref{tab:rmsfeq1} brings together the RMSFE for each of the selection
methods at the different horizons. Both the variable selection and forecasting
exercises highlight the importance of taking account of persistence and
foreign inflation for UK inflation forecasting. Lasso selects $\pi_{t}$ and
$\pi_{t}^{\ast}$ in all three forecast horizon models. ARX which includes UK
and foreign inflation as pre-selected variables tends to perform best in
forecasting but in the present application the OCMT component does not seem to
add much once the pre-selected variables are included. However, Lasso performs
rather poorly when it is conditioned on the preselected variables.%

\begin{table}[H]%
\caption
{Root mean square forecast errors by forecast horizons, h=1,2 and 4  over the period 2020q1-2023q1}%

\begin{center}%
\begin{tabular}
[c]{llccccc}\hline\hline
\textbf{Forecast Source} &  & \textbf{h=1} &  & \textbf{h=2} &  &
\textbf{h=4}\\\hline
AR2 &  & $0.9141$ &  & $1.6039$ &  & $3.2524$\\
ARX &  & $\mathbf{0.7884}$ &  & $\mathbf{1.3813}$ &  & $2.9883$\\
Lasso &  & $0.9696$ &  & $1.4109$ &  & $3.0131$\\
AR2--Lasso &  & $0.9617$ &  & $2.8719$ &  & $4.3440$\\
ARX--Lasso &  & $0.8750$ &  & $2.9231$ &  & $4.5021$\\
AR2--OCMT &  & $0.9233$ &  & $1.6800$ &  & $\mathbf{2.4643}$\\
ARX--OCMT &  & $\mathbf{0.7884}$ &  & $1.4217$ &  & $3.2470$\\
Bank of England &  & $1.3288$ &  & $1.6959$ &  & $3.0042$\\\hline\hline
\end{tabular}

\end{center}

{\footnotesize \emph{{}Note}: The RMSFE figures are taken from online
supplement Tables \ref{Forecasth=1}, \ref{Forecasth=2}, and \ref{Forecasth=4}.
The least value for RMSFE for each forecast horizon is shown in bold.}

\label{tab:rmsfeq1}%

\end{table}%
\bigskip

\subsection{Contemporaneous drivers}

Our forecasts of $\pi_{t+h}$ are based on variables observed at time $t$, and
do not depend on any conditioning. But, as noted above with respect to the
Bank of England, it is common to condition on contemporaneous values of
variables which are considered as \textit{proximate} causes of the variable to
be forecast. Even if such causal variables can be identified, however, it does
not mean that they help with forecasting - often such causal variables are
themselves difficult to forecast. The Bank of England over-estimated inflation
in 2022q4, correctly anticipating higher energy prices but not anticipating
the government energy price guarantees. This illustrates the dangers of
conditioning on variables that cannot be forecast. Understanding does not
necessarily translate into better forecasts. For example, knowing the causes
of earthquakes does not necessarily help in predicting them in a timely manner.

This point can be illustrated by including contemporaneous changes in oil
prices, in the UK inflation equation (over the period pre Covid-19 and the
full sample) for the case $h=1$. For both samples ${\footnotesize \Delta
poil}_{t+1}$ are highly statistically significant, but their lagged values
${\footnotesize \Delta poil}_{t}$ are not.

\bigskip%

\begin{table}[H]%
\caption
{Contemporaneous and lagged effects of oil price changes on UK inflation $(\pi
_{t+1})$}%

\begin{center}%
\begin{tabular}
[c]{ccrrcrr}\hline\hline
\textbf{Covariates} &  & \multicolumn{2}{c}{\textbf{1979q2--2019q4}} &  &
\multicolumn{2}{c}{\textbf{1979q2--2022q4}}\\\hline
$\pi_{t}$ &  & 0.936 & 0.936 & \multicolumn{1}{r}{} & 0.957 & 0.962\\
&  & (13.40) & (12.97) & \multicolumn{1}{r}{} & (13.70) & (13.23)\\
$\pi_{t-1}$ &  & --0.134 & --0.136 & \multicolumn{1}{r}{} & --0.145 &
--0.151\\
&  & (--2.05) & (--2.02) & \multicolumn{1}{r}{} & (--2.20) & (--2.21)\\
$\pi_{t}^{\ast}$ &  & 0.480 & 0.512 & \multicolumn{1}{r}{} & 0.600 & 0.524\\
&  & (4.14) & (3.31) & \multicolumn{1}{r}{} & (5.50) & (4.39)\\
$\pi_{t-1}^{\ast}$ &  & --0.381 & --0.316 & \multicolumn{1}{r}{} & --0.504 &
--0.432\\
&  & (--3.28) & (--2.52) & \multicolumn{1}{r}{} & (--4.64) & (--3.61)\\
$\Delta poil_{t+1}$ &  & \textbf{0.013} & \multicolumn{1}{c}{$\cdot$} &
\multicolumn{1}{r}{} & \textbf{0.012} & \multicolumn{1}{c}{$\cdot$}\\
&  & \textbf{(3.40)} & \multicolumn{1}{c}{$\cdot$} & \multicolumn{1}{r}{} &
\textbf{(3.91)} & \multicolumn{1}{c}{$\cdot$}\\
$\Delta poil_{t}$ &  & \multicolumn{1}{c}{$\cdot$} & 0.006 &
\multicolumn{1}{r}{} & \multicolumn{1}{c}{$\cdot$} & 0.005\\
&  & \multicolumn{1}{c}{$\cdot$} & (1.47) & \multicolumn{1}{r}{} &
\multicolumn{1}{c}{$\cdot$} & (1.37)\\
$\bar{R}^{2}$ &  & 0.953 & 0.950 & \multicolumn{1}{r}{} & 0.952 & 0.948\\
SER &  & 0.612 & 0.631 & \multicolumn{1}{r}{} & 0.626 & 0.651\\\hline\hline
\end{tabular}
\ 
\end{center}

\label{tab:oilpriceuk}%

\end{table}%
\bigskip

\section{Conclusion\label{Conc}}

High dimensional data are not a panacea; the data must have some predictive
content which might come from spatial or temporal sequential patterns.
\ Forecasting is particularly challenging either if there are unknown unknowns
(factors that are not even thought about) or if there are known factors that
are falsely believed to be important. When there are new global factors, like
Covid-19, or when a relevant variable has shown little variation over the
sample period, forecasting their effect is going to be problematic.

Many forecasting problems require a hierarchical structure where latent
factors at local and global levels are explicitly taken into account. This is
particularly relevant for macro forecasting in an increasingly inter-connected
world. It is important that we allow for global factors in national
forecasting exercises - and GVAR was an attempt in this direction.

A number of key methodological issues were illustrated with a simple approach
to forecasting UK inflation which has become a topic of public discussion.
This example showed both the power of parsimonious models and the importance
of global factors. There remain many challenges. How to allow for regime
change and parameter instability in the case of high-dimensional data
analysis? How to choose data samples? Our recent research suggests that it is
best to use long time series samples for variable selections, but consider
carefully what sample to use for forecasting. Given a set of selected
variables, parameter estimation can be based on different window sizes, or
down-weighting. Should we use ensemble or forecast averaging? Forecast
averaging will only work if the covariates used to forecast the target
variable are driven by strong common factors, otherwise one will be averaging
over noise.

There are some more general lessons. Econometric and statistical models must
not become a straightjacket. Forecasters should be open minded about factors
not included in their model, and acknowledge that forecasts are likely to be
wrong if unexpected shocks hit.

\bigskip
\bibliographystyle{econometrica}
\bibliography{PSForecasting}

\pagebreak

\label{Appendix}%

\numberwithin{equation}{section}%
%

\numberwithin{lemma}{section}%
%

\setcounter{page}{1}%
\setcounter{table}{0}%
\setcounter{figure}{0}%
%

\renewcommand{\thepage}{A.\arabic{page}}%
%

\renewcommand{\thetable}{A.\arabic{table}}
\renewcommand{\thefigure}{A.\arabic{figure}}%

\renewcommand{\theequation}{A.\arabic{equation}}%

\onehalfspacing
%

\section*{Appendix}%
%

\renewcommand{\thesection}{\Alph{section}}%

\appendix

\textbf{The Lasso procedure with a set of preselected variables}%
\protect\footnotemark
\footnotetext
{We are grateful to Dr. Mahrad Sharifvaghefi for providing the proofs in this appendix.}%

Let $\mathbf{y}=\left(  y_{1},y_{2},\cdots,y_{T}\right)  ^{\prime}$ be the
vector of observations for the target variable. Suppose we have a vector of
pre-selected covariates denoted by $\mathbf{z}_{t}=\left(  z_{1t}%
,z_{2t},\cdots,z_{mt}\right)  ^{\prime}$. Additionally, there is a vector of
covariates denoted by $\mathbf{x}_{t}=\left(  x_{1t},x_{2t},\cdots
,x_{nt}\right)  ^{\prime}$, from which we aim to select the relevant ones for
the target variable using the Lasso procedure. We can further stack the
observations for $\mathbf{z}_{t}$ and $\mathbf{x}_{t}$ in matrices
$\mathbf{Z}=\left(  \mathbf{z}_{1},\mathbf{z}_{2},\cdots,\mathbf{z}%
_{T}\right)  ^{\prime}$ and $\mathbf{X}=\left(  \mathbf{x}_{1},\mathbf{x}%
_{2},\cdots,\mathbf{x}_{T}\right)  ^{\prime}$, respectively. For a given value
of the tuning parameter, $\lambda$, the Lasso problem can be written as:
\[
\left(  \hat{\boldsymbol{\delta}}(\lambda),\hat{\boldsymbol{\beta}}%
(\lambda)\right)  ^{\prime}=\text{argmin}_{\mathbf{b}_{z},\mathbf{b}}\left\{
\left(  \mathbf{y}-\mathbf{Z}\mathbf{b}_{z}-\mathbf{X}\mathbf{b}\right)
^{\prime}\left(  \mathbf{y}-\mathbf{Z}\mathbf{b}_{z}-\mathbf{X}\mathbf{b}%
\right)  +\lambda\Vert\mathbf{b}\Vert_{1}\right\}  .
\]
Partition $\mathbf{X}=\left(  \mathbf{X}_{1},\mathbf{X}_{2}\right)  $ where
$\mathbf{X}_{1}$ is the matrix of covariates with the corresponding vector of
estimated coefficients, $\boldsymbol{\hat{\beta}}_{1}(\lambda)$, different
from zero and $\mathbf{X}_{2}$ is the matrix of covariates with the
corresponding vector of estimated coefficients, $\boldsymbol{\hat{\beta}}%
_{2}(\lambda)$, equal to zero. So, $\mathbf{X}\boldsymbol{\hat{\beta}}%
(\lambda)=\mathbf{X}_{1}\boldsymbol{\hat{\beta}}_{1}(\lambda)$. By the first
order conditions we have:
\begin{equation}
\mathbf{X}_{1}^{\prime}\left(  \mathbf{y}-\mathbf{Z}\hat{\boldsymbol{\delta}%
}(\lambda)-\mathbf{X}_{1}\hat{\boldsymbol{\beta}}_{1}(\lambda)\right)
-\lambda\text{sign}\left(  \hat{\boldsymbol{\beta}}_{1}(\lambda)\right)  =0,
\label{eq:foc1}%
\end{equation}%
\begin{equation}
\mathbf{Z}^{\prime}\left(  \mathbf{y}-\mathbf{Z}\hat{\boldsymbol{\delta}%
}(\lambda)-\mathbf{X}_{1}\hat{\boldsymbol{\beta}}_{1}(\lambda)\right)  =0.
\label{eq:foc2}%
\end{equation}
and
\begin{equation}
-\lambda\mathbf{1}\leq\mathbf{X}_{2}^{\prime}\left(  \mathbf{y}-\mathbf{Z}%
\hat{\boldsymbol{\delta}}(\lambda)-\mathbf{X}_{1}\hat{\boldsymbol{\beta}}%
_{1}(\lambda)\right)  \leq\lambda\mathbf{1}, \label{eq:foc3}%
\end{equation}
Where $\mathbf{1}$ represents a vector of ones. We can further conclude from
Equation (\ref{eq:foc2}) that:
\begin{equation}
\hat{\boldsymbol{\delta}}(\lambda)=\left(  \mathbf{Z}^{\prime}\mathbf{Z}%
\right)  ^{-1}\mathbf{Z}^{\prime}\left(  \mathbf{y}-\mathbf{X}_{1}%
\hat{\boldsymbol{\beta}}_{1}(\lambda)\right)  . \label{eq:foc2_cont}%
\end{equation}
By substituting $\hat{\boldsymbol{\delta}}(\lambda)$ from (\ref{eq:foc2_cont})
into (\ref{eq:foc1}), we have
\[
\mathbf{X}_{1}^{\prime}\left(  \mathbf{y}-\mathbf{X}_{1}\hat{\boldsymbol{\beta
}}_{1}(\lambda)-\mathbf{Z}\left(  \mathbf{Z}^{\prime}\mathbf{Z}\right)
^{-1}\mathbf{Z}^{\prime}\left(  \mathbf{y}-\mathbf{X}_{1}\hat
{\boldsymbol{\beta}}_{1}(\lambda)\right)  \right)  -\lambda\text{sign}\left(
\hat{\boldsymbol{\beta}}_{1}(\lambda)\right)  =0.
\]
We can further write this as:
\[
\mathbf{X}_{1}^{\prime}\left(  \mathbf{I}-\mathbf{Z}\left(  \mathbf{Z}%
^{\prime}\mathbf{Z}\right)  ^{-1}\mathbf{Z}^{\prime}\right)  \left(
\mathbf{y}-\mathbf{X}_{1}\hat{\boldsymbol{\beta}}_{1}(\lambda)\right)
-\lambda\text{sign}\left(  \hat{\boldsymbol{\beta}}_{1}(\lambda)\right)  =0.
\]
Therefore,
\begin{equation}
\tilde{\mathbf{X}}_{1}^{\prime}\left(  \tilde{\mathbf{y}}-\tilde{\mathbf{X}%
}_{1}\hat{\boldsymbol{\beta}}_{1}(\lambda)\right)  -\lambda\text{sign}\left(
\hat{\boldsymbol{\beta}}_{1}(\lambda)\right)  =0, \label{eq:foc1_v2}%
\end{equation}
where $\tilde{\mathbf{X}}_{1}=\mathbf{M}_{Z}\mathbf{X}_{1}$, $\tilde
{\mathbf{y}}=\mathbf{M}_{Z}\mathbf{y}$ and $\mathbf{M}_{Z}=\mathbf{I}%
-\mathbf{Z}\left(  \mathbf{Z}^{\prime}\mathbf{Z}\right)  ^{-1}\mathbf{Z}%
^{\prime}$.

Similarly, by substituting $\hat{\boldsymbol{\delta}}(\lambda)$ from
(\ref{eq:foc2_cont}) into (\ref{eq:foc3}), we have
\begin{equation}
-\lambda\mathbf{1}\leq\tilde{\mathbf{X}}_{2}^{\prime}\left(  \tilde
{\mathbf{y}}-\tilde{\mathbf{X}}_{1}\hat{\boldsymbol{\beta}}_{1}(\lambda
)\right)  \leq\lambda\mathbf{1}. \label{eq:foc3_v2}%
\end{equation}
Note that (\ref{eq:foc1_v2}) and (\ref{eq:foc3_v2}) are the first order
conditions of the following Lasso problem:
\begin{equation}
\hat{\boldsymbol{\beta}}(\lambda)=\text{argmin}_{\mathbf{b}}\left\{  \left(
\tilde{\mathbf{y}}-\tilde{\mathbf{X}}\mathbf{b}\right)  ^{\prime}\left(
\tilde{\mathbf{y}}-\tilde{\mathbf{X}}\mathbf{b}\right)  +\lambda
||\mathbf{b}||_{1}\right\}  . \label{eq:lasso_v2}%
\end{equation}
Therefore, we can first obtain the estimator of the vector coefficients for
$\mathbf{X}$, $\hat{\boldsymbol{\beta}}(\lambda)$, by solving the Lasso
problem given by (\ref{eq:lasso_v2}) and then estimate the vector of
coefficients for $\mathbf{Z}$, $\hat{\boldsymbol{\delta}}(\lambda)$, by using
Equation (\ref{eq:foc2_cont}).

\pagebreak\bigskip%

\normalsize

\setcounter{section}{0} \renewcommand{\thesection}{S-\arabic{section}}

\setcounter{table}{0} \renewcommand{\thetable}{S.\arabic{table}}

\setcounter{page}{0} \renewcommand{\thepage}{S.\arabic{page}}

\setcounter{equation}{0} \renewcommand{\theequation}{S.\arabic{equation}}

\setcounter{figure}{0} \renewcommand{\thefigure}{S.\arabic{figure}}

\ \quad\vspace{0.05in}

\begin{center}
\textbf{Online Supplement}

to

\textbf{High-dimensional forecasting with known knowns and known
unknowns}\newline

\bigskip

\bigskip by

M. Hashem Pesaran

University of Southern California, USA and Trinity College, Cambridge,
UK\bigskip

Ron P. Smith

Birkbeck, University of London, United Kingdom\newline

\bigskip\bigskip

\bigskip

\today\bigskip
\end{center}

%

\thispagestyle{empty}%
\newpage%

\renewcommand{\thepage}{S.\arabic{page}}%
\pagestyle{myheader}

\section{Introduction}

This online supplement presents the details of the methods used to forecast UK
inflation, the variables selected and the results obtained. For this purpose
the framework set out in the paper can be summarised as:
\begin{equation}
\pi_{t+h}=c_{h}+\mathbf{a}_{h}^{\prime}\mathbf{z}_{t}+\sum_{j=1}^{K}\beta
_{jh}I(j\in DGP)x_{jt}+u_{h,t+h}. \label{DGP2}%
\end{equation}

The target variable is average annual UK inflation, labelled DPUK4, defined as
$\pi_{t+h}=100\times\log(p_{t+h}/p_{t+h-4}),$ where $p_{t}$ is the UK consumer
price index taken from the IMF International Financial Statistics.\footnote{In
terms of the GVAR data, which report $dp_{t}=\log(p_{t}/p_{t-1})$ where
$p_{t}$ is the CPI, $\pi_{t}$ is defined by $\pi_{t}=100\times(dp_{t}%
+dp_{t-1}+dp_{t-2}+dp_{t-3})$.} $\mathbf{z}_{t}$ is the vector of pre-selected
variables which we set to ($\pi_{t},\Delta\pi_{t})$ or ($\pi_{t},\Delta\pi
_{t}$,$\pi_{t}^{\ast}$, and $\Delta\pi_{t}^{\ast}$). The active set, $x_{jt},$
$j=1,2,...,K,$ is the list of variables from which selection is made.

The active set consists of the 26 variables detailed in Table
\ref{tab:Active_Set_Variables} of the paper plus their first differences (in
total $K=52$). As noted in the paper, whereas including a current and lagged
value is equivalent to including a current value and its change in estimation,
it is not in the selection. We denote $\mathcal{A}_{m}^{h}$ as the active set
corresponding to model $m$ among the 9 models that are examined. For each
horizon, selection proceeds recursively using an expanding window starting in
1979q2 initially ending in 2019q1, for $h=4,$ 2019q3 for $h=2,$ and 2019q4 for
$h=4.$ The window then expands one quarter at a time to 2023q1. We end up with
$13$ point forecasts for the quarters $2020q1-2023q1$. Since the model is
re-estimated for each forecast period and horizon, the variables selected may
change by quarter and horizon.

\section{Models}

The first two models set all the $\beta_{jh}=0$ and differ in what is included
in $\mathbf{z}_{t}.$

The $AR2$ model is:%

\[
\pi_{t+h}=c_{h}+a_{1h}\pi_{t}+a_{2h}\Delta\pi_{t}+u_{h,t+h}.
\]

The $ARX$ adds UK-specific foreign inflation, $\pi_{t}^{\ast}$ and its
change$:$
\[
\pi_{t+h}=c_{h}+a_{1h}\pi_{t}+a_{2h}\Delta\pi_{t}+a_{3h}\pi_{t}^{\ast}%
+a_{4h}\Delta\pi_{t}^{\ast}+u_{h,t+h}%
\]

The next model sets $\mathbf{a}_{h}^{\prime}=0$ and just selects using Lasso,
which performs a variable selection from the whole active set of $52$
variables inclusive of $\pi_{t},\Delta\pi_{t}$,$\pi_{t}^{\ast}$, and
$\Delta\pi_{t}^{\ast}$. All covariates are standardized before solving the
following minimization problem:
\[
\min_{\boldsymbol{\beta}\in\mathbb{R}^{|\mathcal{A}_{\text{Lasso}}^{h}|}}%
\frac{1}{2T}\sum_{t=1}^{T}\left(  \pi_{t+h}-c_{h}-\sum_{j\in\mathcal{A}%
_{\text{Lasso}}^{h}}\tilde{x}_{tj}\beta_{j}\right)  ^{2}+\lambda\sum
_{j\in\mathcal{A}_{\text{Lasso}}^{h}}\left\Vert \beta_{j}\right\Vert _{1},
\]
where $\tilde{x}_{tj}=(x_{jt}-\bar{x}_{jT})/s_{jT}$, $\bar{x}_{jT}=T^{-1}%
\sum_{t=1}^{T}x_{jt}$ and $s_{jT}=T^{-1}\sum_{t=1}^{T}\left(  x_{jt}-\bar
{x}_{jT}\right)  $. $|\mathcal{A}_{\text{Lasso}}^{h}|$ represents the
cardinality of the active set, and $||\cdot||_{1}$ is the $\ell_{1}$-norm. The
regularization parameter $\hat{\lambda}_{hT}$ is chosen via 10-fold
cross-validation (CV). As discussed in the paper we adapt the standard CV
procedures to take account of the persistence and changing variance of time
series data. In the standard procedure the CV subsets (folds) are typically
chosen randomly, but since these are time series where order matters, we
retain the order of the time series data in construction of subsets using all
the data, not dropping observations between subsets. In addition, the standard
10-fold procedure chooses the $\hat{\lambda}_{hT}$ that minimises the pooled
MSE over the ten subsets. But when variances differ substantially over
subsets, pooling is not appropriate, instead we follow Chudik, Kapetanios, and
Pesaran (2018, CKP), and use the average of the $\hat{\lambda}_{hT}$ chosen in
each subset. Full details are provided in the online simulation appendix to
CKP (2018).

We considered a number of Lasso models where we did not apply selection to the
AR2 or ARX variables which were included in the $\boldsymbol{z}_{t}$. Then
Lasso was applied to the active set having removed the effect of
$\boldsymbol{z}_{t}.$ This gives models AR2-LASSO; ARX-LASSO. We now consider
GOCMT-based Models. Like OCMT, GOCMT allows for the multiple testing nature of
the procedure ($K$ separate tests - with $K$ large) by increasing the level of
significance with $K$. But it also includes the principal component of the
standardised values of the active set, $\boldsymbol{\hat{\varkappa}}_{t},$ to
allow for the correlations between the variables. In the first stage, $K$
separate OLS regressions are computed entering the variables from the active
set one at a time, together with any $\boldsymbol{z}_{t}$ and
$\boldsymbol{\hat{\varkappa}}_{t}$:
\[
y_{t+h}=c_{h}+\mathbf{a}_{h}^{\prime}\boldsymbol{z}_{t}+\mathbf{b}_{h}%
^{\prime}\boldsymbol{\hat{\varkappa}}_{t}+\phi_{jh}x_{jt}+e_{j,h,t+h}\text{,
}t=1,2,...,T\text{, for }j=1,2,...,K,
\]

Denote the $t$-ratio of $\phi_{jh}$ by $t_{\hat{\phi}_{j,\left(  1\right)  }}%
$. Then variable $j$ is selected if%
\[
\widehat{\mathcal{J}}_{j,(1)}=I\left[  \left\vert t_{\hat{\phi}_{j,\left(
1\right)  }}\right\vert >c_{p}(K)\right]  ,\text{ for }j=1,2,...,K,
\]
where the critical value is given by
\[
c_{p}(K,\delta)=\Phi^{-1}\left(  1-\frac{p}{2K^{\delta}}\right)  .
\]
$p$ is the nominal size set to $5\%$, $\Phi^{-1}(\cdot)$ is the inverse of a
standard normal distribution function and $\delta$ is a fixed constant set
close to $1$. We experiment with two values for $\delta,$ $\delta=1$ and
$\delta=1.5$.

The principal component $\boldsymbol{\hat{\varkappa}}_{t}$ is used just to
control for correlations across the variables in the active set at the
selection stage, it is not included in the final forecast regression. The
AR2-OCMT model includes the AR2 in the $\boldsymbol{z}_{t}$ and applies GOCMT
to the rest of the active set, similarly for ARX-OCMT.

Thus we have a set of 9 models: (1) AR2, (2) ARX, (3) LASSO, (4) AR2-LASSO,
(5) ARX-LASSO, (6) AR2-OCMT ($\delta=1$), (7) ARX-OCMT ($\delta=1$), (8)
AR2-OCMT ($\delta=1.5$), (9) ARX-OCMT ($\delta=1.5$).

\section{Variable Selection\label{Select}}

This section sets out the variables chosen by each selection method for each
horizon by the quarter in which the sample ended. The abbreviations for the
variables are given in Table \ref{tab:Active_Set_Variables}, in the paper.

\subsection{Variables selected by OCMT\label{SelectOCMT}}

\subsubsection{For h=1 quarter ahead models\label{SelOCMTh1}}

This sub-section sets out the variables chosen by OCMT for $h=1$.

AR2-OCMT selects DWUK for samples ending in 2021q2, 2021q4 and 2022q1 and no
other additional variables. ARX-OCMT ($\delta=1$), AR2-OCMT ($\delta=1.5$),
and ARX-OCMT ($\delta=1.5$) do not select any additional variables.

\subsubsection{For h=2 quarter ahead models\label{SelOCMTh2}}

This sub-section sets out the variables chosen by OCMT for $h=2$.%

\begin{table}[H]%
%

\caption{Selected Variables by AR2-OCMT ($\delta=1$) 2q Ahead}%

\begin{center}
{\footnotesize {}%
\begin{tabular}
[c]{cccccccc}\hline\hline
{}\# &  & \multicolumn{6}{c}{{}Ending Year}\\\hline
&  & {}2019q3 & {}2019q4 & {}2020q1 & {}2020q2 & {}2020q3 & {}2020q4\\\hline
{}1 &  & {}DMUK4 & {}DMUK4 & {}DMUK4 & {}DMUK4 & {}DMUK4 & {}DMUK4\\
{}2 &  & {}DEPUK4 & {}DEPUK4 & {}DEPUK4 & {}DEPUK4 & {}DEPUK4 & {}DEPUK4\\
{}3 &  & {}$\cdot$ & {}$\cdot$ & {}$\cdot$ & {}$\cdot$ & {}$\cdot$ & {}DWUK4\\
&  &  &  &  &  &  & \\\hline
&  & {}2021q1 & {}2021q2 & {}2021q3 & {}2021q4 & {}2022q1 & {}2022q2\\\hline
{}1 &  & {}DMUK4 & {}DMUK4 & {}DMUK4 & {}DMUK4 & {}DMUK4 & {}DMUK4\\
{}2 &  & {}DEPUK4 & {}DEPUK4 & {}DEPUK4 & {}DEPUK4 & {}DEPUK4 & {}DEPUK4\\
{}3 &  & {}DWUK4 & {}DWUK4 & {}DWUK4 & {}DWUK4 & {}DWUK4 & {}DWUK4\\
&  &  &  &  &  &  & \\\hline
&  & {}2022q3 & {}2022q4 & {}2023q1 &  &  & \\\hline
{}1 &  & {}DMUK4 & {}DMUK4 & {}DMUK4 &  &  & \\
{}2 &  & {}DWUK4 & {}DWUK4 & {}DWUK4 &  &  & \\\hline
&  & \multicolumn{6}{c}{{}$\delta=1$}\\\hline\hline
\end{tabular}
}
\end{center}

%

\end{table}%

ARX-OCMT {}$\delta=1$ selects just DMUK4 for every sample.%

\begin{table}[H]%
%

\caption{Selected Variables by AR2-OCMT ($\delta=1.5$) 2q Ahead}%

\begin{center}
{\footnotesize {}%
\begin{tabular}
[c]{cccccccc}\hline\hline
{}\# &  & \multicolumn{6}{c}{{}Ending Year}\\\hline
&  & {}2019q3 & {}2019q4 & {}2020q1 & {}2020q2 & {}2020q3 & {}2020q4\\\hline
{}1 &  & {}$\cdot$ & {}$\cdot$ & {}$\cdot$ & {}$\cdot$ & {}$\cdot$ & {}DMUK4\\
&  &  &  &  &  &  & \\\hline
&  & {}2021q1 & {}2021q2 & {}2021q3 & {}2021q4 & {}2022q1 & {}2022q2\\\hline
{}1 &  & {}DMUK4 & {}DMUK4 & {}DMUK4 & {}DMUK4 & {}DMUK4 & {}DMUK4\\
{}2 &  & {}$\cdot$ & {}DWUK4 & {}DWUK4 & {}DWUK4 & {}DWUK4 & {}DWUK4\\
&  &  &  &  &  &  & \\\hline
&  & {}2022q3 & {}2022q4 & {}2023q1 &  &  & \\\hline
{}1 &  & {}DWUK4 & {}$\cdot$ & {}$\cdot$ &  &  & \\
{}2 &  & {}$\cdot$ & {}$\cdot$ & {}$\cdot$ &  &  & \\\hline
&  & \multicolumn{6}{c}{{}$\delta=1.5$}\\\hline\hline
\end{tabular}
}
\end{center}

%

\end{table}%

ARX-OCMT $\delta=1.5$ selects DMUK4 for 2020q4-2022q2.

\subsubsection{For h=4 quarter ahead models\label{SelOCMTh4}}

This sub-section sets out the variables chosen by OCMT selection method for
$h=4$.%

\begin{table}[H]%
%

\caption{Selected Variables by AR2-OCMT ($\delta=1$) 4q Ahead}%

\begin{center}
{\footnotesize {}%
\begin{tabular}
[c]{cccccccc}\hline\hline
{}\# &  & \multicolumn{6}{c}{{}Ending Year}\\\hline
&  & {}2019q1 & {}2019q2 & {}2019q3 & {}2019q4 & {}2020q1 & {}2020q2\\\hline
{}1 &  & {}DMUK4 & {}DMUK4 & {}DMUK4 & {}DMUK4 & {}DMUK4 & {}DMUK4\\
{}2 &  & {}DEPUK4 & {}DEPUK4 & {}DEPUK4 & {}DEPUK4 & {}DEPUK4 & {}DEPUK4\\
{}3 &  & {}DPSUK4 & {}DPSUK4 & {}DPSUK4 & {}DPSUK4 & {}DPSUK4 & {}DPSUK4\\
&  &  &  &  &  &  & \\\hline
&  & {}2020q3 & {}2020q4 & {}2021q1 & {}2021q2 & {}2021q3 & {}2021q4\\\hline
{}1 &  & {}DMUK4 & {}DMUK4 & {}DMUK4 & {}DDYUK4 & {}DEMUK4 & {}LRUK4\\
{}2 &  & {}DEPUK4 & {}DEPUK4 & {}DEPUK4 & {}DGAPUK8 & {}LRUK4 & {}DMUK4\\
{}3 &  & {}DPSUK4 & {}DPSUK4 & {}DPSUK4 & {}DGAPUK12 & {}DMUK4 & {}DPMAT4\\
{}4 &  & {}$\cdot$ & {}$\cdot$ & {}$\cdot$ & {}DEMUK4 & {}DPMAT4 & {}DPSUK4\\
{}5 &  & {}$\cdot$ & {}$\cdot$ & {}$\cdot$ & {}RUK4 & {}DPSUK4 & {}LRUS4\\
{}6 &  & {}$\cdot$ & {}$\cdot$ & {}$\cdot$ & {}LRUK4 & {}LRUS4 & {}DWUK4\\
{}7 &  & {}$\cdot$ & {}$\cdot$ & {}$\cdot$ & {}DMUK4 & {}DWUK4 & {}$\cdot$\\
{}8 &  & {}$\cdot$ & {}$\cdot$ & {}$\cdot$ & {}DPMAT4 & {}$\cdot$ & {}$\cdot
$\\
{}9 &  & {}$\cdot$ & {}$\cdot$ & {}$\cdot$ & {}DPSUK4 & {}$\cdot$ & {}$\cdot
$\\
{}10 &  & {}$\cdot$ & {}$\cdot$ & {}$\cdot$ & {}RUS4 & {}$\cdot$ & {}$\cdot$\\
{}11 &  & {}$\cdot$ & {}$\cdot$ & {}$\cdot$ & {}LRUS4 & {}$\cdot$ & {}$\cdot
$\\
{}12 &  & {}$\cdot$ & {}$\cdot$ & {}$\cdot$ & {}DWUK4 & {}$\cdot$ & {}$\cdot
$\\
&  &  &  &  &  &  & \\\hline
&  & {}2022q1 & {}2022q2 & {}2022q3 & {}2022q4 & {}2023q1 & \\\hline
{}1 &  & {}DMUK4 & {}DMUK4 & {}DMUK4 & {}DMUK4 & {}DMUK4 & \\
{}2 &  & {}DPMAT4 & {}DPMAT4 & {}DPMAT4 & {}DPMAT4 & {}DPMAT4 & \\
{}3 &  & {}DPSUK4 & {}DPSUK4 & {}DPSUK4 & {}DPSUK4 & {}DWUK4 & \\
{}4 &  & {}LRUS4 & {}DWUK4 & {}DWUK4 & {}DWUK4 & {}$\cdot$ & \\
{}5 &  & {}DWUK4 & {}$\cdot$ & {}$\cdot$ & {}$\cdot$ & {}$\cdot$ & \\\hline
&  & \multicolumn{6}{c}{{}$\delta=1$}\\\hline\hline
\end{tabular}
}
\end{center}

%

\end{table}%
%

\begin{table}[H]%
%

\caption{Selected Variables by ARX-OCMT ($\delta=1$) 4q Ahead}%

\begin{center}
{\footnotesize {}%
\begin{tabular}
[c]{cccccccc}\hline\hline
{}\# &  & \multicolumn{6}{c}{{}Ending Year}\\\hline
&  & {}2019q1 & {}2019q2 & {}2019q3 & {}2019q4 & {}2020q1 & {}2020q2\\\hline
{}1 &  & {}DEMUK4 & {}DEMUK4 & {}DEMUK4 & {}DEMUK4 & {}DEMUK4 & {}DEMUK4\\
{}2 &  & {}DMUK4 & {}DMUK4 & {}DMUK4 & {}DMUK4 & {}DMUK4 & {}DMUK4\\
&  &  &  &  &  &  & \\\hline
&  & {}2020q3 & {}2020q4 & {}2021q1 & {}2021q2 & {}2021q3 & {}2021q4\\\hline
{}1 &  & {}DEMUK4 & {}DEMUK4 & {}DEMUK4 & {}DEMUK4 & {}DEMUK4 & {}DEMUK4\\
{}2 &  & {}DMUK4 & {}DMUK4 & {}DMUK4 & {}DMUK4 & {}DMUK4 & {}DMUK4\\
&  &  &  &  &  &  & \\\hline
&  & {}2022q1 & {}2022q2 & {}2022q3 & {}2022q4 & {}2023q1 & \\\hline
{}1 &  & {}DEMUK4 & {}DMUK4 & {}DMUK4 & {}DMUK4 & {}DMUK4 & \\
{}2 &  & {}DMUK4 & {}$\cdot$ & {}$\cdot$ & {}$\cdot$ & {}$\cdot$ & \\
{}3 &  & {}DPMAT4 & {}$\cdot$ & {}$\cdot$ & {}$\cdot$ & {}$\cdot$ & \\\hline
&  & \multicolumn{6}{c}{{}$\delta=1$}\\\hline\hline
\end{tabular}
}
\end{center}

%

\end{table}%
%

\begin{table}[H]%
%

\caption{Selected Variables by AR2-OCMT ($\delta=1.5$) 4q Ahead}%

\begin{center}
{\footnotesize {}%
\begin{tabular}
[c]{cccccccc}\hline\hline
{}\# &  & \multicolumn{6}{c}{{}Ending Year}\\\hline
&  & {}2019q1 & {}2019q2 & {}2019q3 & {}2019q4 & {}2020q1 & {}2020q2\\\hline
{}1 &  & {}DMUK4 & {}DMUK4 & {}DMUK4 & {}DMUK4 & {}DMUK4 & {}DMUK4\\
&  &  &  &  &  &  & \\\hline
&  & {}2020q3 & {}2020q4 & {}2021q1 & {}2021q2 & {}2021q3 & {}2021q4\\\hline
{}1 &  & {}DMUK4 & {}DMUK4 & {}DMUK4 & {}DGAPUK8 & {}DMUK4 & {}DMUK4\\
{}2 &  & {}$\cdot$ & {}$\cdot$ & {}$\cdot$ & {}DGAPUK12 & {}DPSUK4 &
{}DPMAT4\\
{}3 &  & {}$\cdot$ & {}$\cdot$ & {}$\cdot$ & {}DEMUK4 & {}LRUS4 & {}DPSUK4\\
{}4 &  & {}$\cdot$ & {}$\cdot$ & {}$\cdot$ & {}LRUK4 & {}$\cdot$ & {}$\cdot$\\
{}5 &  & {}$\cdot$ & {}$\cdot$ & {}$\cdot$ & {}DMUK4 & {}$\cdot$ & {}$\cdot$\\
{}6 &  & {}$\cdot$ & {}$\cdot$ & {}$\cdot$ & {}DPMAT4 & {}$\cdot$ & {}$\cdot
$\\
{}7 &  & {}$\cdot$ & {}$\cdot$ & {}$\cdot$ & {}DPSUK4 & {}$\cdot$ & {}$\cdot
$\\
{}8 &  & {}$\cdot$ & {}$\cdot$ & {}$\cdot$ & {}LRUS4 & {}$\cdot$ & {}$\cdot$\\
&  &  &  &  &  &  & \\\hline
&  & {}2022q1 & {}2022q2 & {}2022q3 & {}2022q4 & {}2023q1 & \\\hline
{}1 &  & {}DMUK4 & {}DMUK4 & {}DMUK4 & {}DMUK4 & {}DMUK4 & \\
{}2 &  & {}DPMAT4 & {}DPMAT4 & {}DPMAT4 & {}DWUK4 & {}DWUK4 & \\
{}3 &  & {}DPSUK4 & {}DWUK4 & {}DWUK4 & {}$\cdot$ & {}$\cdot$ & \\
{}4 &  & {}DWUK4 & {}$\cdot$ & {}$\cdot$ & {}$\cdot$ & {}$\cdot$ & \\\hline
&  & \multicolumn{6}{c}{{}$\delta=1.5$}\\\hline\hline
\end{tabular}
}
\end{center}

%

\end{table}%
%

\begin{table}[H]%
%

\caption{Selected Variables by ARX-OCMT ($\delta=1.5$) 4q AheadA}%

\begin{center}
{\footnotesize {}%
\begin{tabular}
[c]{cccccccc}\hline\hline
{}\# &  & \multicolumn{6}{c}{{}Ending Year}\\\cline{3-8}
&  & {}2019q1 & {}2019q2 & {}2019q3 & {}2019q4 & {}2020q1 & {}2020q2\\\hline
{}1 &  & {}DEMUK4 & {}DEMUK4 & {}DEMUK4 & {}DEMUK4 & {}DEMUK4 & {}DEMUK4\\
{}2 &  & {}DMUK4 & {}DMUK4 & {}DMUK4 & {}DMUK4 & {}DMUK4 & {}DMUK4\\
&  &  &  &  &  &  & \\\hline
&  & {}2020q3 & {}2020q4 & {}2021q1 & {}2021q2 & {}2021q3 & {}2021q4\\\hline
{}1 &  & {}DEMUK4 & {}DEMUK4 & {}DEMUK4 & {}DEMUK4 & {}DEMUK4 & {}DEMUK4\\
{}2 &  & {}DMUK4 & {}DMUK4 & {}DMUK4 & {}DMUK4 & {}DMUK4 & {}DMUK4\\
&  &  &  &  &  &  & \\\hline
&  & {}2022q1 & {}2022q2 & {}2022q3 & {}2022q4 & {}2023q1 & \\\hline
{}1 &  & {}DMUK4 & {}DMUK4 & {}DMUK4 & {}DMUK4 & {}DMUK4 & \\\hline
&  & \multicolumn{6}{c}{{}$\delta=1.5$}\\\hline\hline
\end{tabular}
}
\end{center}

%

\end{table}%

\pagebreak

\subsection{Variables selected by Lasso\label{SelectedLasso}}

\subsubsection{For h=1 quarter ahead models}

This sub-section sets out the variables chosen by Lasso for $h=1$.%

\begin{table}[H]%
%

\caption{Selected Variables by Lasso  1q Ahead}%

\begin{center}
{\footnotesize
\begin{tabular}
[c]{ccccccccc}\hline\hline
{} &  & \multicolumn{7}{c}{{}Ending Year}\\\hline
\# &  & {}2019q4 & {}2020q1 & {}2020q2 & {}2020q3 & {}2020q4 & {}2021q1 &
{}2021q2\\\hline
{}1 &  & {\scriptsize {}DPUK4} & {\scriptsize {}DPUK4} & {\scriptsize {}DPUK4}
& {\scriptsize {}DPUK4} & {\scriptsize {}DPUK4} & {\scriptsize {}DPUK4} &
{\scriptsize {}DPUK4}\\
{}2 &  & {\scriptsize {}DDPUK4} & {\scriptsize {}DDPUK4} & {\scriptsize {}%
DDPUK4} & {\scriptsize {}DDPUK4} & {\scriptsize {}DDPUK4} & {\scriptsize {}%
DDPUK4} & {\scriptsize {}DDPUK4}\\
{}3 &  & {\scriptsize {}DLRUK4} & {\scriptsize {}DLRUK4} & {\scriptsize {}%
DRUK4} & {\scriptsize {}DLRUK4} & {\scriptsize {}DLRUK4} & {\scriptsize {}%
DEMUK4} & {\scriptsize {}DLRUK4}\\
{}4 &  & {\scriptsize {}DMUK4} & {\scriptsize {}DMUK4} & {\scriptsize {}%
DLRUK4} & {\scriptsize {}DMUK4} & {\scriptsize {}DMUK4} & {\scriptsize {}%
DRUK4} & {\scriptsize {}DMUK4}\\
{}5 &  & {\scriptsize {}DPMUK4} & {\scriptsize {}DPMUK4} & {\scriptsize {}%
DMUK4} & {\scriptsize {}DPMUK4} & {\scriptsize {}DPMUK4} & {\scriptsize {}%
DLRUK4} & {\scriptsize {}DPMUK4}\\
{}6 &  & {\scriptsize {}DPSUK4} & {\scriptsize {}DPSUK4} & {\scriptsize {}%
DDPOIL4} & {\scriptsize {}DPSUK4} & {\scriptsize {}DPSUK4} & {\scriptsize {}%
DMUK4} & {\scriptsize {}DPSUK4}\\
{}7 &  & {\scriptsize {}DDPSUK4} & {\scriptsize {}DDPSUK4} & {\scriptsize {}%
DPMUK4} & {\scriptsize {}DDPSUK4} & {\scriptsize {}DWUK4} & {\scriptsize {}%
DDEQUK4} & {\scriptsize {}DDPSUK4}\\
{}8 &  & {\scriptsize {}DWUK4} & {\scriptsize {}DWUK4} & {\scriptsize {}%
DPSUK4} & {\scriptsize {}DWUK4} & {\scriptsize {}}$\cdot$ & {\scriptsize {}%
DPMUK4} & {\scriptsize {}DWUK4}\\
{}9 &  & {\scriptsize {}}$\cdot$ & {\scriptsize {}}$\cdot$ & {\scriptsize {}%
DDPSUK4} & {\scriptsize {}}$\cdot$ & {\scriptsize {}}$\cdot$ & {\scriptsize {}%
DPSUK4} & {\scriptsize {}}$\cdot$\\
{}10 &  & {\scriptsize {}}$\cdot$ & {\scriptsize {}}$\cdot$ & {\scriptsize {}%
DDYSUK4} & {\scriptsize {}}$\cdot$ & {\scriptsize {}}$\cdot$ & {\scriptsize {}%
DDPSUK4} & {\scriptsize {}}$\cdot$\\
{}11 &  & {\scriptsize {}}$\cdot$ & {\scriptsize {}}$\cdot$ & {\scriptsize {}%
DGAPSUK12} & {\scriptsize {}}$\cdot$ & {\scriptsize {}}$\cdot$ &
{\scriptsize {}DWUK4} & {\scriptsize {}}$\cdot$\\
{}12 &  & {\scriptsize {}}$\cdot$ & {\scriptsize {}}$\cdot$ & {\scriptsize {}%
DWUK4} & {\scriptsize {}}$\cdot$ & {\scriptsize {}}$\cdot$ & {\scriptsize {}%
}$\cdot$ & {\scriptsize {}}$\cdot$\\\hline
{}$\lambda$ &  & {\scriptsize {}0.1094} & {\scriptsize {}0.1093} &
{\scriptsize {}0.0736} & {\scriptsize {}0.1098} & {\scriptsize {}0.1101} &
{\scriptsize {}0.0741} & {\scriptsize {}0.1099}\\\hline
&  &  &  &  &  &  &  & \\\hline
\# &  & {\scriptsize {}2021q3} & {\scriptsize {}2021q4} & {\scriptsize {}%
2022q1} & {\scriptsize {}2022q2} & {\scriptsize {}2022q3} & {\scriptsize {}%
2022q4} & {\scriptsize {}2023q1}\\\hline
{}1 &  & {\scriptsize {}DPUK4} & {\scriptsize {}DPUK4} & {\scriptsize {}DPUK4}
& {\scriptsize {}DPUK4} & {\scriptsize {}DPUK4} & {\scriptsize {}DPUK4} &
{\scriptsize {}DPUK4}\\
{}2 &  & {\scriptsize {}DDPUK4} & {\scriptsize {}DDPUK4} & {\scriptsize {}%
DDPUK4} & {\scriptsize {}DDPUK4} & {\scriptsize {}DDPUK4} & {\scriptsize {}%
DDPUK4} & {\scriptsize {}DDPUK4}\\
{}3 &  & {\scriptsize {}DLRUK4} & {\scriptsize {}DLRUK4} & {\scriptsize {}%
DLRUK4} & {\scriptsize {}DLRUK4} & {\scriptsize {}DLRUK4} & {\scriptsize {}%
DLRUK4} & {\scriptsize {}DLRUK4}\\
{}4 &  & {\scriptsize {}DMUK4} & {\scriptsize {}DMUK4} & {\scriptsize {}DMUK4}
& {\scriptsize {}DPMUK4} & {\scriptsize {}DPMUK4} & {\scriptsize {}DPMUK4} &
{\scriptsize {}DPMUK4}\\
{}5 &  & {\scriptsize {}DPMUK4} & {\scriptsize {}DPMUK4} & {\scriptsize {}%
DPMUK4} & {\scriptsize {}DPSUK4} & {\scriptsize {}DPSUK4} & {\scriptsize {}%
DPSUK4} & {\scriptsize {}DPSUK4}\\
{}6 &  & {\scriptsize {}DPSUK4} & {\scriptsize {}DPSUK4} & {\scriptsize {}%
DPSUK4} & {\scriptsize {}DDPSUK4} & {\scriptsize {}DDPSUK4} & {\scriptsize {}%
DDPSUK4} & {\scriptsize {}DDPSUK4}\\
{}7 &  & {\scriptsize {}DDPSUK4} & {\scriptsize {}DDPSUK4} & {\scriptsize {}%
DDPSUK4} & {\scriptsize {}DWUK4} & {\scriptsize {}DWUK4} & {\scriptsize {}%
DWUK4} & {\scriptsize {}DWUK4}\\
{}8 &  & {\scriptsize {}DWUK4} & {\scriptsize {}DWUK4} & {\scriptsize {}DWUK4}
& {\scriptsize {}}$\cdot$ & {\scriptsize {}}$\cdot$ & {\scriptsize {}}$\cdot$
& {\scriptsize {}}$\cdot$\\\hline
{}$\lambda$ &  & {}0.1096 & {}0.1091 & {}0.1091 & {}0.1104 & {}0.1128 &
{}0.1159 & {}0.1184\\\hline\hline
\end{tabular}
}
\end{center}

{\footnotesize \emph{{}Note}{}: }${\footnotesize \hat{\lambda}}_{hT}$
{\footnotesize is the estimate of the Lasso penalty parameter computed by
10-fold cross-validation.}%

\end{table}%
%

\begin{table}[H]%
%

\caption{Selected Variables by AR2-Lasso 1q Ahead}%

\begin{center}
{\footnotesize
\begin{tabular}
[c]{ccccccccc}\hline\hline
{} &  & \multicolumn{7}{c}{{}Ending Year}\\\hline
\# &  & {}2019q4 & {}2020q1 & {}2020q2 & {}2020q3 & {}2020q4 & {}2021q1 &
{}2021q2\\\hline
{}1 &  & {\scriptsize {}DEMUK4} & {\scriptsize {}DMUK4} & {\scriptsize {}%
DEMUK4} & {\scriptsize {}DEMUK4} & {\scriptsize {}DMUK4} & {\scriptsize {}%
DPSUK4} & {\scriptsize {}DMUK4}\\
{}2 &  & {\scriptsize {}DMUK4} & {\scriptsize {}DPSUK4} & {\scriptsize {}%
DLRUK4} & {\scriptsize {}DDWUK4} & {\scriptsize {}DPSUK4} & {\scriptsize {}%
DDPSUK4} & {\scriptsize {}DDEQUK4}\\
{}3 &  & {\scriptsize {}DPSUK4} & {\scriptsize {}DDPSUK4} & {\scriptsize {}%
DMUK4} & {\scriptsize {}DLRUK4} & {\scriptsize {}DDPSUK4} & {\scriptsize {}%
DWUK4} & {\scriptsize {}DPSUK4}\\
{}4 &  & {\scriptsize {}DDPSUK4} & {\scriptsize {}DGAPSUK8} & {\scriptsize {}%
DPMUK4} & {\scriptsize {}DMUK4} & {\scriptsize {}DWUK4} & {\scriptsize {}%
}$\cdot$ & {\scriptsize {}DDPSUK4}\\
{}5 &  & {\scriptsize {}DGAPSUK8} & {\scriptsize {}DGAPSUK12} &
{\scriptsize {}DPSUK4} & {\scriptsize {}DDEQUK4} & {\scriptsize {}}$\cdot$ &
{\scriptsize {}}$\cdot$ & {\scriptsize {}DWUK4}\\
{}6 &  & {\scriptsize {}DGAPSUK12} & {\scriptsize {}DWUK4} & {\scriptsize {}%
DDPSUK4} & {\scriptsize {}DPMUK4} & {\scriptsize {}}$\cdot$ & {\scriptsize {}%
}$\cdot$ & {\scriptsize {}}$\cdot$\\
{}7 &  & {\scriptsize {}DWUK4} & {\scriptsize {}}$\cdot$ & {\scriptsize {}%
DGAPSUK8} & {\scriptsize {}DPSUK4} & {\scriptsize {}}$\cdot$ & {\scriptsize {}%
}$\cdot$ & {\scriptsize {}}$\cdot$\\
{}8 &  & {\scriptsize {}}$\cdot$ & {\scriptsize {}}$\cdot$ & {\scriptsize {}%
DGAPSUK12} & {\scriptsize {}DDPSUK4} & {\scriptsize {}}$\cdot$ &
{\scriptsize {}}$\cdot$ & {\scriptsize {}}$\cdot$\\
{}9 &  & {\scriptsize {}}$\cdot$ & {\scriptsize {}}$\cdot$ & {\scriptsize {}%
DWUK4} & {\scriptsize {}DWUK4} & {\scriptsize {}}$\cdot$ & {\scriptsize {}%
}$\cdot$ & {\scriptsize {}}$\cdot$\\\hline
{}$\hat{\lambda}_{hT}$ &  & {}0.0883 & {}0.0913 & {}0.0823 & {}0.0767 &
{}0.0899 & {}0.0955 & {}0.0919\\\hline
&  &  &  &  &  &  &  & \\\hline
\# &  & {}2021q3 & {}2021q4 & {}2022q1 & {}2022q2 & {}2022q3 & {}2022q4 &
{}2023q1\\\hline
{}1 &  & {\scriptsize {}DMUK4} & {\scriptsize {}DLRUK4} & {\scriptsize {}%
DLRUK4} & {\scriptsize {}DLRUK4} & {\scriptsize {}DLRUK4} & {\scriptsize {}%
DLRUK4} & {\scriptsize {}DLRUK4}\\
{}2 &  & {\scriptsize {}DDEQUK4} & {\scriptsize {}DMUK4} & {\scriptsize {}%
DMUK4} & {\scriptsize {}DDEQUK4} & {\scriptsize {}DMUK4} & {\scriptsize {}%
DDEQUK4} & {\scriptsize {}DDEQUK4}\\
{}3 &  & {\scriptsize {}DPSUK4} & {\scriptsize {}DDEQUK4} & {\scriptsize {}%
DDEQUK4} & {\scriptsize {}DPMUK4} & {\scriptsize {}DDEQUK4} & {\scriptsize {}%
DPMUK4} & {\scriptsize {}DPMUK4}\\
{}4 &  & {\scriptsize {}DDPSUK4} & {\scriptsize {}DPMUK4} & {\scriptsize {}%
DPMUK4} & {\scriptsize {}DPSUK4} & {\scriptsize {}DPMUK4} & {\scriptsize {}%
DPSUK4} & {\scriptsize {}DPSUK4}\\
{}5 &  & {\scriptsize {}DWUK4} & {\scriptsize {}DPSUK4} & {\scriptsize {}%
DPSUK4} & {\scriptsize {}DDPSUK4} & {\scriptsize {}DPSUK4} & {\scriptsize {}%
DDPSUK4} & {\scriptsize {}DDPSUK4}\\
{}6 &  & {\scriptsize {}}$\cdot$ & {\scriptsize {}DDPSUK4} & {\scriptsize {}%
DDPSUK4} & {\scriptsize {}DWUK4} & {\scriptsize {}DDPSUK4} & {\scriptsize {}%
DWUK4} & {\scriptsize {}DWUK4}\\
{}7 &  & {\scriptsize {}}$\cdot$ & {\scriptsize {}DWUK4} & {\scriptsize {}%
DWUK4} & {\scriptsize {}}$\cdot$ & {\scriptsize {}DWUK4} & {\scriptsize {}%
}$\cdot$ & {\scriptsize {}}$\cdot$\\\hline
{}$\hat{\lambda}_{hT}$ &  & {}0.0952 & {}0.0927 & {}0.0878 & {}0.0874 &
{}0.0787 & {}0.0904 & {}0.0840\\\hline\hline
\end{tabular}
}
\end{center}

{\footnotesize \emph{{}Note}{}: }${\footnotesize \hat{\lambda}}_{hT}$
{\footnotesize is the estimate of the Lasso penalty parameter computed by
10-fold cross-validation.}%

\end{table}%
%

\begin{table}[H]%
%

\caption{Selected Variables by ARX-Lasso 1q Ahead}%

\begin{center}
{\footnotesize
\begin{tabular}
[c]{ccccccccc}\hline\hline
{} &  & \multicolumn{7}{c}{{}Ending Year}\\\hline
\# &  & {}2019q4 & {}2020q1 & {}2020q2 & {}2020q3 & {}2020q4 & {}2021q1 &
{}2021q2\\\hline
{}1 &  & {\scriptsize DEMUK4} & {\scriptsize DEMUK4} & {\scriptsize DEMUK4} &
{\scriptsize DEMUK4} & {\scriptsize DEMUK4} & {\scriptsize DEMUK4} &
{\scriptsize DEMUK4}\\
{}2 &  & {\scriptsize DVUK4} & {\scriptsize DVUK4} & {\scriptsize DVUK4} &
{\scriptsize DVUK4} & {\scriptsize DDWUK4} & {\scriptsize DVUK4} &
{\scriptsize DVUK4}\\
{}3 &  & {\scriptsize DDWUK4} & {\scriptsize DDWUK4} & {\scriptsize DDWUK4} &
{\scriptsize DDWUK4} & {\scriptsize DLRUK4} & {\scriptsize DDWUK4} &
{\scriptsize DDWUK4}\\
{}4 &  & {\scriptsize DLRUK4} & {\scriptsize DLRUK4} & {\scriptsize DLRUK4} &
{\scriptsize DLRUK4} & {\scriptsize DMUK4} & {\scriptsize DLRUK4} &
{\scriptsize DLRUK4}\\
{}5 &  & {\scriptsize DMUK4} & {\scriptsize DMUK4} & {\scriptsize DMUK4} &
{\scriptsize DMUK4} & {\scriptsize DDEQUK4} & {\scriptsize DMUK4} &
{\scriptsize DMUK4}\\
{}6 &  & {\scriptsize DYCHINA4} & {\scriptsize DYCHINA4} &
{\scriptsize DDEQUK4} & {\scriptsize DDEQUK4} & {\scriptsize {}}$\cdot$ &
{\scriptsize DDEQUK4} & {\scriptsize DDEQUK4}\\
{}7 &  & {\scriptsize DGAPSUK8} & {\scriptsize DGAPSUK8} &
{\scriptsize DGAPSUK8} & {\scriptsize {}}$\cdot$ & {\scriptsize {}}$\cdot$ &
{\scriptsize DPMUK4} & {\scriptsize DPUS4}\\
{}8 &  & {\scriptsize DGAPSUK12} & {\scriptsize DGAPSUK12} &
{\scriptsize DGAPSUK12} & {\scriptsize {}}$\cdot$ & {\scriptsize {}}$\cdot$ &
{\scriptsize DPUS4} & {\scriptsize DWUK4}\\
{}9 &  & {\scriptsize {}}$\cdot$ & {\scriptsize {}}$\cdot$ & {\scriptsize {}%
}$\cdot$ & {\scriptsize {}}$\cdot$ & {\scriptsize {}}$\cdot$ &
{\scriptsize DWUK4} & {\scriptsize {}}$\cdot$\\\hline
{}$\hat{\lambda}_{hT}$ &  & 0.0659 & 0.0678 & 0.0665 & 0.0702 & 0.0796 &
0.0562 & 0.0639\\\hline
&  &  &  &  &  &  &  & \\\hline
\# &  & {}2021q3 & {}2021q4 & {}2022q1 & {}2022q2 & {}2022q3 & {}2022q4 &
{}2023q1\\\hline
{}1 &  & {\scriptsize {}DEMUK4} & {\scriptsize {}DEMUK4} & {\scriptsize {}%
DDYUK4} & {\scriptsize {}DDYUK4} & {\scriptsize {}DDYUK4} & {\scriptsize {}%
DEMUK4} & {\scriptsize {}DEMUK4}\\
{}2 &  & {\scriptsize {}DLRUK4} & {\scriptsize {}DLRUK4} & {\scriptsize {}%
DEMUK4} & {\scriptsize {}DEMUK4} & {\scriptsize {}DEMUK4} & {\scriptsize {}%
DDVUK4} & {\scriptsize {}DDVUK4}\\
{}3 &  & {\scriptsize {}DMUK4} & {\scriptsize {}DMUK4} & {\scriptsize {}%
DDVUK4} & {\scriptsize {}DDVUK4} & {\scriptsize {}DDVUK4} & {\scriptsize {}%
DDWUK4} & {\scriptsize {}DLRUK4}\\
{}4 &  & {\scriptsize {}DDEQUK4} & {\scriptsize {}DDEQUK4} & {\scriptsize {}%
DDUUK} & {\scriptsize {}DDUUK} & {\scriptsize {}DDUUK} & {\scriptsize {}%
DLRUK4} & {\scriptsize {}DMUK4}\\
{}5 &  & {\scriptsize {}}$\cdot$ & {\scriptsize {}DWUK4} & {\scriptsize {}%
DLRUK4} & {\scriptsize {}DLRUK4} & {\scriptsize {}DLRUK4} & {\scriptsize {}%
DMUK4} & {\scriptsize {}DDEQUK4}\\
{}6 &  & {\scriptsize {}}$\cdot$ & {\scriptsize {}}$\cdot$ & {\scriptsize {}%
DMUK4} & {\scriptsize {}DMUK4} & {\scriptsize {}DMUK4} & {\scriptsize {}%
DDEQUK4} & {\scriptsize {}DPMUK4}\\
{}7 &  & {\scriptsize {}}$\cdot$ & {\scriptsize {}}$\cdot$ & {\scriptsize {}%
DDEQUK4} & {\scriptsize {}DDEQUK4} & {\scriptsize {}DDEQUK4} & {\scriptsize {}%
DDPMAT4} & {\scriptsize {}DYCHINA4}\\
{}8 &  & {\scriptsize {}}$\cdot$ & {\scriptsize {}}$\cdot$ & {\scriptsize {}%
DDPMAT4} & {\scriptsize {}DDPMAT4} & {\scriptsize {}DDPMAT4} & {\scriptsize {}%
DPMUK4} & {\scriptsize {}DWUK4}\\
{}9 &  & {\scriptsize {}}$\cdot$ & {\scriptsize {}}$\cdot$ & {\scriptsize {}%
DPMUK4} & {\scriptsize {}DPMUK4} & {\scriptsize {}DPMUK4} & {\scriptsize {}%
DYCHINA4} & {\scriptsize {}}$\cdot$\\
{}10 &  & {\scriptsize {}}$\cdot$ & {\scriptsize {}}$\cdot$ & {\scriptsize {}%
DYCHINA4} & {\scriptsize {}DYCHINA4} & {\scriptsize {}DYCHINA4} &
{\scriptsize {}DDPUS4} & {\scriptsize {}}$\cdot$\\
{}11 &  & {\scriptsize {}}$\cdot$ & {\scriptsize {}}$\cdot$ & {\scriptsize {}%
DPUS4} & {\scriptsize {}DDPUS4} & {\scriptsize {}DDPUS4} & {\scriptsize {}%
DWUK4} & {\scriptsize {}}$\cdot$\\
{}12 &  & {\scriptsize {}}$\cdot$ & {\scriptsize {}}$\cdot$ & {\scriptsize {}%
DWUK4} & {\scriptsize {}DWUK4} & {\scriptsize {}DWUK4} & {\scriptsize {}%
}$\cdot$ & {\scriptsize {}}$\cdot$\\\hline
{}$\hat{\lambda}_{hT}$ &  & {}0.0752 & {}0.0714 & {}0.0422 & {}0.0457 &
{}0.0456 & {}0.0550 & {}0.0624\\\hline\hline
\end{tabular}
}
\end{center}

{\footnotesize \emph{{}Note}{}: }${\footnotesize \hat{\lambda}}_{hT}$
{\footnotesize is the estimate of the Lasso penalty parameter computed by
10-fold cross-validation}.%

\end{table}%

\pagebreak

\subsubsection{For h=2 quarter ahead models}

This sub-section sets out the variables chosen by Lasso for $h=2$.%

\begin{table}[H]%
%

\caption{Selected Variables by Lasso  2q Ahead}%

\begin{center}
{\footnotesize
\begin{tabular}
[c]{cccccccccc}\hline\hline
{} &  & \multicolumn{8}{c}{{}Ending Year}\\\hline
\# &  & {}2019q3 & {}2019q4 & {}2020q1 & {}2020q2 & {}2020q3 & {}2020q4 &
{}2021q1 & {}2021q2\\\hline
{}1 &  & {\scriptsize {}DPUK4} & {\scriptsize {}DPUK4} & {\scriptsize {}DPUK4}
& {\scriptsize {}DPUK4} & {\scriptsize {}DPUK4} & {\scriptsize {}DPUK4} &
{\scriptsize {}DPUK4} & {\scriptsize {}DPUK4}\\
{}2 &  & {\scriptsize {}DDPUK4} & {\scriptsize {}DDPUK4} & {\scriptsize {}%
DDPUK4} & {\scriptsize {}DDPUK4} & {\scriptsize {}DDPUK4} & {\scriptsize {}%
DDPUK4} & {\scriptsize {}DDPUK4} & {\scriptsize {}DDPUK4}\\
{}3 &  & {\scriptsize {}DMUK4} & {\scriptsize {}DMUK4} & {\scriptsize {}DMUK4}
& {\scriptsize {}DMUK4} & {\scriptsize {}DMUK4} & {\scriptsize {}DMUK4} &
{\scriptsize {}DMUK4} & {\scriptsize {}DMUK4}\\
{}4 &  & {\scriptsize {}DPSUK4} & {\scriptsize {}DPSUK4} & {\scriptsize {}%
DPSUK4} & {\scriptsize {}DPSUK4} & {\scriptsize {}DPSUK4} & {\scriptsize {}%
DPSUK4} & {\scriptsize {}DPSUK4} & {\scriptsize {}DPMUK4}\\
{}5 &  & {\scriptsize {}DWUK4} & {\scriptsize {}DWUK4} & {\scriptsize {}DWUK4}
& {\scriptsize {}DWUK4} & {\scriptsize {}DWUK4} & {\scriptsize {}DWUK4} &
{\scriptsize {}DWUK4} & {\scriptsize {}DPSUK4}\\
{}6 &  & {\scriptsize {}}$\cdot$ & {\scriptsize {}}$\cdot$ & {\scriptsize {}%
}$\cdot$ & {\scriptsize {}}$\cdot$ & {\scriptsize {}}$\cdot$ & {\scriptsize {}%
}$\cdot$ & {\scriptsize {}}$\cdot$ & {\scriptsize {}DWUK4}\\\hline
{}$\hat{\lambda}_{hT}$ &  & {\scriptsize {}0.3113} & {\scriptsize {}0.3114} &
{\scriptsize {}0.2494} & {\scriptsize {}0.2813} & {\scriptsize {}0.3133} &
{\scriptsize {}0.2521} & {\scriptsize {}0.2529} & {\scriptsize {}%
0.2521}\\\hline
&  &  &  &  &  &  &  &  & \\\hline
\# &  & {\scriptsize 2021q3} & {\scriptsize 2021q4} & {\scriptsize 2022q1} &
{\scriptsize 2022q2} & {\scriptsize 2022q3} & {\scriptsize 2022q4} &
{\scriptsize 2023q1} & \\\hline
1 &  & {\scriptsize {}DPUK4} & {\scriptsize {}DPUK4} & {\scriptsize {}DPUK4} &
{\scriptsize {}DPUK4} & {\scriptsize {}DPUK4} & {\scriptsize {}DPUK4} &
{\scriptsize {}DPUK4} & \\
2 &  & {\scriptsize {}DDPUK4} & {\scriptsize {}DDPUK4} & {\scriptsize {}%
DDPUK4} & {\scriptsize {}DDPUK4} & {\scriptsize {}DDPUK4} & {\scriptsize {}%
DDPUK4} & {\scriptsize {}DDPUK4} & \\
3 &  & {\scriptsize {}DLRUK4} & {\scriptsize {}DMUK4} & {\scriptsize {}DMUK4}
& {\scriptsize {}DMUK4} & {\scriptsize {}DLRUK4} & {\scriptsize {}DLRUK4} &
{\scriptsize {}DLRUK4} & \\
4 &  & {\scriptsize {}DMUK4} & {\scriptsize {}DPMUK4} & {\scriptsize {}DPMUK4}
& {\scriptsize {}DPMUK4} & {\scriptsize {}DMUK4} & {\scriptsize {}DMUK4} &
{\scriptsize {}DMUK4} & \\
5 &  & {\scriptsize {}DPMUK4} & {\scriptsize {}DPSUK4} & {\scriptsize {}%
DPSUK4} & {\scriptsize {}DPSUK4} & {\scriptsize {}DPMUK4} & {\scriptsize {}%
DPMUK4} & {\scriptsize {}DPMUK4} & \\
6 &  & {\scriptsize {}DPSUK4} & {\scriptsize {}DWUK4} & {\scriptsize {}DWUK4}
& {\scriptsize {}DWUK4} & {\scriptsize {}DPSUK4} & {\scriptsize {}DPSUK4} &
{\scriptsize {}DPSUK4} & \\
7 &  & {\scriptsize {}DWUK4} & {\scriptsize {}}$\cdot$ & {\scriptsize {}%
}$\cdot$ & {\scriptsize {}}$\cdot$ & {\scriptsize {}DDPSUK4} & {\scriptsize {}%
DDPSUK4} & {\scriptsize {}DDPSUK4} & \\
8 &  & {\scriptsize {}}$\cdot$ & {\scriptsize {}}$\cdot$ & {\scriptsize {}%
}$\cdot$ & {\scriptsize {}}$\cdot$ & {\scriptsize {}DWUK4} & {\scriptsize {}%
DPUS4} & {\scriptsize {}DPUS4} & \\
9 &  & {\scriptsize {}}$\cdot$ & {\scriptsize {}}$\cdot$ & {\scriptsize {}%
}$\cdot$ & {\scriptsize {}}$\cdot$ & {\scriptsize {}}$\cdot$ & {\scriptsize {}%
DWUK4} & {\scriptsize {}DWUK4} & \\
$\hat{\lambda}_{hT}$ &  & 0.2196 & 0.2490 & 0.2475 & 0.3112 & 0.2221 &
0.2297 & 0.2366 & \\\hline\hline
\end{tabular}
}
\end{center}

{\footnotesize \emph{{}Note}{}: }${\footnotesize \hat{\lambda}}_{hT}$
{\footnotesize is the estimate of the Lasso penalty parameter computed by
10-fold cross-validation.}%

\end{table}%
%

\begin{table}[H]%
%

\caption{Selected Variables by AR2-LASSO  2q Ahead}%

\begin{center}
{\footnotesize
\begin{tabular}
[c]{cccccccc}\hline\hline
{} &  & \multicolumn{6}{c}{{}Ending Year}\\\hline
\# &  & {}2019q3 & {}2019q4 & {}2020q1 & {}2020q2 & {}2020q3 & {}%
2020q4\\\hline
{}1 &  & {\scriptsize {}DEMUK4} & {\scriptsize {}DEMUK4} & {\scriptsize {}%
DEMUK4} & {\scriptsize {}DEMUK4} & {\scriptsize {}DEMUK4} & {\scriptsize {}%
DEMUK4}\\
{}2 &  & {\scriptsize {}DVUK4} & {\scriptsize {}DVUK4} & {\scriptsize {}DVUK4}
& {\scriptsize {}DVUK4} & {\scriptsize {}DVUK4} & {\scriptsize {}DVUK4}\\
{}3 &  & {\scriptsize {}DDWUK4} & {\scriptsize {}DDWUK4} & {\scriptsize {}%
DDWUK4} & {\scriptsize {}DDWUK4} & {\scriptsize {}DDWUK4} & {\scriptsize {}%
DDUUK}\\
{}4 &  & {\scriptsize {}DMUK4} & {\scriptsize {}DMUK4} & {\scriptsize {}DMUK4}
& {\scriptsize {}DMUK4} & {\scriptsize {}DMUK4} & {\scriptsize {}DDWUK4}\\
{}5 &  & {\scriptsize {}DPMAT4} & {\scriptsize {}DPMUK4} & {\scriptsize {}%
DPMAT4} & {\scriptsize {}DPMAT4} & {\scriptsize {}DPMAT4} & {\scriptsize {}%
DLRUK4}\\
{}6 &  & {\scriptsize {}DPMUK4} & {\scriptsize {}DDPMUK4} & {\scriptsize {}%
DPMUK4} & {\scriptsize {}DPMUK4} & {\scriptsize {}DPMUK4} & {\scriptsize {}%
DMUK4}\\
{}7 &  & {\scriptsize {}DDPMUK4} & {\scriptsize {}DPSUK4} & {\scriptsize {}%
DDPMUK4} & {\scriptsize {}DDPMUK4} & {\scriptsize {}DDPMUK4} & {\scriptsize {}%
DPMAT4}\\
{}8 &  & {\scriptsize {}DEPUK4} & {\scriptsize {}DDPSUK4} & {\scriptsize {}%
DEPUK4} & {\scriptsize {}DEPUK4} & {\scriptsize {}DEPUK4} & {\scriptsize {}%
DDPMETAL4}\\
{}9 &  & {\scriptsize {}DPSUK4} & {\scriptsize {}DDYCHINA4} & {\scriptsize {}%
DPSUK4} & {\scriptsize {}DPSUK4} & {\scriptsize {}DPSUK4} & {\scriptsize {}%
DDPMUK4}\\
{}10 &  & {\scriptsize {}DDPSUK4} & {\scriptsize {}DGAPSUK8} & {\scriptsize {}%
DDPSUK4} & {\scriptsize {}DDPSUK4} & {\scriptsize {}DDPSUK4} & {\scriptsize {}%
DEPUK4}\\
{}11 &  & {\scriptsize {}DYCHINA4} & {\scriptsize {}DWUK4} & {\scriptsize {}%
DDYCHINA4} & {\scriptsize {}DDYCHINA4} & {\scriptsize {}DDYCHINA4} &
{\scriptsize {}DPSUK4}\\
{}12 &  & {\scriptsize {}DDYCHINA4} & {\scriptsize {}}$\cdot$ &
{\scriptsize {}DGAPSUK8} & {\scriptsize {}DGAPSUK8} & {\scriptsize {}DGAPSUK8}
& {\scriptsize {}DDPSUK4}\\
{}13 &  & {\scriptsize {}DGAPSUK8} & {\scriptsize {}}$\cdot$ & {\scriptsize {}%
DWUK4} & {\scriptsize {}DWUK4} & {\scriptsize {}DWUK4} & {\scriptsize {}%
DGAPSUK8}\\
{}14 &  & {\scriptsize {}DWUK4} & {\scriptsize {}}$\cdot$ & {\scriptsize {}%
}$\cdot$ & {\scriptsize {}}$\cdot$ & {\scriptsize {}}$\cdot$ & {\scriptsize {}%
DWUK4}\\\hline
{}$\hat{\lambda}_{hT}$ &  & {}0.0966 & {}0.1298 & {}0.1020 & {}0.1084 &
{}0.0955 & {}0.0941\\\hline
&  &  &  &  &  &  & \\\hline
\# &  & {}2021q1 & {}2021q2 & {}2021q3 & {}2021q4 & {}2022q1 & {}%
2022q2\\\hline
1 &  & {\scriptsize {}DEMUK4} & {\scriptsize {}DEMUK4} & {\scriptsize {}%
DEMUK4} & {\scriptsize {}DEMUK4} & {\scriptsize {}DEMUK4} & {\scriptsize {}%
DEMUK4}\\
2 &  & {\scriptsize {}DVUK4} & {\scriptsize {}DVUK4} & {\scriptsize {}DVUK4} &
{\scriptsize {}DDWUK4} & {\scriptsize {}DDUUK4} & {\scriptsize {}DDUUK4}\\
3 &  & {\scriptsize {}DDUUK} & {\scriptsize {}DDUUK} & {\scriptsize {}DDUUK} &
{\scriptsize {}DLRUK4} & {\scriptsize {}DLRUK4} & {\scriptsize {}DLRUK4}\\
4 &  & {\scriptsize {}DDWUK4} & {\scriptsize {}DDWUK4} & {\scriptsize {}%
DDWUK4} & {\scriptsize {}DMUK4} & {\scriptsize {}DMUK4} & {\scriptsize {}%
DMUK4}\\
5 &  & {\scriptsize {}DLRUK4} & {\scriptsize {}DLRUK4} & {\scriptsize {}%
DLRUK4} & {\scriptsize {}DPMAT4} & {\scriptsize {}DPMAT4} & {\scriptsize {}%
DPMAT4}\\
6 &  & {\scriptsize {}DMUK4} & {\scriptsize {}DMUK4} & {\scriptsize {}DMUK4} &
{\scriptsize {}DPMUK4} & {\scriptsize {}DPMUK4} & {\scriptsize {}DPMUK4}\\
7 &  & {\scriptsize {}DPMAT4} & {\scriptsize {}DPMAT4} & {\scriptsize {}%
DPMAT4} & {\scriptsize {}DDPMUK4} & {\scriptsize {}DDPMUK4} & {\scriptsize {}%
DDPMUK4}\\
8 &  & {\scriptsize {}DDPMETAL4} & {\scriptsize {}DDPMETAL4} & {\scriptsize {}%
DDPMUK4} & {\scriptsize {}DPSUK4} & {\scriptsize {}DPSUK4} & {\scriptsize {}%
DPSUK4}\\
9 &  & {\scriptsize {}DDPMUK4} & {\scriptsize {}DDPMUK4} & {\scriptsize {}%
DEPUK4} & {\scriptsize {}DDPSUK4} & {\scriptsize {}DDPSUK4} & {\scriptsize {}%
DDPSUK4}\\
10 &  & {\scriptsize {}DEPUK4} & {\scriptsize {}DEPUK4} & {\scriptsize {}%
DPSUK4} & {\scriptsize {}DWUK4} & {\scriptsize {}DWUK4} & {\scriptsize {}%
DWUK4}\\
11 &  & {\scriptsize {}DPSUK4} & {\scriptsize {}DPSUK4} & {\scriptsize {}%
DDPSUK4} & {\scriptsize {}}$\cdot$ & {\scriptsize {}}$\cdot$ & {\scriptsize {}%
}$\cdot$\\
12 &  & {\scriptsize {}DDPSUK4} & {\scriptsize {}DDPSUK4} & {\scriptsize {}%
DWUK4} & {\scriptsize {}}$\cdot$ & {\scriptsize {}}$\cdot$ & {\scriptsize {}%
}$\cdot$\\
13 &  & {\scriptsize {}DWUK4} & {\scriptsize {}DWUK4} & {\scriptsize {}}%
$\cdot$ & {\scriptsize {}}$\cdot$ & {\scriptsize {}}$\cdot$ & {\scriptsize {}%
}$\cdot$\\\hline
$\hat{\lambda}_{hT}$ &  & 0.1005 & {}0.0955 & {}0.1043 & {}0.1164 & {}0.1127 &
{}0.1237\\\hline
&  &  &  &  &  &  & \\\hline
\# &  & {}2022q3 & {}2022q4 & {}2023q1 &  &  & \\\hline
1 &  & {\scriptsize {}DEMUK4} & {\scriptsize {}DEMUK4} & {\scriptsize {}%
DDUUK4} &  &  & \\
2 &  & {\scriptsize {}DDUUK4} & {\scriptsize {}DDUUK4} & {\scriptsize {}DMUK4}
&  &  & \\
3 &  & {\scriptsize {}DLRUK4} & {\scriptsize {}DLRUK4} & {\scriptsize {}%
DPMUK4} &  &  & \\
4 &  & {\scriptsize {}DMUK4} & {\scriptsize {}DMUK4} & {\scriptsize {}DDPMUK4}
&  &  & \\
5 &  & {\scriptsize {}DPMAT4} & {\scriptsize {}DPMAT4} & {\scriptsize {}%
DPSUK4} &  &  & \\
6 &  & {\scriptsize {}DPMUK4} & {\scriptsize {}DPMUK4} & {\scriptsize {}%
DDPSUK4} &  &  & \\
7 &  & {\scriptsize {}DDPMUK4} & {\scriptsize {}DDPMUK4} & {\scriptsize {}%
DWUK4} &  &  & \\
8 &  & {\scriptsize {}DPSUK4} & {\scriptsize {}DPSUK4} & {\scriptsize {}%
}$\cdot$ &  &  & \\
9 &  & {\scriptsize {}DDPSUK4} & {\scriptsize {}DDPSUK4} & {\scriptsize {}%
}$\cdot$ &  &  & \\
10 &  & {\scriptsize {}DWUK4} & {\scriptsize {}DWUK4} & {\scriptsize {}}%
$\cdot$ &  &  & \\\hline
$\hat{\lambda}_{hT}$ &  & {}0.1232 & {}0.1252 & {}0.1406 &  &  &
\\\hline\hline
\end{tabular}
}
\end{center}

{\footnotesize \emph{{}Note}{}: }${\footnotesize \hat{\lambda}}_{hT}$
{\footnotesize is the estimate of the Lasso penalty parameter computed by
10-fold cross-validation.}%

\end{table}%
%

\begin{table}[H]%
%

\caption{Selected Variables by ARX-Lasso  2q Ahead}%

\begin{center}
{\footnotesize
\begin{tabular}
[c]{cccccccc}\hline\hline
{}\# &  & \multicolumn{6}{c}{{}Ending Year}\\\hline
&  & {}2019q3 & {}2019q4 & {}2020q1 & {}2020q2 & {}2020q3 & {}2020q4\\\hline
{}1 &  & {\scriptsize {}DEMUK4} & {\scriptsize {}DEMUK4} & {\scriptsize {}%
DEMUK4} & {\scriptsize {}DEMUK4} & {\scriptsize {}DEMUK4} & {\scriptsize {}%
DEMUK4}\\
{}2 &  & {\scriptsize {}DVUK4} & {\scriptsize {}DVUK4} & {\scriptsize {}DVUK4}
& {\scriptsize {}DVUK4} & {\scriptsize {}DVUK4} & {\scriptsize {}DVUK4}\\
{}3 &  & {\scriptsize {}DDWUK4} & {\scriptsize {}DDWUK4} & {\scriptsize {}%
DDWUK4} & {\scriptsize {}DDWUK4} & {\scriptsize {}DDWUK4} & {\scriptsize {}%
DDUUK4}\\
{}4 &  & {\scriptsize {}DLRUK4} & {\scriptsize {}DLRUK4} & {\scriptsize {}%
DLRUK4} & {\scriptsize {}DLRUK4} & {\scriptsize {}DLRUK4} & {\scriptsize {}%
DDWUK4}\\
{}5 &  & {\scriptsize {}DMUK4} & {\scriptsize {}DMUK4} & {\scriptsize {}DMUK4}
& {\scriptsize {}DMUK4} & {\scriptsize {}DMUK4} & {\scriptsize {}DLRUK4}\\
{}6 &  & {\scriptsize {}DPMAT4} & {\scriptsize {}DPMAT4} & {\scriptsize {}%
DPMAT4} & {\scriptsize {}DPMAT4} & {\scriptsize {}DDPMUK4} & {\scriptsize {}%
DMUK4}\\
{}7 &  & {\scriptsize {}DDPMAT4} & {\scriptsize {}DDPMUK4} & {\scriptsize {}%
DDPMUK4} & {\scriptsize {}DDPMUK4} & {\scriptsize {}DEPUK4} & {\scriptsize {}%
DPMAT4}\\
{}8 &  & {\scriptsize {}DPMUK4} & {\scriptsize {}DEPUK4} & {\scriptsize {}%
DEPUK4} & {\scriptsize {}DEPUK4} & {\scriptsize {}DYCHINA4} & {\scriptsize {}%
DDPMETAL4}\\
{}9 &  & {\scriptsize {}DDPMUK4} & {\scriptsize {}DYCHINA4} & {\scriptsize {}%
DYCHINA4} & {\scriptsize {}DYCHINA4} & {\scriptsize {}DGAPSUK8} &
{\scriptsize {}DDPMUK4}\\
{}10 &  & {\scriptsize {}DEPUK4} & {\scriptsize {}DDYCHINA4} & {\scriptsize {}%
DDYCHINA4} & {\scriptsize {}DDYCHINA4} & {\scriptsize {}DDPUS4} &
{\scriptsize {}DEPUK4}\\
{}11 &  & {\scriptsize {}DYCHINA4} & {\scriptsize {}DGAPSUK8} &
{\scriptsize {}DGAPSUK8} & {\scriptsize {}DGAPSUK8} & {\scriptsize {}}$\cdot$
& {\scriptsize {}DYCHINA4}\\
{}12 &  & {\scriptsize {}DDYCHINA4} & {\scriptsize {}DDPUS4} & {\scriptsize {}%
DDPUS4} & {\scriptsize {}DDPUS4} & {\scriptsize {}}$\cdot$ & {\scriptsize {}%
DGAPSUK8}\\
{}13 &  & {\scriptsize {}DGAPSUK8} & {\scriptsize {}}$\cdot$ & {\scriptsize {}%
}$\cdot$ & {\scriptsize {}}$\cdot$ & {\scriptsize {}}$\cdot$ & {\scriptsize {}%
DPUS4}\\
{}14 &  & {\scriptsize {}DDPUS4} & {\scriptsize {}}$\cdot$ & {\scriptsize {}%
}$\cdot$ & {\scriptsize {}}$\cdot$ & {\scriptsize {}}$\cdot$ & {\scriptsize {}%
DDPUS4}\\
{}15 &  & {\scriptsize {}DWUK4} & {\scriptsize {}}$\cdot$ & {\scriptsize {}%
}$\cdot$ & {\scriptsize {}}$\cdot$ & {\scriptsize {}}$\cdot$ & {\scriptsize {}%
DWUK4}\\\hline
{}$\hat{\lambda}_{hT}$ &  & {}0.0757 & {}0.0925 & {}0.0920 & {}0.0920 &
{}0.1036 & {}0.0713\\\hline
&  &  &  &  &  &  & \\\hline
\# &  & {}2021q1 & {}2021q2 & {}2021q3 & {}2021q4 & {}2022q1 & {}%
2022q2\\\hline
1 &  & {\scriptsize {}DEMUK4} & {\scriptsize {}DEMUK4} & {\scriptsize {}%
DEMUK4} & {\scriptsize {}DEMUK4} & {\scriptsize {}DEMUK4} & {\scriptsize {}%
DEMUK4}\\
2 &  & {\scriptsize {}DVUK4} & {\scriptsize {}DVUK4} & {\scriptsize {}DVUK4} &
{\scriptsize {}DVUK4} & {\scriptsize {}DDUUK} & {\scriptsize {}DDUUK}\\
3 &  & {\scriptsize {}DDUUK} & {\scriptsize {}DDUUK} & {\scriptsize {}DDUUK} &
{\scriptsize {}DDUUK} & {\scriptsize {}DDWUK4} & {\scriptsize {}DLRUK4}\\
4 &  & {\scriptsize {}DDWUK4} & {\scriptsize {}DDWUK4} & {\scriptsize {}%
DDWUK4} & {\scriptsize {}DDWUK4} & {\scriptsize {}DLRUK4} & {\scriptsize {}%
DMUK4}\\
5 &  & {\scriptsize {}DLRUK4} & {\scriptsize {}DLRUK4} & {\scriptsize {}%
DLRUK4} & {\scriptsize {}DLRUK4} & {\scriptsize {}DMUK4} & {\scriptsize {}%
DPMUK4}\\
6 &  & {\scriptsize {}DMUK4} & {\scriptsize {}DMUK4} & {\scriptsize {}DMUK4} &
{\scriptsize {}DMUK4} & {\scriptsize {}DPMAT4} & {\scriptsize {}DDPMUK4}\\
7 &  & {\scriptsize {}DPMAT4} & {\scriptsize {}DPMAT4} & {\scriptsize {}%
DPMAT4} & {\scriptsize {}DPMAT4} & {\scriptsize {}DPMUK4} & {\scriptsize {}%
DYCHINA4}\\
8 &  & {\scriptsize {}DDPMETAL4} & {\scriptsize {}DDPMETAL4} & {\scriptsize {}%
DDPMETAL4} & {\scriptsize {}DDPMETAL4} & {\scriptsize {}DDPMUK4} &
{\scriptsize {}DWUK4}\\
9 &  & {\scriptsize {}DDPMUK4} & {\scriptsize {}DDPMUK4} & {\scriptsize {}%
DDPMUK4} & {\scriptsize {}DDPMUK4} & {\scriptsize {}DEPUK4} & {\scriptsize {}%
}$\cdot$\\
10 &  & {\scriptsize {}DEPUK4} & {\scriptsize {}DEPUK4} & {\scriptsize {}%
DEPUK4} & {\scriptsize {}DEPUK4} & {\scriptsize {}DYCHINA4} & {\scriptsize {}%
}$\cdot$\\
11 &  & {\scriptsize {}DYCHINA4} & {\scriptsize {}DYCHINA4} & {\scriptsize {}%
DYCHINA4} & {\scriptsize {}DYCHINA4} & {\scriptsize {}DDPUS4} &
{\scriptsize {}}$\cdot$\\
12 &  & {\scriptsize {}DPUS4} & {\scriptsize {}DPUS4} & {\scriptsize {}DPUS4}
& {\scriptsize {}DPUS4} & {\scriptsize {}DWUK4} & {\scriptsize {}}$\cdot$\\
13 &  & {\scriptsize {}DDPUS4} & {\scriptsize {}DDPUS4} & {\scriptsize {}%
DDPUS4} & {\scriptsize {}DWUK4} & {\scriptsize {}}$\cdot$ & {\scriptsize {}%
}$\cdot$\\
14 &  & {\scriptsize {}DWUK4} & {\scriptsize {}DWUK4} & {\scriptsize {}DWUK4}
& {\scriptsize {}}$\cdot$ & {\scriptsize {}}$\cdot$ & {\scriptsize {}}$\cdot
$\\\hline
$\hat{\lambda}_{hT}$ &  & {}0.0699 & {}0.0666 & {}0.0739 & {}0.0921 &
{}0.0964 & {}0.1092\\\hline
&  &  &  &  &  &  & \\\hline
\# &  & {}2022q3 & {}2022q4 & {}2023q1 &  &  & \\\hline
1 &  & {\scriptsize {}DEMUK4} & {\scriptsize {}DEMUK4} & {\scriptsize {}%
DEMUK4} &  &  & \\
2 &  & {\scriptsize {}DDUUK} & {\scriptsize {}DDUUK} & {\scriptsize {}DDUUK} &
&  & \\
3 &  & {\scriptsize {}DDWUK4} & {\scriptsize {}DDWUK4} & {\scriptsize {}%
DDWUK4} &  &  & \\
4 &  & {\scriptsize {}RUK4} & {\scriptsize {}RUK4} & {\scriptsize {}DLRUK4} &
&  & \\
5 &  & {\scriptsize {}DLRUK4} & {\scriptsize {}DLRUK4} & {\scriptsize {}DMUK4}
&  &  & \\
6 &  & {\scriptsize {}DMUK4} & {\scriptsize {}DMUK4} & {\scriptsize {}DPMUK4}
&  &  & \\
7 &  & {\scriptsize {}DPMUK4} & {\scriptsize {}DDPMAT4} & {\scriptsize {}%
DDPMUK4} &  &  & \\
8 &  & {\scriptsize {}DDPMUK4} & {\scriptsize {}DPMUK4} & {\scriptsize {}%
DYCHINA4} &  &  & \\
9 &  & {\scriptsize {}DYCHINA4} & {\scriptsize {}DDPMUK4} & {\scriptsize {}%
DWUK4} &  &  & \\
10 &  & {\scriptsize {}DDPUS4} & {\scriptsize {}DYCHINA4} & {\scriptsize {}%
}$\cdot$ &  &  & \\
11 &  & {\scriptsize {}DWUK4} & {\scriptsize {}DDPUS4} & {\scriptsize {}%
}$\cdot$ &  &  & \\
12 &  & {\scriptsize {}}$\cdot$ & {\scriptsize {}DWUK4} & {\scriptsize {}%
}$\cdot$ &  &  & \\\hline
$\hat{\lambda}_{hT}$ &  & {}0.1046 & {}0.0974 & {}0.1191 &  &  &
\\\hline\hline
\end{tabular}
}
\end{center}

{\footnotesize \emph{{}Note}{}: }${\footnotesize \hat{\lambda}}_{hT}$
{\footnotesize is the estimate of the Lasso penalty parameter computed by
10-fold cross-validation.}%

\end{table}%

\pagebreak

\subsubsection{For h=4 quarter ahead models}

This sub-section sets out the variables chosen by Lasso for $h=4$.%

\begin{table}[H]%
%

\caption{Selected Variables by Lasso  4q Ahead}%

\begin{center}
{\footnotesize
\begin{tabular}
[c]{cccccccc}\hline\hline
{}\# &  & \multicolumn{6}{c}{{}Ending Year}\\\hline
&  & {}2019q1 & {}2019q2 & {}2019q3 & {}2019q4 & {}2020q1 & {}2020q2\\\hline
{}1 &  & {}DPUK4 & {}DPUK4 & {}DPUK4 & {}DPUK4 & {}DPUK4 & {}DPUK4\\
{}2 &  & {}DEMUK4 & {}DEMUK4 & {}DEMUK4 & {}DEMUK4 & {}DEMUK4 & {}DEMUK4\\
{}3 &  & {}DMUK4 & {}DMUK4 & {}DMUK4 & {}DMUK4 & {}DMUK4 & {}DMUK4\\
{}4 &  & {}DPMAT4 & {}DPMAT4 & {}DPMAT4 & {}DPMAT4 & {}DPMAT4 & {}DPMAT4\\
{}5 &  & {}DPSUK4 & {}DPSUK4 & {}DPSUK4 & {}DPSUK4 & {}DPSUK4 & {}DPSUK4\\
{}6 &  & {}DWUK4 & {}DWUK4 & {}DWUK4 & {}DWUK4 & {}DWUK4 & {}DWUK4\\\hline
{}$\hat{\lambda}_{hT}$ &  & {}0.3345 & {}0.3346 & {}0.3823 & {}0.3824 &
{}0.3829 & {}0.3610\\\hline
&  &  &  &  &  &  & \\\hline
&  & {}2020q3 & {}2020q4 & {}2021q1 & {}2021q2 & {}2021q3 & {}2021q4\\\hline
{}1 &  & {}DPUK4 & {}DPUK4 & {}DPUK4 & {}DPUK4 & {}DPUK4 & {}DPUK4\\
{}2 &  & {}DMUK4 & {}DMUK4 & {}DMUK4 & {}DMUK4 & {}DMUK4 & {}DMUK4\\
{}3 &  & {}DPMAT4 & {}DPMAT4 & {}DPMAT4 & {}DPMAT4 & {}DPSUK4 & {}DPMAT4\\
{}4 &  & {}DPSUK4 & {}DPSUK4 & {}DPSUK4 & {}DPSUK4 & {}$\cdot$ & {}DPSUK4\\
{}5 &  & {}DWUK4 & {}$\cdot$ & {}DWUK4 & {}$\cdot$ & {}$\cdot$ &
{}DWUK4\\\hline
{}$\hat{\lambda}_{hT}$ &  & {}0.3874 & {}0.3897 & {}0.4401 & {}0.4394 &
{}0.4613 & {}0.4296\\\hline
&  &  &  &  &  &  & \\\hline
&  & {}2022q1 & {}2022q2 & {}2022q3 & {}2022q4 & {}2023q1 & \\\hline
{}1 &  & {}DPUK4 & {}DPUK4 & {}DPUK4 & {}DPUK4 & {}DPUK4 & \\
{}2 &  & {}DMUK4 & {}DMUK4 & {}DMUK4 & {}DMUK4 & {}DMUK4 & \\
{}3 &  & {}DPMAT4 & {}DPMAT4 & {}DPMAT4 & {}DPMAT4 & {}DPMAT4 & \\
{}4 &  & {}DPSUK4 & {}DPSUK4 & {}DPSUK4 & {}DPSUK4 & {}DPSUK4 & \\
{}5 &  & {}DWUK4 & {}DWUK4 & {}DWUK4 & {}DPUS4 & {}DPUS4 & \\
{}6 &  & {}$\cdot$ & {}$\cdot$ & {}$\cdot$ & {}DWUK4 & {}DWUK4 & \\\hline
{}$\hat{\lambda}_{hT}$ &  & {}0.4506 & {}0.4632 & {}0.4669 & {}0.4163 &
{}0.4240 & \\\hline\hline
\end{tabular}
}
\end{center}

{\footnotesize \emph{{}Note}{}: }${\footnotesize \hat{\lambda}}_{hT}$
{\footnotesize is the estimate of the Lasso penalty parameter computed by
10-fold cross-validation.}%

\end{table}%
%

\begin{table}[H]%
%

\caption{Selected Variables by AR2-Lasso 4q Ahead}%

\begin{center}
{\footnotesize
\begin{tabular}
[c]{cccccccc}\hline\hline
{}\# &  & \multicolumn{6}{c}{{}Ending Year}\\\hline
&  & {}2019q1 & {}2019q2 & {}2019q3 & {}2019q4 & {}2020q1 & {}2020q2\\\hline
{}1 &  & {}DEMUK4 & {}DEMUK4 & {}DEMUK4 & {}DEMUK4 & {}DEMUK4 & {}DEMUK4\\
{}2 &  & {}DMUK4 & {}DMUK4 & {}DMUK4 & {}DMUK4 & {}DMUK4 & {}DMUK4\\
{}3 &  & {}DPMAT4 & {}DPMAT4 & {}DPMAT4 & {}DPMAT4 & {}DPMAT4 & {}DPMAT4\\
{}4 &  & {}DDPMUK4 & {}DDPMUK4 & {}DDPMUK4 & {}DDPMUK4 & {}DDPMUK4 &
{}DDPMUK4\\
{}5 &  & {}DPSUK4 & {}DEPUK4 & {}DEPUK4 & {}DEPUK4 & {}DEPUK4 & {}DEPUK4\\
{}6 &  & {}DDPSUK4 & {}DPSUK4 & {}DPSUK4 & {}DPSUK4 & {}DPSUK4 & {}DPSUK4\\
{}7 &  & {}DGAPSUK8 & {}DDPSUK4 & {}DDPSUK4 & {}DDPSUK4 & {}DDPSUK4 &
{}DDPSUK4\\
{}8 &  & {}$\cdot$ & {}DGAPSUK8 & {}DGAPSUK8 & {}DDYCHINA4 & {}DDYCHINA4 &
{}DDYCHINA4\\
{}9 &  & {}$\cdot$ & {}$\cdot$ & {}$\cdot$ & {}DGAPSUK8 & {}DGAPSUK8 &
{}DGAPSUK8\\\hline
{}$\hat{\lambda}_{hT}$ &  & {}0.2357 & {}0.2233 & {}0.2237 & {}0.1999 &
{}0.1998 & {}0.2014\\\hline
&  &  &  &  &  &  & \\\hline
&  & {}2020q3 & {}2020q4 & {}2021q1 & {}2021q2 & {}2021q3 & {}2021q4\\\hline
{}1 &  & {}DEMUK4 & {}DEMUK4 & {}DEMUK4 & {}DEMUK4 & {}DEMUK4 & {}DEMUK4\\
{}2 &  & {}DMUK4 & {}DMUK4 & {}DMUK4 & {}DMUK4 & {}DMUK4 & {}DMUK4\\
{}3 &  & {}DPMAT4 & {}DPMAT4 & {}DPMAT4 & {}DPMAT4 & {}DPMAT4 & {}DPMAT4\\
{}4 &  & {}DDPMUK4 & {}DDPMUK4 & {}DDPMUK4 & {}DDPMUK4 & {}DDPMUK4 &
{}DDPMUK4\\
{}5 &  & {}DEPUK4 & {}DEPUK4 & {}DEPUK4 & {}DPSUK4 & {}DPSUK4 & {}DPSUK4\\
{}6 &  & {}DPSUK4 & {}DPSUK4 & {}DPSUK4 & {}DDPSUK4 & {}DDPSUK4 & {}DDPSUK4\\
{}7 &  & {}DDPSUK4 & {}DDPSUK4 & {}DDPSUK4 & {}$\cdot$ & {}$\cdot$ & {}$\cdot
$\\
{}8 &  & {}DDYCHINA4 & {}DDYCHINA4 & {}DGAPSUK8 & {}$\cdot$ & {}$\cdot$ &
{}$\cdot$\\
9 &  & {}DGAPSUK8 & {}DGAPSUK8 & {}$\cdot$ & {}$\cdot$ & {}$\cdot$ & {}$\cdot
$\\\hline
$\hat{\lambda}_{hT}$ &  & {}0.1901 & {}0.1902 & {}0.1780 & {}0.2158 &
{}0.2170 & {}0.2106\\\hline
&  &  &  &  &  &  & \\\hline
&  & {}2022q1 & {}2022q2 & {}2022q3 & {}2022q4 & {}2023q1 & \\\hline
{}1 &  & {}DEMUK4 & {}DMUK4 & {}DMUK4 & {}DDUUK & {}DDUUK & \\
{}2 &  & {}DMUK4 & {}DPMAT4 & {}DPMAT4 & {}DMUK4 & {}DMUK4 & \\
{}3 &  & {}DPMAT4 & {}DDPMUK4 & {}DDPMUK4 & {}DPMAT4 & {}DPMAT4 & \\
{}4 &  & {}DDPMUK4 & {}DPSUK4 & {}DPSUK4 & {}DDPMUK4 & {}DDPMUK4 & \\
{}5 &  & {}DPSUK4 & {}DDPSUK4 & {}DDPSUK4 & {}DPSUK4 & {}DPSUK4 & \\
{}6 &  & {}DDPSUK4 & {}DDYSUK4 & {}DWUK4 & {}DDPSUK4 & {}DDPSUK4 & \\
{}7 &  & {}$\cdot$ & {}DWUK4 & {}$\cdot$ & {}DYUS4 & {}DWUK4 & \\
{}8 &  & {}$\cdot$ & {}$\cdot$ & {}$\cdot$ & {}DWUK4 & {}$\cdot$ & \\\hline
$\hat{\lambda}_{hT}$ &  & {}0.2735 & {}0.2513 & {}0.2736 & {}0.2493 &
{}0.2522 & \\\hline\hline
\end{tabular}
}
\end{center}

{\footnotesize \emph{{}Note}{}: }${\footnotesize \hat{\lambda}}_{hT}$
{\footnotesize is the estimate of the Lasso penalty parameter computed by
10-fold cross-validation.}%

\end{table}%
%

\begin{table}[H]%
%

\caption{Selected Variables by ARX-Lasso 4q Ahead}%

\begin{center}
{\footnotesize
\begin{tabular}
[c]{cccccccc}\hline\hline
{}\# &  & \multicolumn{6}{c}{{}Ending Year}\\\hline
&  & {}2019q1 & {}2019q2 & {}2019q3 & {}2019q4 & {}2020q1 & {}2020q2\\\hline
{}1 &  & {}DEMUK4 & {}DEMUK4 & {}DEMUK4 & {}DEMUK4 & {}DEMUK4 & {}DEMUK4\\
{}2 &  & {}DMUK4 & {}DMUK4 & {}DVUK4 & {}DMUK4 & {}DMUK4 & {}DVUK4\\
{}3 &  & {}DPMAT4 & {}DPMAT4 & {}DMUK4 & {}DPMAT4 & {}DPMAT4 & {}DMUK4\\
{}4 &  & {}DGAPSUK8 & {}DDPMUK4 & {}DPMAT4 & {}DDPMUK4 & {}DDPMUK4 &
{}DPMAT4\\
{}5 &  & {}$\cdot$ & {}DEPUK4 & {}DDPMUK4 & {}DEPUK4 & {}DEPUK4 & {}DDPMUK4\\
{}6 &  & {}$\cdot$ & {}DGAPSUK8 & {}DEPUK4 & {}DGAPSUK8 & {}DGAPSUK8 &
{}DEPUK4\\
{}7 &  & {}$\cdot$ & {}$\cdot$ & {}DGAPSUK8 & {}$\cdot$ & {}$\cdot$ &
{}DDYCHINA4\\
{}8 &  & {}$\cdot$ & {}$\cdot$ & {}$\cdot$ & {}$\cdot$ & {}$\cdot$ &
{}DGAPSUK8\\
{}9 &  & {}$\cdot$ & {}$\cdot$ & {}$\cdot$ & {}$\cdot$ & {}$\cdot$ & {}DPUS4\\
{}10 &  & {}$\cdot$ & {}$\cdot$ & {}$\cdot$ & {}$\cdot$ & {}$\cdot$ &
{}DDPUS4\\\hline
{}$\hat{\lambda}_{hT}$ &  & {}0.2349 & {}0.2238 & {}0.2020 & {}0.2139 &
{}0.2127 & {}0.1579\\\hline
&  &  &  &  &  &  & \\\hline
&  & {}2020q3 & {}2020q4 & {}2021q1 & {}2021q2 & {}2021q3 & {}2021q4\\\hline
{}1 &  & {}DEMUK4 & {}DEMUK4 & {}DEMUK4 & {}DEMUK4 & {}DEMUK4 & {}DEMUK4\\
{}2 &  & {}DVUK4 & {}DVUK4 & {}DVUK4 & {}DMUK4 & {}DVUK4 & {}DVUK4\\
{}3 &  & {}DMUK4 & {}DMUK4 & {}DMUK4 & {}DPMAT4 & {}DMUK4 & {}DMUK4\\
{}4 &  & {}DPMAT4 & {}DPMAT4 & {}DPMAT4 & {}DDPMUK4 & {}DPMAT4 & {}DPMAT4\\
{}5 &  & {}DDPMUK4 & {}DDPMUK4 & {}DDPMUK4 & {}$\cdot$ & {}DDPMUK4 &
{}DDPMUK4\\
{}6 &  & {}DEPUK4 & {}DEPUK4 & {}DEPUK4 & {}$\cdot$ & {}DEPUK4 & {}$\cdot$\\
{}7 &  & {}DDYCHINA4 & {}DDYCHINA4 & {}DGAPSUK8 & {}$\cdot$ & {}DPUS4 &
{}$\cdot$\\
{}8 &  & {}DGAPSUK8 & {}DGAPSUK8 & {}$\cdot$ & {}$\cdot$ & {}$\cdot$ &
{}$\cdot$\\
{}9 &  & {}DPUS4 & {}DPUS4 & {}$\cdot$ & {}$\cdot$ & {}$\cdot$ & {}$\cdot$\\
{}10 &  & {}DDPUS4 & {}DDPUS4 & {}$\cdot$ & {}$\cdot$ & {}$\cdot$ & {}$\cdot
$\\\hline
{}$\hat{\lambda}_{hT}$ &  & {}0.1471 & {}0.1583 & {}0.1923 & {}0.2166 &
{}0.1847 & {}0.1930\\\hline
&  &  &  &  &  &  & \\\hline
&  & {}2022q1 & {}2022q2 & {}2022q3 & {}2022q4 & {}2023q1 & \\\hline
{}1 &  & {}DEMUK4 & {}DEMUK4 & {}DMUK4 & {}DMUK4 & {}DMUK4 & \\
{}2 &  & {}DVUK4 & {}DMUK4 & {}DPMAT4 & {}DPMAT4 & {}DPMAT4 & \\
{}3 &  & {}DMUK4 & {}DPMAT4 & {}$\cdot$ & {}$\cdot$ & {}$\cdot$ & \\
{}4 &  & {}DPMAT4 & {}DDPMUK4 & {}$\cdot$ & {}$\cdot$ & {}$\cdot$ & \\
{}5 &  & {}DDPMUK4 & {}DDYSUK4 & {}$\cdot$ & {}$\cdot$ & {}$\cdot$ & \\\hline
{}$\hat{\lambda}_{hT}$ &  & {}0.2017 & {}0.2330 & {}0.2932 & {}0.2719 &
{}0.3251 & \\\hline\hline
\end{tabular}
}
\end{center}

{\footnotesize \emph{{}Note}{}: }${\footnotesize \hat{\lambda}}_{hT}$
{\footnotesize is the estimate of the Lasso penalty parameter computed by
10-fold cross-validation.}%

\end{table}%

\section{Realizations and Forecasts\label{Forecasts}}
\subsection{\textbf{h=1} quarter ahead forecasts\label{Forecasth=1}}%

\begin{table}[H]%

\caption{One Quarter Ahead Forecast Results}%

\begin{center}%
\begin{tabular}
[c]{ccccccccccc}\hline\hline
{\footnotesize Models} &  & {\footnotesize Actual} &  & {\footnotesize AR2} &
& {\footnotesize ARX} &  & {\footnotesize Lasso} &  &
{\footnotesize Lasso-AR2}\\\hline
{\footnotesize 2020q1} &  & {\footnotesize 1.6808} &  & {\footnotesize 1.4930}
&  & {\footnotesize 1.5005} &  & {\footnotesize 1.4411} &  &
{\footnotesize 1.1366}\\
{\footnotesize 2020q2} &  & {\footnotesize 0.7819} &  & {\footnotesize 1.9051}
&  & {\footnotesize 1.8708} &  & {\footnotesize 1.7274} &  &
{\footnotesize 1.2898}\\
{\footnotesize 2020q3} &  & {\footnotesize 0.7678} &  & {\footnotesize 0.7901}
&  & {\footnotesize 0.3358} &  & {\footnotesize --1.6184} &  &
{\footnotesize --1.7781}\\
{\footnotesize 2020q4} &  & {\footnotesize 0.7147} &  & {\footnotesize 1.0091}
&  & {\footnotesize 0.9193} &  & {\footnotesize 1.0883} &  &
{\footnotesize 0.9987}\\
{\footnotesize 2021q1} &  & {\footnotesize 0.9211} &  & {\footnotesize 0.9482}
&  & {\footnotesize 0.8014} &  & {\footnotesize 1.2458} &  &
{\footnotesize 1.0556}\\
{\footnotesize 2021q2} &  & {\footnotesize 2.0533} &  & {\footnotesize 1.1989}
&  & {\footnotesize 1.3852} &  & {\footnotesize 1.7410} &  &
{\footnotesize 1.4759}\\
{\footnotesize 2021q3} &  & {\footnotesize 2.6215} &  & {\footnotesize 2.4556}
&  & {\footnotesize 2.7557} &  & {\footnotesize 3.1343} &  &
{\footnotesize 3.0982}\\
{\footnotesize 2021q4} &  & {\footnotesize 4.3095} &  & {\footnotesize 2.8118}
&  & {\footnotesize 2.9337} &  & {\footnotesize 3.1241} &  &
{\footnotesize 3.0738}\\
{\footnotesize 2022q1} &  & {\footnotesize 5.3960} &  & {\footnotesize 4.6293}
&  & {\footnotesize 4.8813} &  & {\footnotesize 4.8406} &  &
{\footnotesize 5.0132}\\
{\footnotesize 2022q2} &  & {\footnotesize 7.6259} &  & {\footnotesize 5.4458}
&  & {\footnotesize 5.7835} &  & {\footnotesize 5.8914} &  &
{\footnotesize 6.0747}\\
{\footnotesize 2022q3} &  & {\footnotesize 8.3877} &  & {\footnotesize 7.8385}
&  & {\footnotesize 8.1711} &  & {\footnotesize 8.0610} &  &
{\footnotesize 8.3580}\\
{\footnotesize 2022q4} &  & {\footnotesize 8.9856} &  & {\footnotesize 8.0723}
&  & {\footnotesize 8.2705} &  & {\footnotesize 8.6325} &  &
{\footnotesize 8.7409}\\
{\footnotesize 2023q1} &  & {\footnotesize 8.6111} &  & {\footnotesize 8.5777}
&  & {\footnotesize 8.3988} &  & {\footnotesize 8.7458} &  &
{\footnotesize 9.0455}\\
{\footnotesize 2023q2} &  &  &  & {\footnotesize 7.9164} &  &
{\footnotesize 7.2945} &  & {\footnotesize 7.6445} &  & {\footnotesize 7.8217}%
\\
{\footnotesize RMSFE} &  &  &  & {\footnotesize 0.9141} &  &
{\footnotesize 0.7884} &  & {\footnotesize 0.9696} &  & {\footnotesize 0.9617}%
\\\hline
&  & {\footnotesize Lasso-ARX} &  & {\footnotesize OCMT-AR2} &  &
{\footnotesize OCMT-ARX} &  & {\footnotesize OCMT-AR2} &  &
{\footnotesize OCMT-ARX}\\
&  &  &  & ${\footnotesize \delta=1}$ &  & ${\footnotesize \delta=1}$ &  &
${\footnotesize \delta=1.5}$ &  & ${\footnotesize \delta=1.5}$\\\hline
{\footnotesize 2020q1} &  & {\footnotesize 1.3148} &  & {\footnotesize 1.4930}
&  & {\footnotesize 1.5005} &  & {\footnotesize 1.4930} &  &
{\footnotesize 1.5005}\\
{\footnotesize 2020q2} &  & {\footnotesize 1.8793} &  & {\footnotesize 1.9051}
&  & {\footnotesize 1.8708} &  & {\footnotesize 1.9051} &  &
{\footnotesize 1.8708}\\
{\footnotesize 2020q3} &  & {\footnotesize --0.9434} &  &
{\footnotesize 0.7901} &  & {\footnotesize 0.3358} &  & {\footnotesize 0.7901}
&  & {\footnotesize 0.3358}\\
{\footnotesize 2020q4} &  & {\footnotesize 1.4325} &  & {\footnotesize 1.0091}
&  & {\footnotesize 0.9193} &  & {\footnotesize 1.0091} &  &
{\footnotesize 0.9193}\\
{\footnotesize 2021q1} &  & {\footnotesize 1.1040} &  & {\footnotesize 0.9482}
&  & {\footnotesize 0.8014} &  & {\footnotesize 0.9482} &  &
{\footnotesize 0.8014}\\
{\footnotesize 2021q2} &  & {\footnotesize 1.7652} &  & {\footnotesize 1.1989}
&  & {\footnotesize 1.3852} &  & {\footnotesize 1.1989} &  &
{\footnotesize 1.3852}\\
{\footnotesize 2021q3} &  & {\footnotesize 2.6691} &  & {\footnotesize 3.0457}
&  & {\footnotesize 2.7557} &  & {\footnotesize 2.4556} &  &
{\footnotesize 2.7557}\\
{\footnotesize 2021q4} &  & {\footnotesize 3.0872} &  & {\footnotesize 2.8118}
&  & {\footnotesize 2.9337} &  & {\footnotesize 2.8118} &  &
{\footnotesize 2.9337}\\
{\footnotesize 2022q1} &  & {\footnotesize 4.9254} &  & {\footnotesize 4.5468}
&  & {\footnotesize 4.8813} &  & {\footnotesize 4.6293} &  &
{\footnotesize 4.8813}\\
{\footnotesize 2022q2} &  & {\footnotesize 5.9782} &  & {\footnotesize 5.4612}
&  & {\footnotesize 5.7835} &  & {\footnotesize 5.4458} &  &
{\footnotesize 5.7835}\\
{\footnotesize 2022q3} &  & {\footnotesize 8.6872} &  & {\footnotesize 7.8385}
&  & {\footnotesize 8.1711} &  & {\footnotesize 7.8385} &  &
{\footnotesize 8.1711}\\
{\footnotesize 2022q4} &  & {\footnotesize 9.1017} &  & {\footnotesize 8.0723}
&  & {\footnotesize 8.2705} &  & {\footnotesize 8.0723} &  &
{\footnotesize 8.2705}\\
{\footnotesize 2023q1} &  & {\footnotesize 9.3315} &  & {\footnotesize 8.5777}
&  & {\footnotesize 8.3988} &  & {\footnotesize 8.5777} &  &
{\footnotesize 8.3988}\\
{\footnotesize 2023q2} &  & {\footnotesize 7.6670} &  & {\footnotesize 7.9164}
&  & {\footnotesize 7.2945} &  & {\footnotesize 7.9164} &  &
{\footnotesize 7.2945}\\
{\footnotesize RMSFE} &  & {\footnotesize 0.8750} &  & {\footnotesize 0.9233}
&  & {\footnotesize 0.7884} &  & {\footnotesize 0.9141} &  &
{\footnotesize 0.7884}\\\hline\hline
\end{tabular}

\end{center}

{\footnotesize { \emph{Notes}: }The "Actual" column gives the realized
inflation rate for quarter }$t,$ {\footnotesize the other columns the
forecasts for each model. RMSFE is the Root Mean Square Forecast Error. Lasso
models are estimated using the code of Chudik, Kapetanios, and Pesaran
(2018).}%

\end{table}%

\begin{center}%
\begin{figure}[ph]%
\centering

\includegraphics[height=4.7049in,width=6.262in]{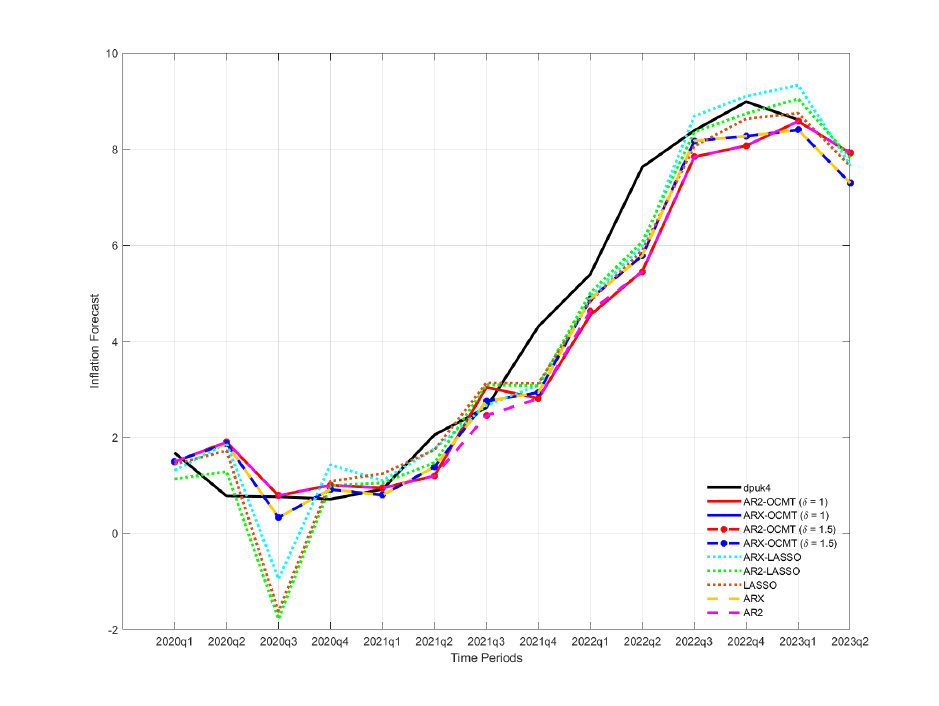}

\caption{\textbf{One Quarter Ahead Forecast Plots}}%
\end{figure}

\end{center}

\subsection{\textbf{h=2} quarters ahead forecasts\label{Forecasth=2}}%

\begin{table}[H]%
%

\caption{Two Quarters Ahead Forecast Results}%

\begin{center}
{\footnotesize
\begin{tabular}
[c]{ccccccccccc}\hline\hline
Models &  & Actual &  & AR2 &  & ARX &  & Lasso &  & Lasso-AR2\\\hline
2020q1 &  & 1.6808 &  & 1.9854 &  & 1.6773 &  & 1.7494 &  & 1.8994\\
2020q2 &  & 0.7819 &  & 1.6115 &  & 1.5979 &  & 1.4380 &  & 0.8428\\
2020q3 &  & 0.7678 &  & 2.0705 &  & 1.9977 &  & 1.8217 &  & 2.0103\\
2020q4 &  & 0.7147 &  & 0.8941 &  & 0.1859 &  & 0.5350 &  & --4.7147\\
2021q1 &  & 0.9211 &  & 1.2182 &  & 1.0423 &  & 1.3422 &  & 7.7188\\
2021q2 &  & 2.0533 &  & 1.1590 &  & 0.9049 &  & 1.5683 &  & 1.6895\\
2021q3 &  & 2.6215 &  & 1.4199 &  & 1.6725 &  & 1.8149 &  & 1.9367\\
2021q4 &  & 4.3095 &  & 2.6839 &  & 3.0931 &  & 3.6585 &  & 2.0096\\
2022q1 &  & 5.3960 &  & 2.9218 &  & 3.1026 &  & 3.3691 &  & 2.2883\\
2022q2 &  & 7.6259 &  & 4.7324 &  & 5.1079 &  & 4.7079 &  & 4.9191\\
2022q3 &  & 8.3877 &  & 5.3660 &  & 5.8973 &  & 5.6290 &  & 5.9572\\
2022q4 &  & 8.9856 &  & 7.7515 &  & 8.2645 &  & 7.4422 &  & 7.9237\\
2023q1 &  & 8.6111 &  & 7.7117 &  & 8.0551 &  & 8.1659 &  & 8.4389\\
2023q2 &  &  &  & 8.1422 &  & 7.8532 &  & 7.9158 &  & 7.4686\\
2023q3 &  &  &  & 7.3340 &  & 6.2890 &  & 6.1918 &  & 5.7513\\
RMSFE &  &  &  & 1.6039 &  & 1.3813 &  & 1.4109 &  & 2.8719\\\hline
&  & Lasso-ARX &  & OCMT-AR2 &  & OCMT-ARX &  & OCMT-AR2 &  & OCMT-ARX\\
&  &  &  & $\delta=1$ &  & $\delta=1$ &  & $\delta=1.5$ &  & $\delta
=1.5$\\\hline
2020q1 &  & 1.8915 &  & 1.7638 &  & 1.4778 &  & 1.9854 &  & 1.6773\\
2020q2 &  & 1.0312 &  & 1.4532 &  & 1.5288 &  & 1.6115 &  & 1.5979\\
2020q3 &  & 2.2963 &  & 1.9038 &  & 1.8882 &  & 2.0705 &  & 1.9977\\
2020q4 &  & --3.5975 &  & 1.1311 &  & 0.3795 &  & 0.8941 &  & 0.1859\\
2021q1 &  & 8.4157 &  & 1.6797 &  & 1.5628 &  & 1.2182 &  & 1.0423\\
2021q2 &  & 2.1084 &  & 1.5683 &  & 1.3765 &  & 1.6344 &  & 1.3765\\
2021q3 &  & 1.9160 &  & 1.8149 &  & 2.1371 &  & 1.8554 &  & 2.1371\\
2021q4 &  & 1.9221 &  & 3.3262 &  & 3.3351 &  & 3.5873 &  & 3.3351\\
2022q1 &  & 2.2122 &  & 3.0279 &  & 3.0481 &  & 3.1239 &  & 3.0481\\
2022q2 &  & 4.0165 &  & 4.4689 &  & 4.9381 &  & 4.4676 &  & 4.9381\\
2022q3 &  & 6.3907 &  & 5.2246 &  & 5.6954 &  & 5.2253 &  & 5.6954\\
2022q4 &  & 8.4832 &  & 7.0244 &  & 7.8871 &  & 6.9260 &  & 7.8871\\
2023q1 &  & 9.1288 &  & 6.8038 &  & 7.5799 &  & 6.9298 &  & 8.0551\\
2023q2 &  & 8.0515 &  & 7.2864 &  & 7.4267 &  & 8.1422 &  & 7.8532\\
2023q3 &  & 5.8963 &  & 6.3556 &  & 5.6673 &  & 7.3340 &  & 6.2890\\
RMSFE &  & 2.9231 &  & 1.6880 &  & 1.4217 &  & 1.6662 &  &
1.4037\\\hline\hline
\end{tabular}
}
\end{center}

{\footnotesize \emph{{}}{\emph{Notes}: }The "Actual" column gives the realized
inflation rate at period }$t,$ {\footnotesize the other columns the forecasts
for each model. RMSFE is the Root Mean Square Forecast Error. Lasso models are
estimated using the code of Chudik, Kapetanios, and Pesaran (2018).}%

\end{table}%

\begin{center}%
\begin{figure}[ph]%
\centering

\includegraphics[height=4.7046in,width=6.2621in]{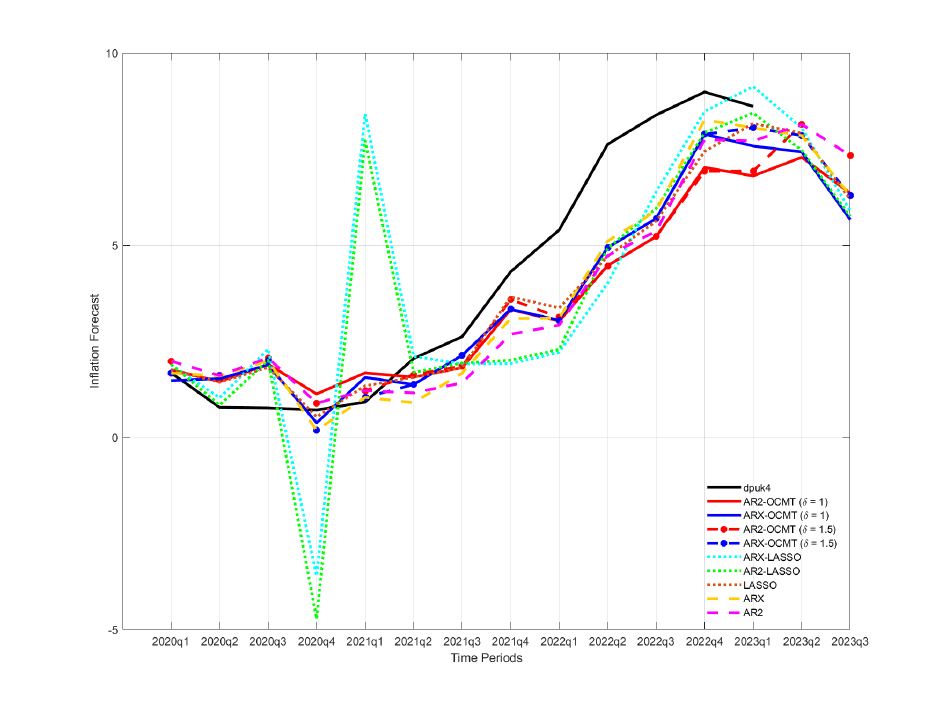}

\caption{\textbf{Two Quarters Ahead Forecast Plots}}%
\end{figure}

\end{center}

\subsection{\textbf{h=4} quarters ahead forecasts\label{Forecasth=4}}%

\begin{table}[H]%
%

\caption{Four Quarters Ahead Forecast Results}%

\begin{center}
{\footnotesize
\begin{tabular}
[c]{ccccccccccc}\hline\hline
Models &  & Actual &  & AR2 &  & ARX &  & Lasso &  & Lasso-AR2\\\hline
2020q1 &  & 1.6808 &  & 2.3062 &  & 1.6093 &  & 1.6462 &  & 1.8149\\
2020q2 &  & 0.7819 &  & 2.3417 &  & 2.2183 &  & 1.8204 &  & 1.9359\\
2020q3 &  & 0.7678 &  & 2.2620 &  & 1.7780 &  & 1.5496 &  & 1.4110\\
2020q4 &  & 0.7147 &  & 2.0491 &  & 2.0380 &  & 1.5778 &  & 0.8652\\
2021q1 &  & 0.9211 &  & 2.1876 &  & 2.0822 &  & 1.5795 &  & 1.4660\\
2021q2 &  & 2.0533 &  & 1.6754 &  & 0.5406 &  & 0.5873 &  & --7.2149\\
2021q3 &  & 2.6215 &  & 1.6277 &  & 1.3694 &  & 2.0021 &  & 8.3843\\
2021q4 &  & 4.3095 &  & 1.5822 &  & 1.1737 &  & 2.3696 &  & 1.1065\\
2022q1 &  & 5.3960 &  & 1.6856 &  & 2.1203 &  & 2.7836 &  & 1.4914\\
2022q2 &  & 7.6259 &  & 2.3353 &  & 2.9872 &  & 3.3813 &  & 2.8290\\
2022q3 &  & 8.3877 &  & 2.6974 &  & 2.9742 &  & 2.5152 &  & 2.8806\\
2022q4 &  & 8.9856 &  & 3.6744 &  & 4.2203 &  & 3.4427 &  & 3.5490\\
2023q1 &  & 8.6111 &  & 4.3493 &  & 5.1543 &  & 4.2743 &  & 4.6459\\
2023q2 &  &  &  & 5.7823 &  & 6.6329 &  & 4.7789 &  & 5.1198\\
2023q3 &  &  &  & 6.2075 &  & 6.7753 &  & 4.7462 &  & 5.4047\\
2023q4 &  &  &  & 6.6574 &  & 6.2714 &  & 4.8914 &  & 4.0569\\
2024q1 &  &  &  & 6.3149 &  & 4.6141 &  & 3.7130 &  & 2.6772\\
RMSFE &  &  &  & 3.2524 &  & 2.9883 &  & 3.0131 &  & 4.3440\\\hline
&  & Lasso-ARX &  & OCMT-AR2 &  & OCMT-ARX &  & OCMT-AR2 &  & OCMT-ARX\\
&  &  &  & $\delta=1$ &  & $\delta=1$ &  & $\delta=1.5$ &  & $\delta
=1.5$\\\hline
2020q1 &  & 1.7270 &  & 1.6437 &  & 1.3401 &  & 1.7873 &  & 1.3401\\
2020q2 &  & 1.9359 &  & 1.6977 &  & 1.9417 &  & 1.8175 &  & 1.9417\\
2020q3 &  & 1.6059 &  & 1.5965 &  & 1.4484 &  & 1.7514 &  & 1.4484\\
2020q4 &  & 0.9909 &  & 1.6540 &  & 1.9467 &  & 1.7858 &  & 1.9467\\
2021q1 &  & --0.1663 &  & 1.7455 &  & 1.7464 &  & 1.8576 &  & 1.7464\\
2021q2 &  & --4.6085 &  & 1.7584 &  & 0.4562 &  & 2.0774 &  & 0.4562\\
2021q3 &  & 8.6921 &  & 2.2141 &  & 1.3667 &  & 2.5984 &  & 1.3667\\
2021q4 &  & 1.5211 &  & 2.0870 &  & 0.7506 &  & 2.4724 &  & 0.7506\\
2022q1 &  & 1.8425 &  & 2.2670 &  & 1.5652 &  & 2.5388 &  & 1.5652\\
2022q2 &  & 2.8290 &  & 8.3146 &  & 2.4832 &  & 3.6200 &  & 2.4832\\
2022q3 &  & 0.6644 &  & 3.5159 &  & 2.7159 &  & 2.9876 &  & 2.7159\\
2022q4 &  & 2.1661 &  & 4.2029 &  & 3.7504 &  & 3.1511 &  & 3.7504\\
2023q1 &  & 3.1894 &  & 4.8976 &  & 4.6815 &  & 4.0785 &  & 4.7542\\
2023q2 &  & 4.9977 &  & 4.5467 &  & 5.7679 &  & 4.4758 &  & 5.7679\\
2023q3 &  & 5.4954 &  & 4.6832 &  & 5.6688 &  & 4.6143 &  & 5.6688\\
2023q4 &  & 4.9713 &  & 4.7016 &  & 5.2734 &  & 4.9634 &  & 5.2734\\
2024q1 &  & 2.9576 &  & 3.6366 &  & 3.1912 &  & 4.2441 &  & 3.1912\\
RMSFE &  & 4.5021 &  & 2.4643 &  & 3.2470 &  & 2.9795 &  &
3.2402\\\hline\hline
\end{tabular}
}
\end{center}

{\footnotesize \emph{{}}{\emph{Notes}: }The "Actual" column gives the realized
inflation rate at period }$t,$ {\footnotesize the other columns the forecasts
for each model. RMSFE is the Root Mean Square Forecast Error. Lasso models are
estimated using the code of Chudik, Kapetanios, and Pesaran (2018).}%

\end{table}%

\begin{center}%
\begin{figure}[ph]%
\centering

\includegraphics[height=4.7046in,width=6.2612in]{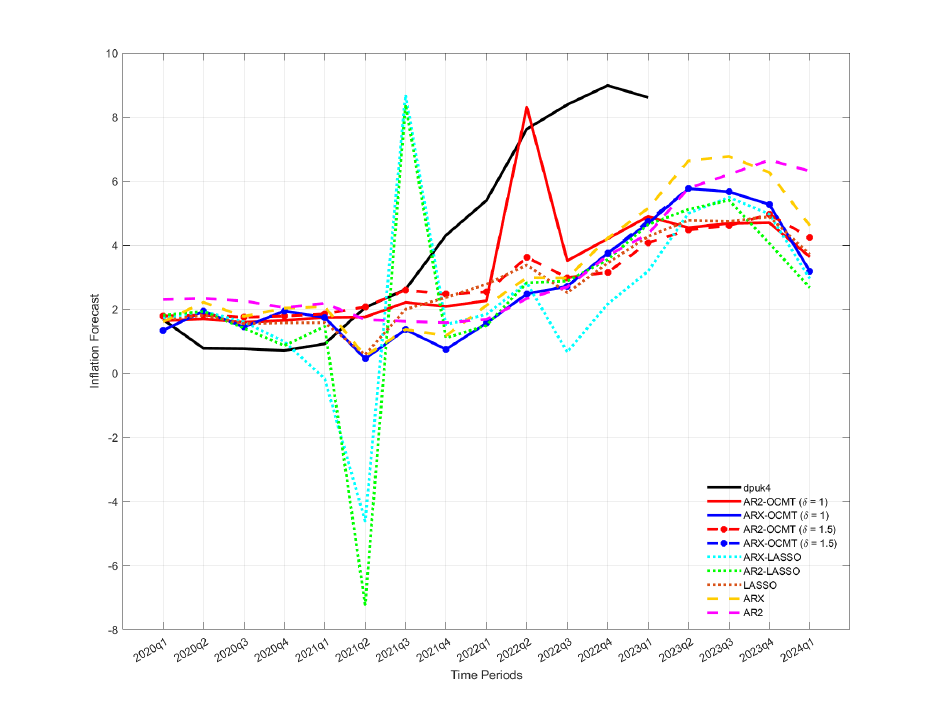}

\caption{\textbf{Four Quarters Ahead Forecast Plots}}%
\end{figure}

\end{center}

\section*{References}%

{\footnotesize Chudik, Kapetanios, and Pesaran (2018) \textquotedblleft A One
Covariate at a Time, Multiple Testing Approach to Variable Selection in
High-Dimensional Linear Regression Models,\textquotedblright%
\ \emph{Econometrica}, Vol. 86, No. 4, 1479-1512. }

\end{document}